\documentclass[useAMS,usenatbib]{mn2e}

\usepackage{graphicx}
\usepackage{amsmath}
\usepackage{url}

\title[Absorption effects in blazar's $\gamma$-ray spectra]
{Absorption effects in the blazar's $\gamma$-ray spectra due to luminous stars crossing the jet}
\author[W. Bednarek \& J. Sitarek]
{W. Bednarek \& J. Sitarek\\ 
University of \L \'od\'z, Department of Astrophysics, Faculty of Physics and Applied Informatics,
ul. Pomorska 149/153, 90-236 \L \'od\'z, Poland,\\
wlodzimierz.bednarek@uni.lodz.pl; julian.sitarek@uni.lodz.pl\\}
\begin{document}

\date{Accepted . Received ; in original form }

\pagerange{\pageref{firstpage}--\pageref{lastpage}} \pubyear{2015}

\maketitle

\label{firstpage}

\begin{abstract}
$\gamma$-ray emission in active galaxies is likely produced within the inner jet, or in the close vicinity
of the supermassive black hole (SMBH) at sub-parsec distances.  $\gamma$ rays have to pass
through the surrounding massive stellar cluster which luminous stars can accidentally appear
close to the observer's line of sight. In such a case, soft radiation of massive stars can
create enough target for transient absorption of the $\gamma$ rays in multi-GeV to TeV energy range. 
We consider the effect of such stellar encounters on the $\gamma$-ray spectrum produced 
within the massive stellar cluster surrounding a central SMBH. We predict characteristic, time-dependent
effects on the $\gamma$-ray spectra due to the encounter with the single luminous star and also stellar
binary system. We conclude that during the encounter, the $\gamma$-ray spectrum of an active galaxy should
steepen at tens of GeV and harden in the range of hundreds of GeV. 
As an example, we consider such effects
on the spectra observed from a typical blazar, 1ES\ 1959+650 (in an active state) and also in the case of 
a radio galaxy M87 (in a low state). It is shown that observation of such transient characteristic features 
in the $\gamma$-ray spectra, observed from blazars and radio galaxies, lays within the sensitivity of 
the future Cherenkov Telescope Array.    
\end{abstract}

\section{Introduction}

It is supposed that high-energy $\gamma$ rays from active galaxies 
are produced within the inner jet or in 
the magnetosphere of the supermassive black hole (SMBH) due to their extremely short variability time
scale. In fact, in the case of some blazars, the TeV $\gamma$-ray emission can change 
significantly on a time scale as short as minutes (e.g. Mrk~501, see Albert et al.~2007; 
PKS~2155-304, see Aharonian et al.~2007 or IC\,310, see Aleksic et al. 2014),
suggesting that the emission site has to be close to the SMBH, even if a mild
relativistic boosting is taken into account. Also theoretical models for the high-energy $\gamma$-ray emission 
usually locate relativistically moving emission region at sub-parsec distances from the SMBH in order to 
correctly describe the spectral features (e.g. Bednarek \& Protheroe 1997a, Tavecchio et al.~1998). 
Collimated $\gamma$-ray emission (a $\gamma$-ray beam) has to propagate through the surrounding region of the SMBH
in which many compact objects are expected. The angular extend of such a beam
in the blazar type of active galaxies, $\alpha$, is believed to be determined by the Lorentz factor of the inner 
jet which for 
$\gamma_{\rm b}\sim 10$ is $\alpha \sim 1/\gamma_{\rm b}\sim 0.1$ rad. However, in the case of 
TeV $\gamma$-ray radio galaxies, the beams have to be much broader, or the jet has to be composed of a faster, 
more compact, and slower, more extend, part, since the 
inclination angles of such galaxies, in respect to the observer's line of sight, are typically much 
larger than $\sim 0.1$ rad. In fact, the inner radio jets in 
these sources are inclined at large angles to 
the line of sight, e.g. in Cen~A $i\sim (50-80)^\circ$ (Tingay et al.~2001) and $i\sim (12-45)^\circ$ (M\"uller et al.~2014),
in NGC 1275 $i\sim (30 - 55)^\circ$ 
(Walker et al.~1994, Vermeulen et al.~1994) or $i\sim (65\pm 16)^\circ$ (Fujita \& Nagai~2017), 
and in M87 $i\sim (10-19)^\circ$ (Biretta et al.~1999). Therefore, their TeV $\gamma$-ray beam 
has to be much wider.

\begin{figure}
\vskip 5.5truecm
\includegraphics{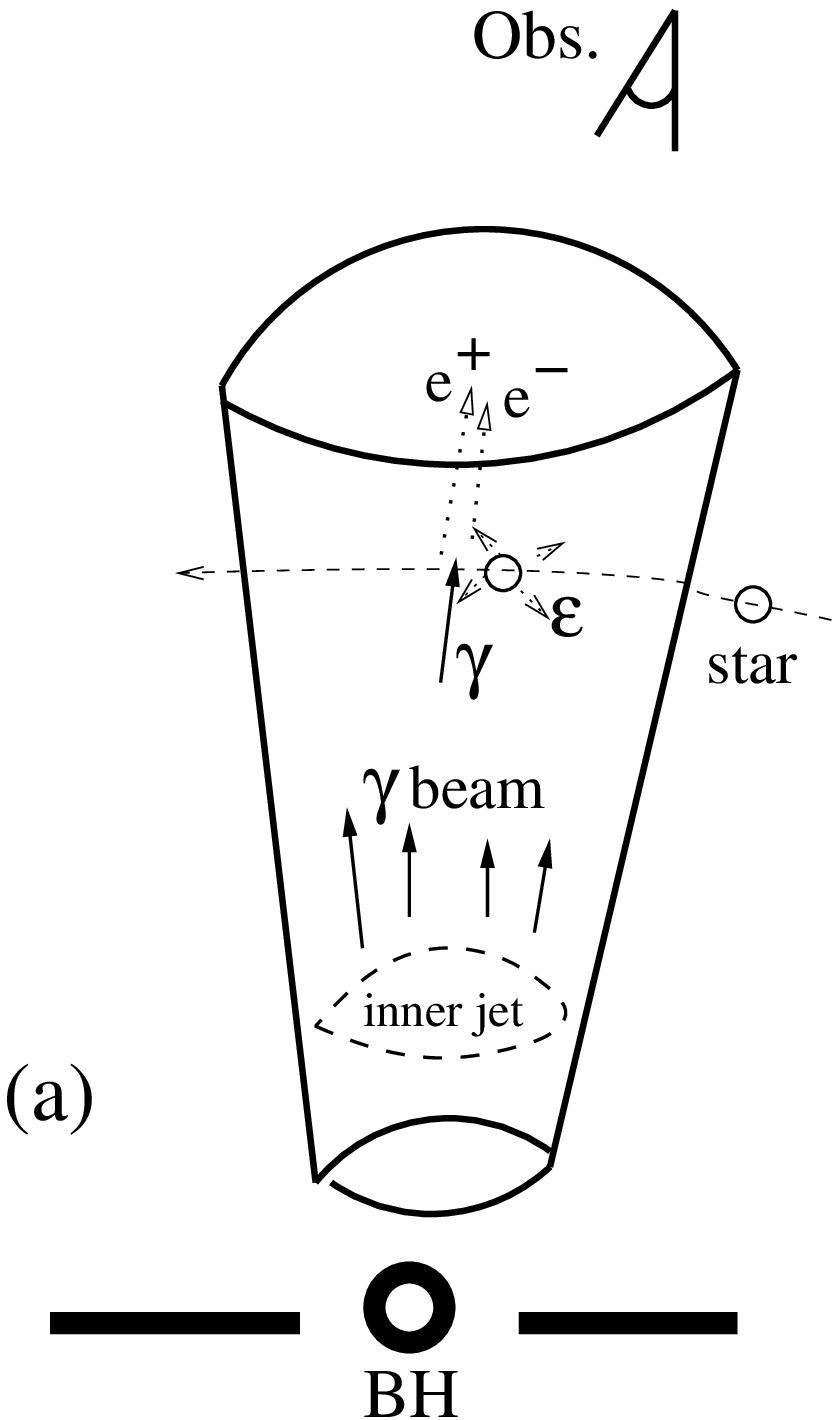}
\includegraphics{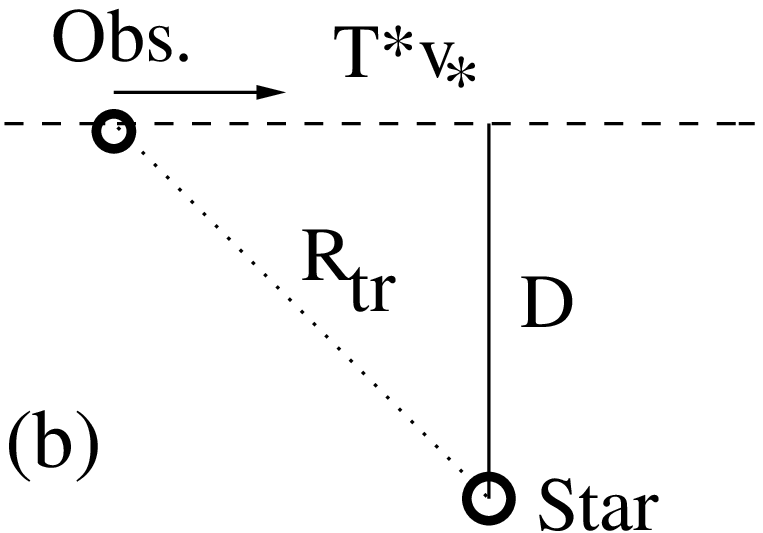}
\caption{Schematic representation of the considered scenario for the production of specific absorption 
feature in the spectrum of GeV-TeV emitting active galaxies as seen in the active galaxy (figure a) and 
in the star reference frames (b). In (a): a $\gamma$-ray beam is produced in the inner part of the jet. 
It meets on its way the soft radiation of a luminous star which entered the jet. $\gamma$ rays passing 
close by to the star are absorbed in soft radiation ($\gamma + \varepsilon\rightarrow e^+e^-$). 
As a result the absorption feature should appear in the broad band $\gamma$-ray spectrum produced in 
the inner jet. In (b): The observer's (Obs) line of sight moves along the straight (dashed) line with 
the impact parameter $D$. At the time $T$ (measured from the closest approach) the distance between 
the star and the observer is $R_{\rm tr}$. The Observer location is defined by the product of the time $T$ 
and the velocity of star $v_\star$. The direction of the $\gamma$-ray beam is perpendicular to the plane of 
the drawing. 
}
\label{fig1}
\end{figure}

We propose that the beam of energetic $\gamma$ rays, produced in the central regions of active galactic nuclei 
(AGNs), encounters luminous stars from time to time. Those stars form a quasi-spherical halo around 
the central super-massive black hole (see Fig.~\ref{fig1}). 
As already noted in Bednarek et al. (2016), the current star formation rate in the radio galaxy Cen~A 
($\sim$2 M$_\odot$ yr$^{-1}$) is responsible for the formation  of 
$(6 -12)\times 10^7$ M$_\odot$ of young stars. 
This star formation rate is expected to provide $\sim$3$\times 10^5$ stars with masses above 20 M$_\odot$ (Wykes et al. (2014). 
Significant part of these stars should cross once in a while the jet pointing towards the observer.
For the jet with simple conical geometry, the number of stars within the jet is estimated to be 
of the order of $\sim$10$^4\theta_{20}^{-2}$, where the jet half opening angle is normalized to 
$\theta = 20\,\theta_{20}$ degrees.
Therefore, the interaction of the $\gamma$-ray beam, produced in the central part of the active galaxy
with the radiation field of luminous stars around an active galaxy seems to be possible.

The number of transiting events clearly depend on the opening angle of the $\gamma$-ray beam from 
the vicinity of SMBH.
The theoretical modelling of the $\gamma$-ray emission from the blazar type active galaxies usually postulate 
that it is 
collimated along the direction of a relativistic conical jet within the angle which is of the order of 
the inverse 
of the Lorentz factor of the jet (typically close to ten). However, it may not always be the case.
The degree of collimation of the $\gamma$-ray emission is expected 
to be lower in the case of jets with other geometry (e.g. parabolic) or jets showing significant curvature, 
e.g. head-tail radio galaxies. 
Moreover, the off-axis $\gamma$-ray emission in radio galaxies is expected in the presence 
of two emission regions which exchange the radiation field (see Georganopoulos et al.~2005 and Banasi\'nski \& Bednarek~2018)
and also in the inverse Compton (IC)
e$^\pm$-pair cascade models in which the ordered magnetic field can re-distribute directions of 
secondary $\gamma$-rays outside the opening angle of the 
jet (e.g. Sitarek \& Bednarek 2010, or Roustazadeh \& B\"ottcher 2011).
Off-axis TeV $\gamma$-ray emission is also expected to occur in the magnetosphere of the rotating super massive black hole (SMBH) 
(e.g. Rieger \& Mannheim 2002, Neronov \& Aharonian 2007, Rieger \& Aharonian 2008, Levinson \& Rieger 2011,
Aleksi\'c et al. 2014).

A class of scenarios for the high energy processes, in which jets of active galaxies collide with 
compact objects, have been recently studied in detail. Different types of stellar objects can enter the jet 
from the massive stellar cluster surrounding SMBH (e.g. stars, clouds or even globular clusters, fragments of 
supernova remnants or pulsar wind nebulae). Collisions of the jet plasma with above-mentioned compact 
objects provide good conditions for acceleration of particles which can be responsible for the high-energy 
$\gamma$-ray emission 
(see e.g. Bednarek \& Protheroe~1997b, Barkov et al.~2010, Bosch-Ramon et al.~2012, Araudo et al.~2013,
Wykes et al.~2014, Bednarek \& Banasi\'nski~2015, Bosch-Ramon~2015, de la Cita et al.~2016, 
Banasi\'nski et al.~2016, Vieyro et al.~2017, Torres-Alba et al.~2019).

We consider another aspect of the general encounter scenario between the $\gamma$-ray beam and luminous stars.
It is assumed that the collimated beam of $\gamma$ rays is already produced in the close vicinity of SMBH
in one of the scenarios mentioned above. This $\gamma$-ray beam occasionally encounters compact 
luminous stars which pass close to the observer's line of sight.
As a result, a transient, broad feature is expected to 
appear in the continuum $\gamma$-ray spectrum due to the partial absorption of those $\gamma$ rays
in the thermal radiation of the star. Such transient feature is expected in the multi-GeV to 
sub-TeV $\gamma$-ray energies. It manifests itself as a 
hardening of the spectrum below $\sim$TeV energies and as a softening of $\gamma$-ray spectrum above
$\sim$10 GeV.
We propose that such effects can be transiently detected in the sub-TeV $\gamma$-ray 
spectra observed from active galaxies by Cherenkov telescopes (e.g. future Cherenkov Telescope Array, CTA).  

Note that similar, but persistent absorption features in the multi-GeV $\gamma$-ray spectra 
of some optically violently variable type blazars, are also expected to appear in the case of very strong 
soft radiation field 
formed by the broad emission line regions around SMBHs (e.g. Poutanen \& Stern 2010, Stern \& Poutanen 2011,Abolmasov \& Poutanen 2017). 
They might also effect the TeV $\gamma$-ray emission from BL Lac type active galaxies due to the absorption
of $\gamma$ rays in the infra-red radiation of molecular toruses (Protheroe \&  Biermann~1997, 
Donea \& Protheroe~2003).

\section{Massive stars through the $\gamma$ beam}

We assume that the $\gamma$-ray production region in the vicinity of the SMBH (either the inner
jet or the black hole
magnetosphere) is surrounded by a young massive cluster of stars (see Fig.~1). 
The $\gamma$-ray spectra from active galaxies are at first order well described by a simple power-law, or 
a log-parabola function through the GeV-TeV energy range. It is likely that
some luminous stars pass close to the line of sight of a distant observer with a given impact parameter 
$D$, defined as the shortest distance between the line of sight and the centre of the star.
During such passage, the dense radiation field of the star can partly absorb $\gamma$ rays in a specific 
energy range. 
Since the stars move around SMBH with a significant velocity, the absorption effects should have transient 
nature.  

We scale the typical parameters of the early type luminous stars with the surface temperature 
$T_\star\sim 3\times 10^4T_{4.5}$~K and the stellar radius $R_\star\sim 10^{12}R_{12}$~cm. 
Then, the optical depth is the largest for the $\gamma$-ray photons with energies, 
$E_\gamma\sim 2m_{\rm e}^2/\varepsilon\sim 67/T_{4.5}$~{\rm GeV} in the case of head on collisions. 
It can be approximated by 
\begin{eqnarray}
\tau_{\gamma-\gamma}(r)\sim D n_{\rm ph} \sigma_{\gamma-\gamma}\sim 
110 R_{12}T_{4.5}^3/r,
\label{eq1}
\end{eqnarray} 
\noindent
where $\sigma_{\gamma-\gamma}$ is the cross section for $e^\pm$ pair production in collision of two photons,
$n_{\rm ph}$ is the density of stellar photons 
and $r = D/R_\star$. This rough estimation shows that
$\gamma$ rays moving at significant distances from the stellar surface can be absorbed.

The duration of the absorption effect on the $\gamma$-ray spectrum can be estimated for
the known velocity of the luminous star on a circular orbit around the supermassive black hole. 
This velocity depends on the distance, $R = 1R_{1}$~pc, of the star from the SMBH. It is given by,
\begin{eqnarray}
v_\star\approx 9.4\times 10^7 (M_8/R_{1})^{1/2}~~~{\rm cm~s^{-1}}.
\label{eq2}
\end{eqnarray}
\noindent
where $M_{\rm SMBH} = 10^8M_8$~$M_\odot$ is the mass of the SMBH. 
Then, the characteristic minimum time scale of 
the stellar transit is of the order of  $\sim D/v_\star\sim 3rR_{12}(R_{1}/M_8)^{1/2}$~hours.
Therefore, we expect that the $\gamma$-ray spectrum observed from active galaxies
can be sporadically significantly modified in the multi-GeV to sub-TeV $\gamma$-ray energy range.  
We investigate the details of such effects on the $\gamma$-ray spectrum in the case of  
a transition of either a single or a binary system of two luminous stars close to the line of sight of 
a distant observer.   

\begin{figure*}
\vskip 4.truecm
\includegraphics{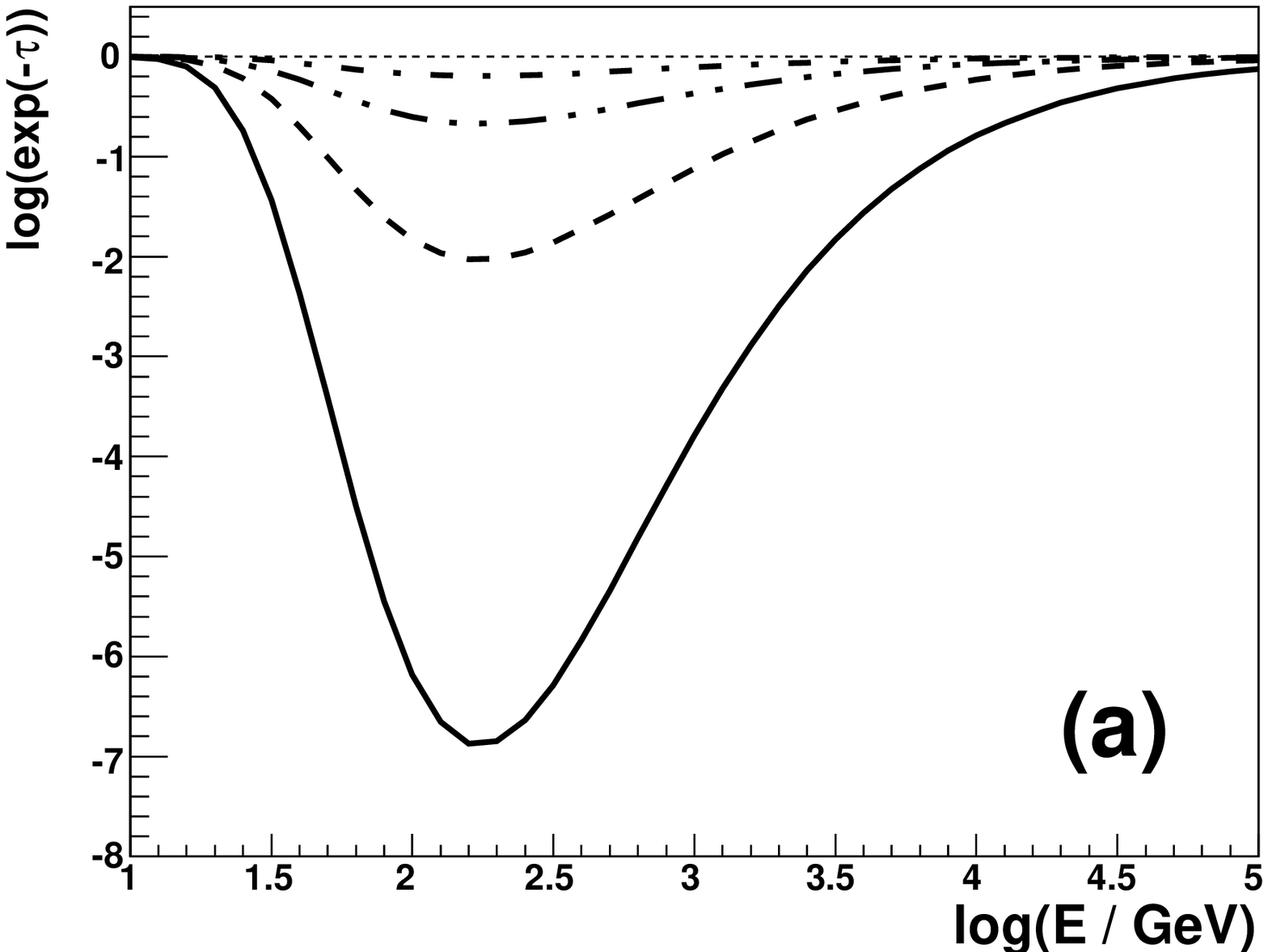}
\includegraphics{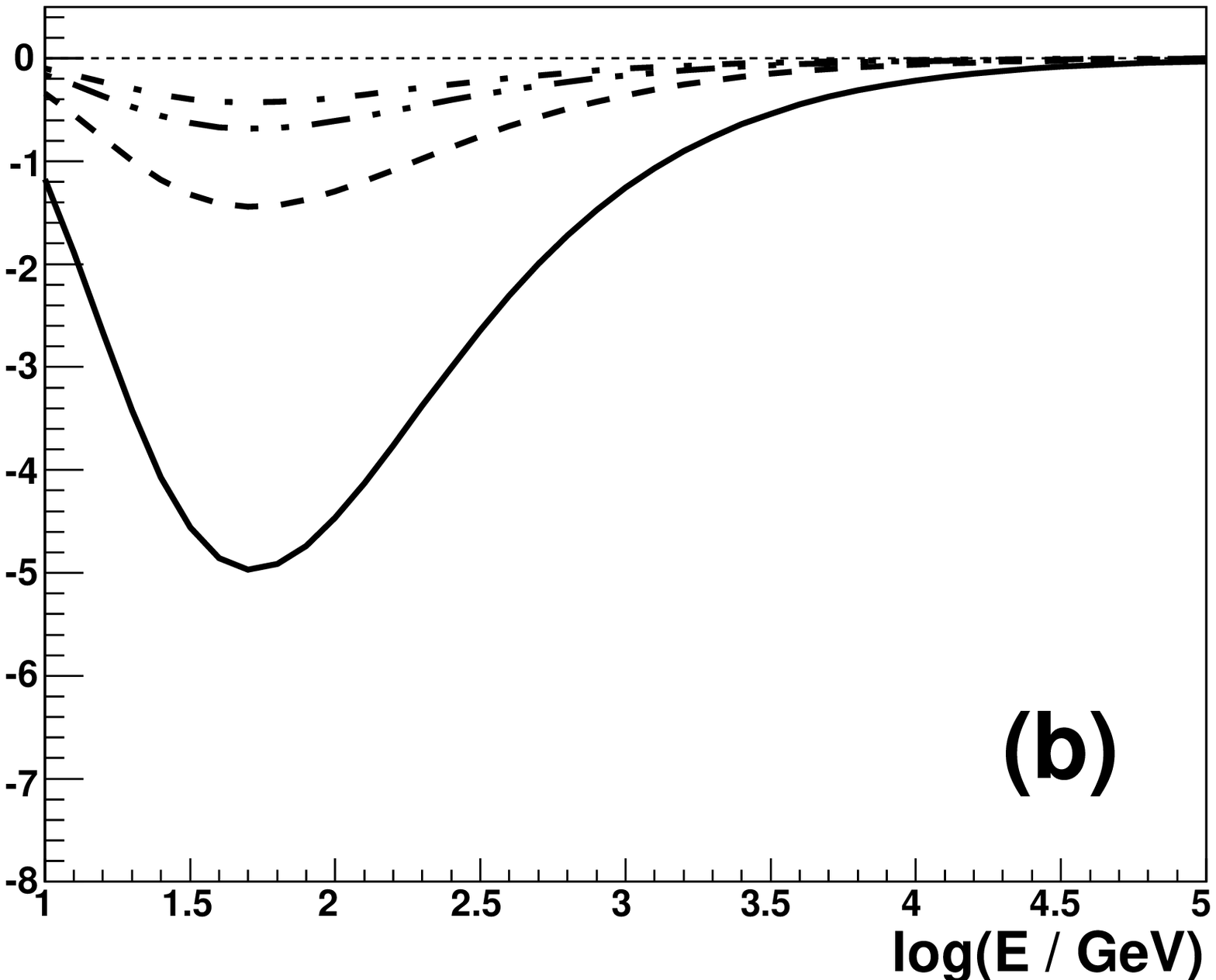}
\includegraphics{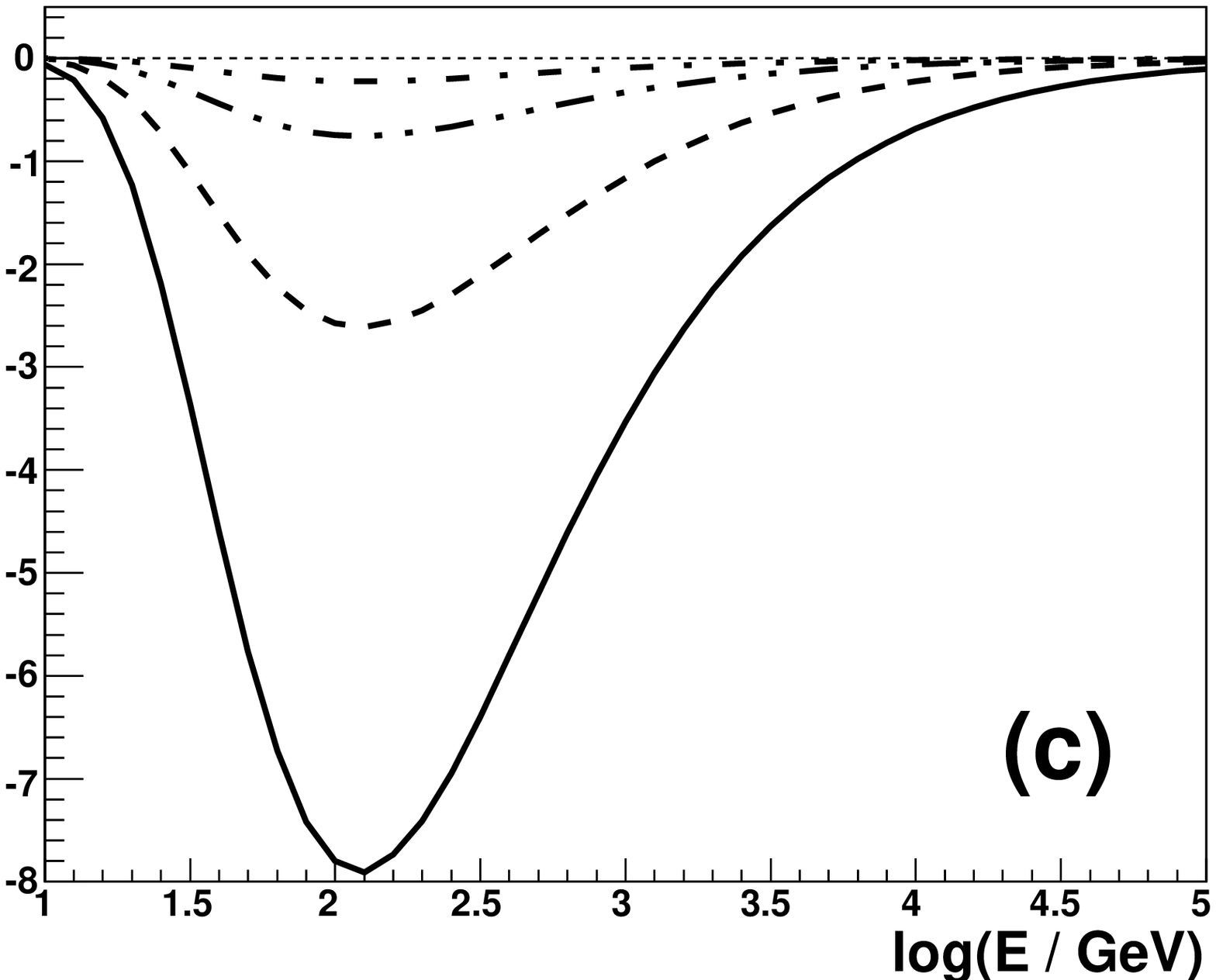}
\caption{Optical depths for $\gamma$ rays in the radiation field of stars with the parameters 
$R_\star = 10^{12}$~cm and $T_\star = 3\times 10^4$~K (a) 
and $R_\star = 2\times 10^{11}$~cm and $T_\star = 10^5$~K (b). In figure (a) $\gamma$-rays propagate along 
the straight line with the impact parameters are $D = 3R_\star$ (solid curve), 10$R_\star$ (dashed), 
30$R_\star$ (dot-dot-dashed), 100$R_\star$ (dot-dashed).
In figure (b) $D = 30R_\star$ (solid), 100$R_\star$ (dashed), 
200$R_\star$ (dot-dot-dashed), 300$R_\star$ (dot-dashed). In figure (c), parameters of the star are those 
for WR 20a star ($R_{\rm WR20a} = 1.35\times 10^{12}$~cm, $T_{\rm WR20a} = 4.3\times 10^4$~K) and 
$D = 10R_\star$ (solid), 30$R_\star$ (dashed), 100$R_\star$ (dot-dot-dashed), 200$R_\star$ (dot-dashed).} 
\label{fig2}
\end{figure*}
\section{Absorption effects due to passage of a single star}
 
To analyse the effect of passage of a single star, we first compute the optical depth for $\gamma$-ray photons 
passing at different impact parameters to it. In the calculations we apply typical parameters of the O type and WR type stars. 
The optical depth is integrated along the direction towards the observer, which closest approach 
to the star is defined by the impact parameter $D$.
We show that for typical parameters of luminous stars and impact parameters 
as large as $D = (30-100)R_\star$, the absorption of $\gamma$ rays is still significant (see Fig.~2).
For such large impact parameters, the simple power law $\gamma$-ray spectrum should still show a broad 
dip in the spectrum by a factor of a few at energies around $\sim$50-200 GeV.
The resulting softening and hardening of the source spectrum can be detected by Cherenkov telescopes 
sensitive in energy range of tens of GeV (e.g. such as MAGIC and future CTA).
 
The instantaneous distance of the $\gamma$-ray photons  from the star can be calculated
$R_{\rm tr} = \sqrt{D^2 + v_\star^2T^2}$, where $T$ is the transit time 
measured from the closest approach of the star from the observer's line of sight (see Fig.~1b). 

The absorption effect  of the $\gamma$-ray spectrum in the radiation field of the star, during 
specific transit, can be evaluated by introducing the so-called "reduction factor" ($RF$) which
determines the ratio of integral $\gamma$-ray photon fluxes in specific range of energies 
$E_{\rm min}$ -- $E_{\rm max}$: with the effect of the absorption to the non-absorbed (intrinsic) one.
The $RF$ factor is defined as, 
\begin{eqnarray}
RF_{E_{\rm min}} = {{\int_{E_{\rm min}}^{E_{\rm max}}
(dN_\gamma/dE_\gamma)e^{-\tau}dE_\gamma}\over{\int_{E_{\rm min}}^{E_{\rm max}}
(dN_\gamma/dE_\gamma)dE_\gamma}}.
\label{eq3}
\end{eqnarray} 
\noindent
For the first calculations, we assume a differential $\gamma$-ray spectrum (produced either in the inner 
jet or in the vicinity of the SMBH) of a power-law type with an index of $-2$, i.e. 
$dN_\gamma / dE_\gamma\propto E_\gamma^{-2}$, and the maximum energy at which the absorption effect of 
$\gamma$ rays is still important. 
$\tau$ is the optical depth for $\gamma$ rays in the radiation field of the star as shown in Fig.~2.
As an example, $\log RF$ is calculated for the specific parameters of the luminous star (see legend in Fig.~3) 
and fixed 
impact parameters but for different velocities of the luminous stars $v_\star$ and as a function of
the transit time $T$, for two threshold energies $E_{\rm min} = 30$ GeV (Fig.~3a) and 300 GeV 
(Fig.~3b). The dependence of $RF$ on $E_{\rm min}$, for fixed stellar velocity and impact parameter,
are shown in Fig.~3c. The typical transits, resulting in a significant absorption
of $\gamma$ rays, last for a few to several days for the considered parameters.
In the case of a transit of a single star, the effect of $\gamma$-ray absorption in the stellar radiation is 
symmetrical.

\begin{figure*}
\vskip 4.5truecm
\includegraphics{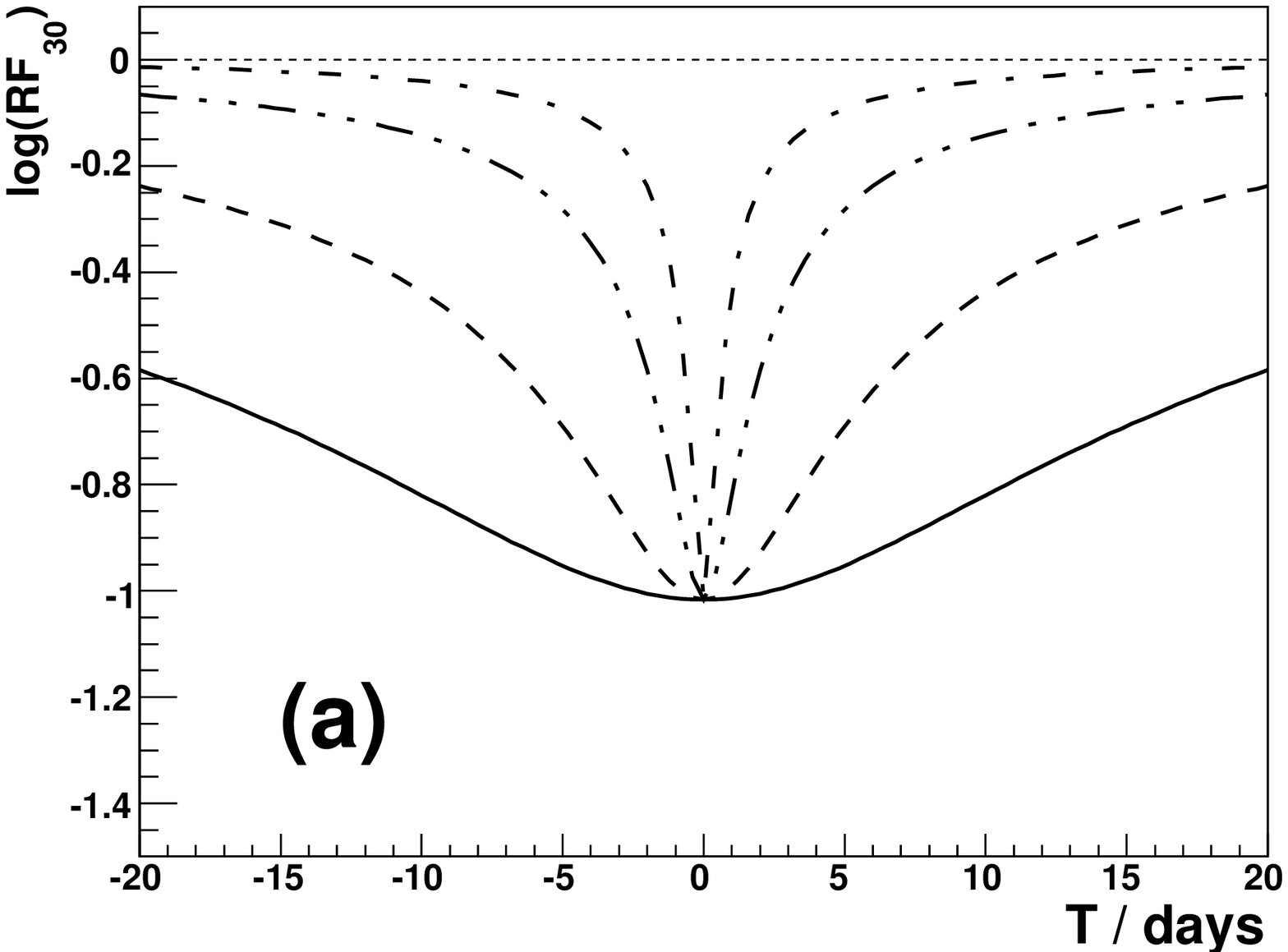}
\includegraphics{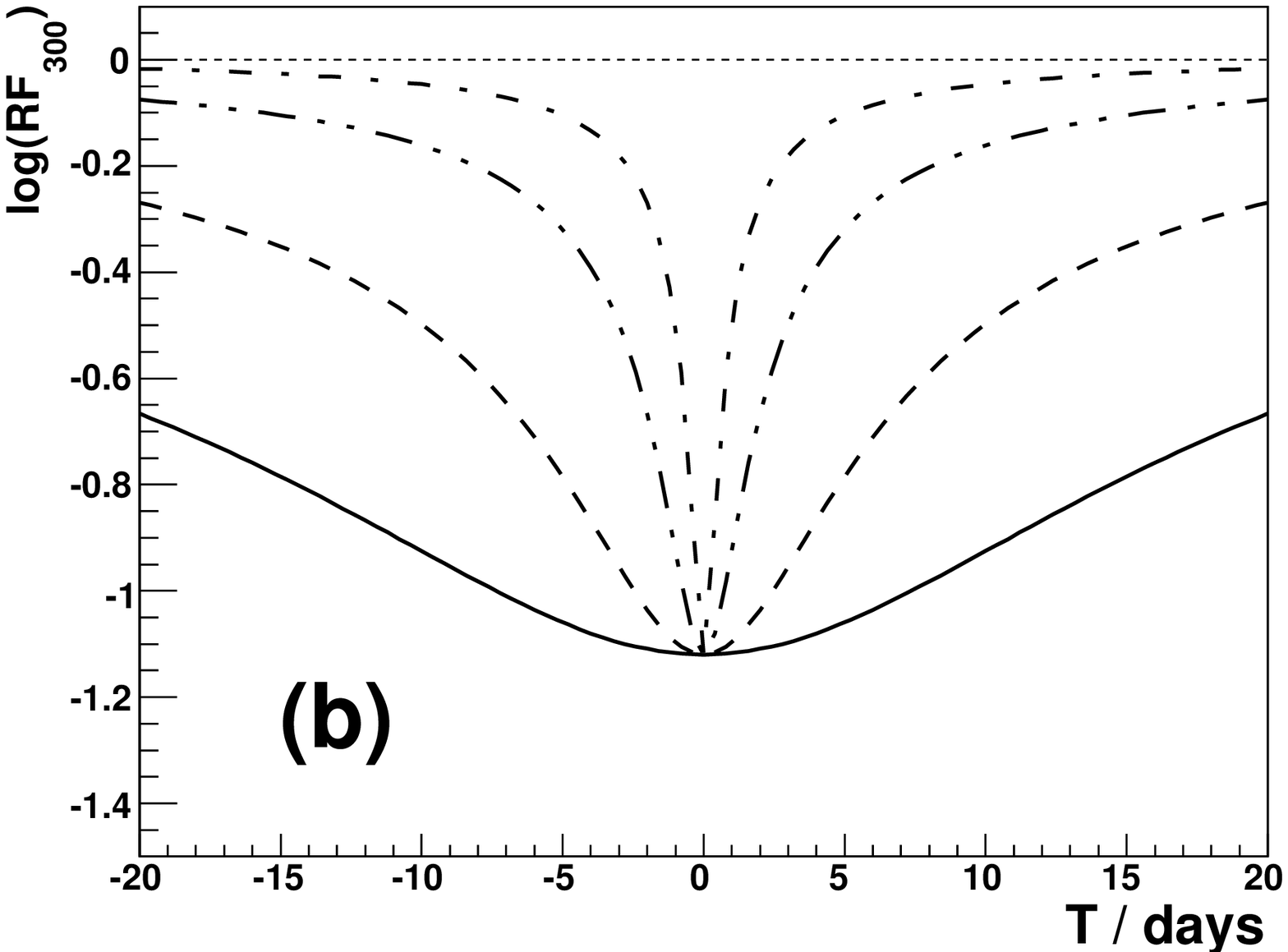}
\includegraphics{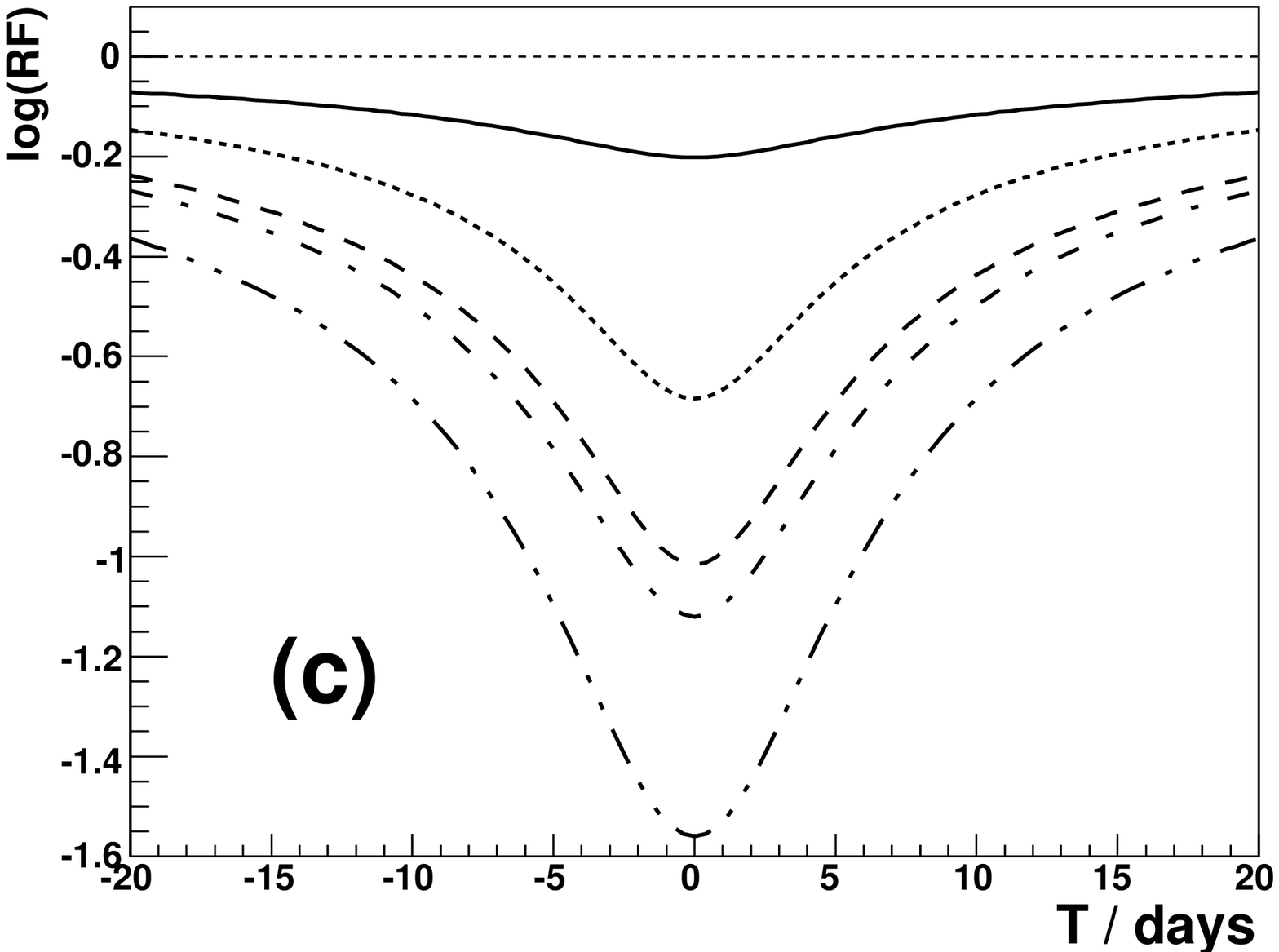}
\caption{Log of the reduction factor (RF) of the $\gamma$-ray flux above 30 GeV (a) and 300 GeV (b) 
in the case of a star with the radius $R_\star = 10^{12}$~cm and the 
surface temperature $T_\star = 3\times 10^4$~K. The star crosses the $\gamma$-ray beam with the impact 
parameter $D = 10^{13}$~cm and with the velocity $v_\star = 10^7$~cm~s$^{-1}$ (solid curve), 
$3\times 10^7$~cm~s$^{-1}$ (dashed), $10^8$~cm~s$^{-1}$ (dot-dot-dashed), and $3\times 10^8$~cm~s$^{-1}$ 
(dot-dashed). The reduction factors for the star with the same parameters as in (a) and velocity 
$v_\star = 3\times 10^7$~cm~s$^{-1}$ but for 
the $\gamma$-ray flux above 10 GeV (solid), 30 GeV (dashed), 
100 GeV (dot-dashed), 300 GeV (dot-dot-dashed), and 1 TeV (dotted) are shown in (c).} 
\label{fig3}
\end{figure*}
\begin{figure*}
\vskip 4.truecm
\includegraphics{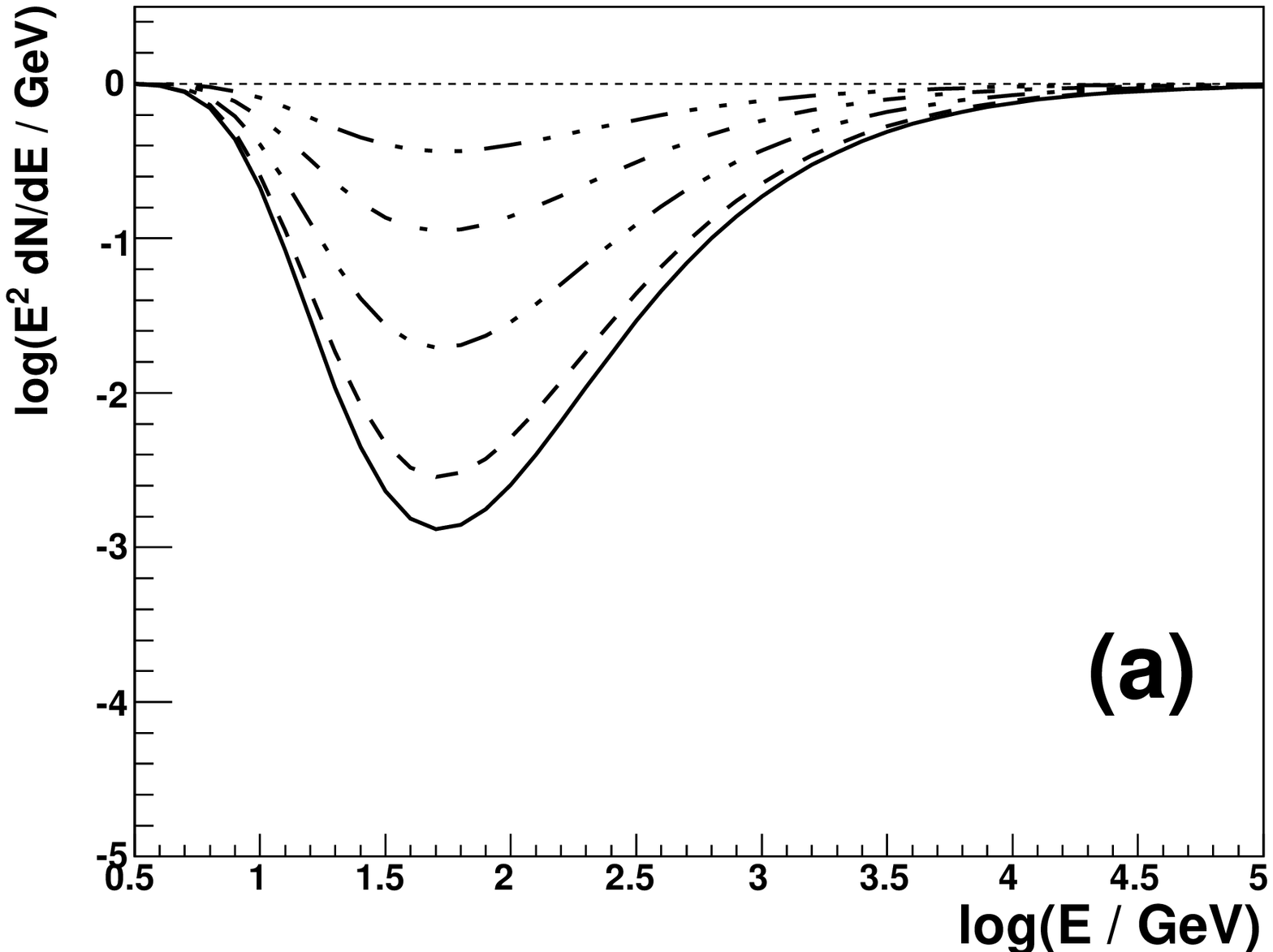}
\includegraphics{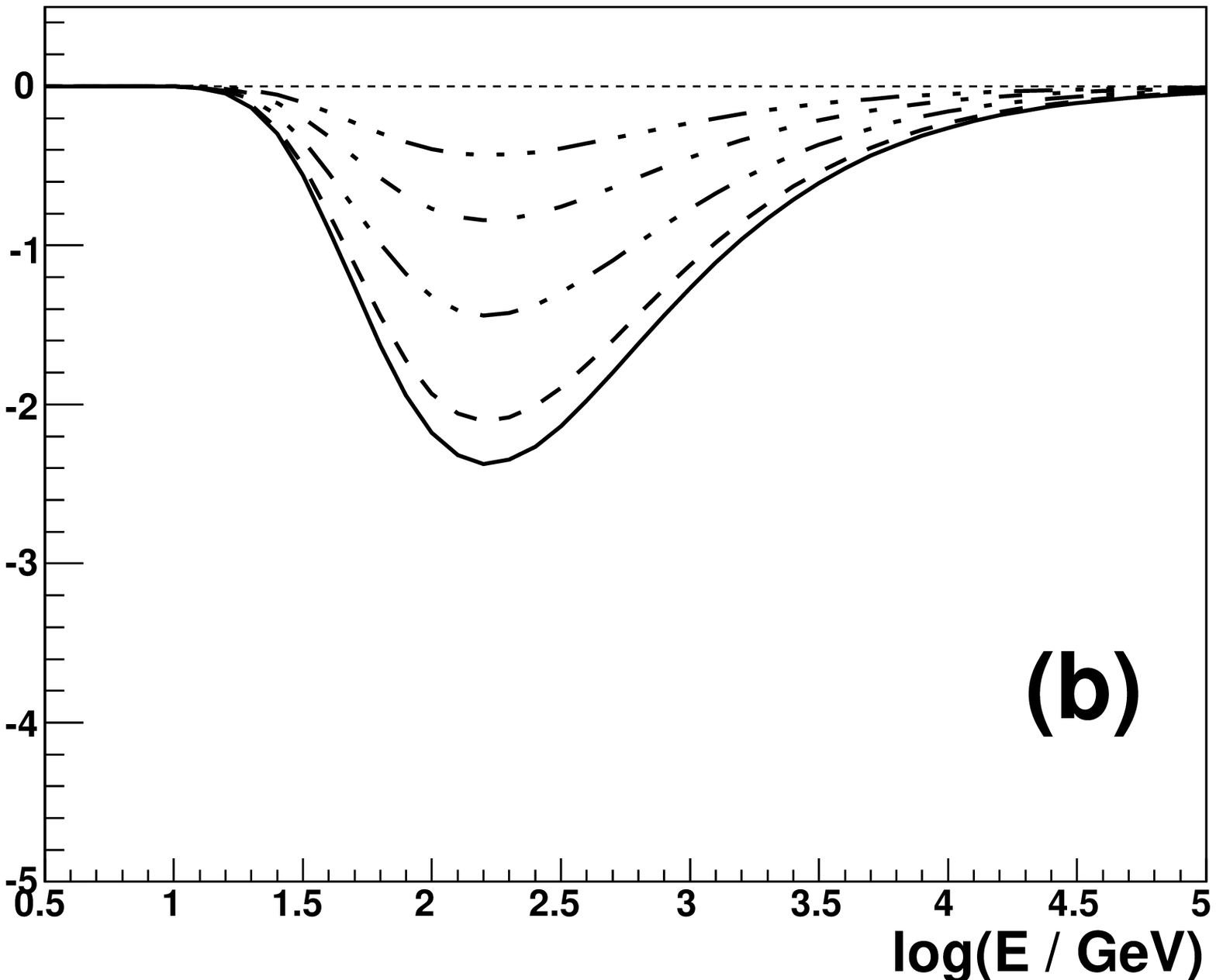}
\includegraphics{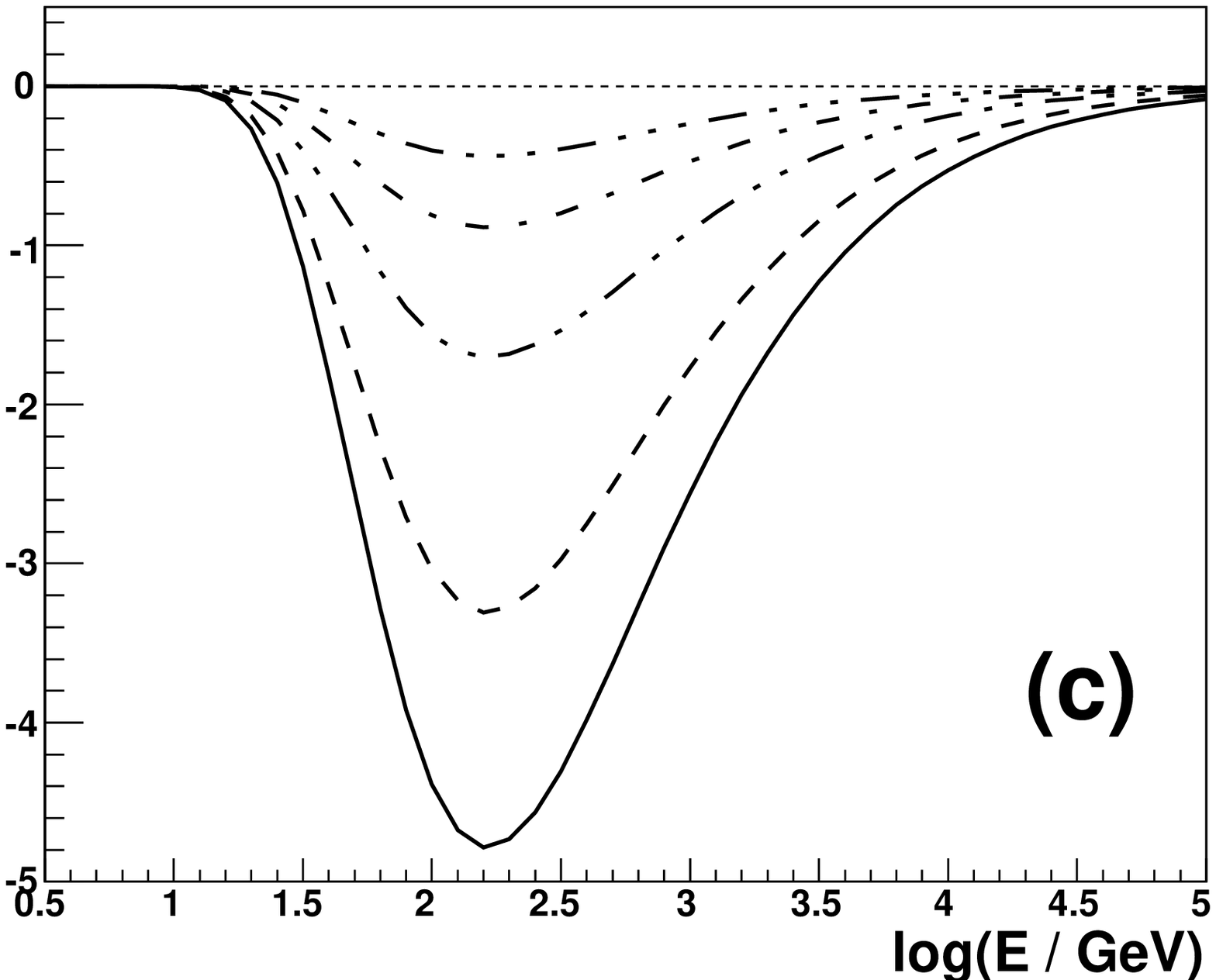}
\caption{Gamma-ray spectra at different transit times T = -20~days (dot-dot-dot-dashed curve), 
-10 days (dot-dashed), -5 days (dot-dot-dashed), -2 days (dashed), 
0 days (solid), in the case of a transit due to the single star with the parameters: the stellar radius 
$R_\star = 2\times 10^{11}$ cm, the surface temperature $T_\star = 10^5$ K, and the observer transiting 
with the impact parameter $D = 10^{13}$ cm and the velocity
$v_\star = 3\times 10^7$ cm s$^{-1}$ (figure a); $R_\star = 10^{12}$ cm, $T_\star = 3\times 10^4$ K, 
$D = 10^{13}$ cm and $v_\star = 3\times 10^7$ cm s$^{-1}$ (b), and for the parameters as  in (b) but for 
$D = 5\times 10^{12}$ cm (c).} 
\label{fig4}
\end{figure*}

We also show how the $\gamma$-ray spectra are modified  at specific 
transit times by the absorption effects in the case of transiting single star (see Fig.~4).
The broad absorption dip appears between a few tens of GeV and a few hundreds of GeV.
As a result a part of the spectrum 
above $\sim$10 GeV should clearly steepen and a part of the spectrum in the sub-TeV energy range
 hardens. Such transient features, with characteristic time scales lasting from 
 a few days up to a few tens of days, are predicted to appear in
the observations of active galaxies using Cherenkov telescopes.  

\begin{figure}
\vskip 6.5truecm
\includegraphics{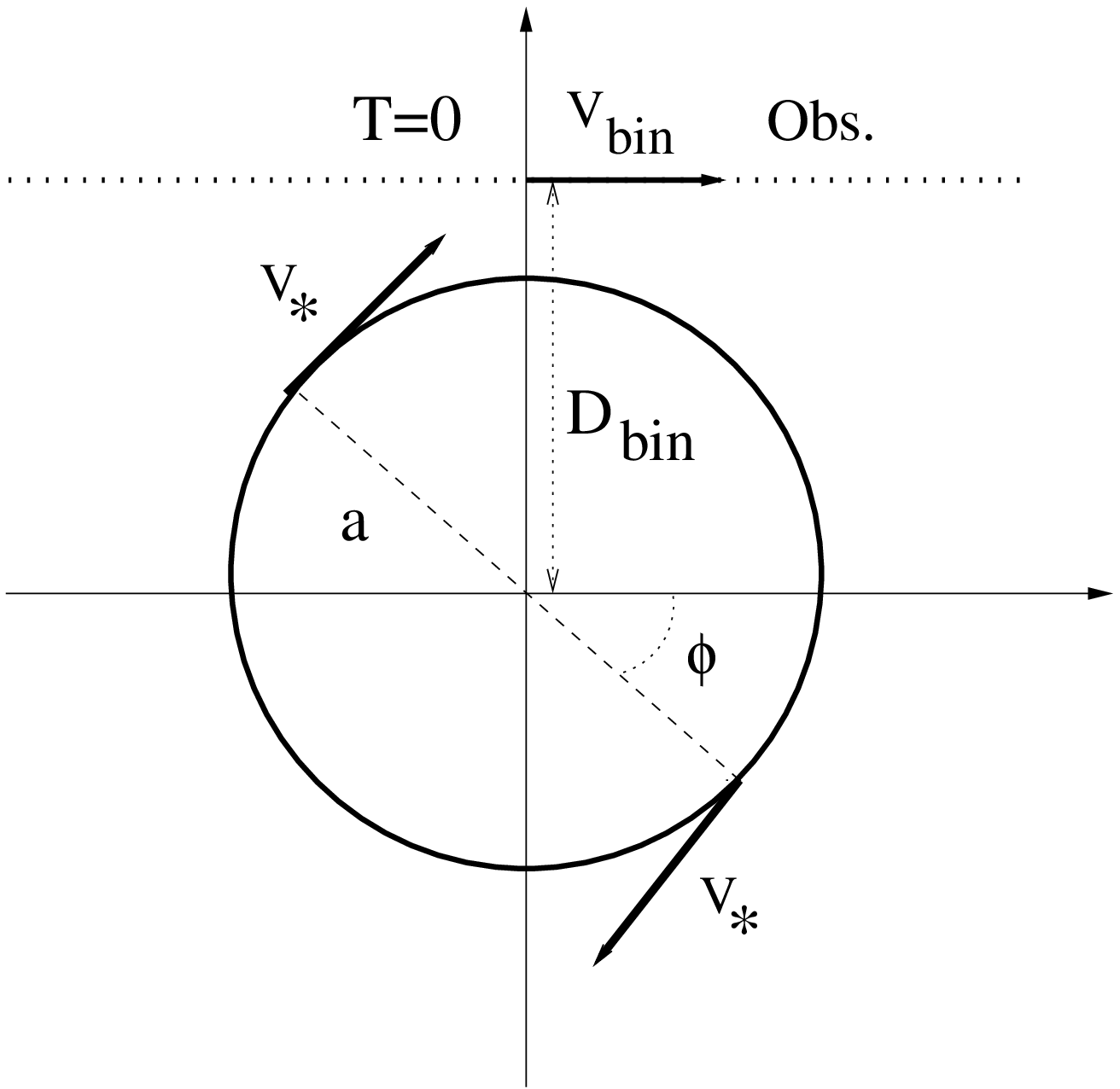}
\caption{Schematic representation of the passage of a binary system through the $\gamma$-ray
beam with the impact parameter $D$ in the reference frame of the binary system centre of mass. 
The observer (Obs.) moves with a velocity $v_{\rm bin}$ along a straight line (dotted). 
The binary system contains two equal mass stars which move with velocity 
$v_\star$ on a 
circular orbit with the radius $a$. The orbit of the binary system lays in the plane perpendicular to the 
direction of the observer. The phase of the stars is marked by $\phi$. The time is measured from the closest 
distance between the observer and the centre of the binary system.}
\label{fig5}
\end{figure}
\section{Absorption effects due to passage of a binary system}

Half of the stars is expected to form binary systems. Therefore, we also consider the passage
of luminous binary stellar system close to the observer's line of sight. In such a case, the
effect of absorption of primary beam of $\gamma$ rays can have much more complicated time 
dependence since the distances between the observer's  line of sight and each of the stars
is additionally modulated by the movement of stars within the binary system (for the schematic 
geometry see Fig.~5). For simplicity, we consider that the binary system contains two stars of
equal mass. Then, the velocities of specific stars within the binary system are  
\begin{eqnarray}
v_\star = ({{GM_\star}\over{4a}})^{1/2}\approx 4.1\times 10^6 ({{M_1}\over{a_{13}}})^{1/2}~~~
{\rm cm~s^{-1}},
\label{eq4}
\end{eqnarray}
\noindent
where the radius of the binary system is $a = 10^{13}a_{13}$~cm, and the masses of 
stars are $M_\star = 10M_1$~M$_\odot$. Note that for some parameters they can be of similar order as the transit 
velocities of the stars moving around the SMBH (see Eq.~2). 
In fact, the transit velocities of binary systems are limited by the condition of disruption of 
the binary system in the gravitational field of SMBH. We estimate the minimum distance of 
the binary system from the SMBH at which the tidal forces on stellar companions are balanced 
by gravitational force of companion stars on, 
\begin{eqnarray}
L\approx (16M_8/M_1)^{1/3}a\approx 1.8\times 10^{-3}a_{13}(M_8/M_1)^{1/3}~~~{\rm pc}.
\label{eq5}
\end{eqnarray}
\noindent
We assume that most of the stellar binary systems in the central stellar cluster stay
at distances from the SMBH that are larger than the above estimate. 

The absorption effects of $\gamma$ rays in the stellar radiation field strongly depend on the 
closest distance between the observer's line of sight and 
the centre of mass of luminous stars, i.e. the impact parameter $D_{\rm bin}$.
Therefore, we calculate the distance between the line of sight and the centres of stars, $D$, for some specific 
parameters of the stars and transition event. 
The results are shown as a function of the transition time for the example
values of the impact parameter, 
$D_{\rm bin}$, different velocities of transiting binary system, $v_{\star}$, and radii of the binary system, 
$a$, for fixed phase of the binary $\phi = 0^\circ$ (see Fig.~6). On the other hand, the dependence of
this distances on the phase of the binary, $\phi$, is shown in Fig.~7.
Note the fast dependence of the impact parameters of individual stars on the transit time $T$. 
During several days, 
those distances can change by an order of magnitude. They show double-peak structure due to the transiting 
two stars within the binary systems. 
Similar interesting effects are expected to appear in the $\gamma$-ray spectra 
influenced by the absorption of the $\gamma$-ray beam in the radiation field of such a binary system.  

\begin{figure*}
\vskip 10.truecm
\includegraphics{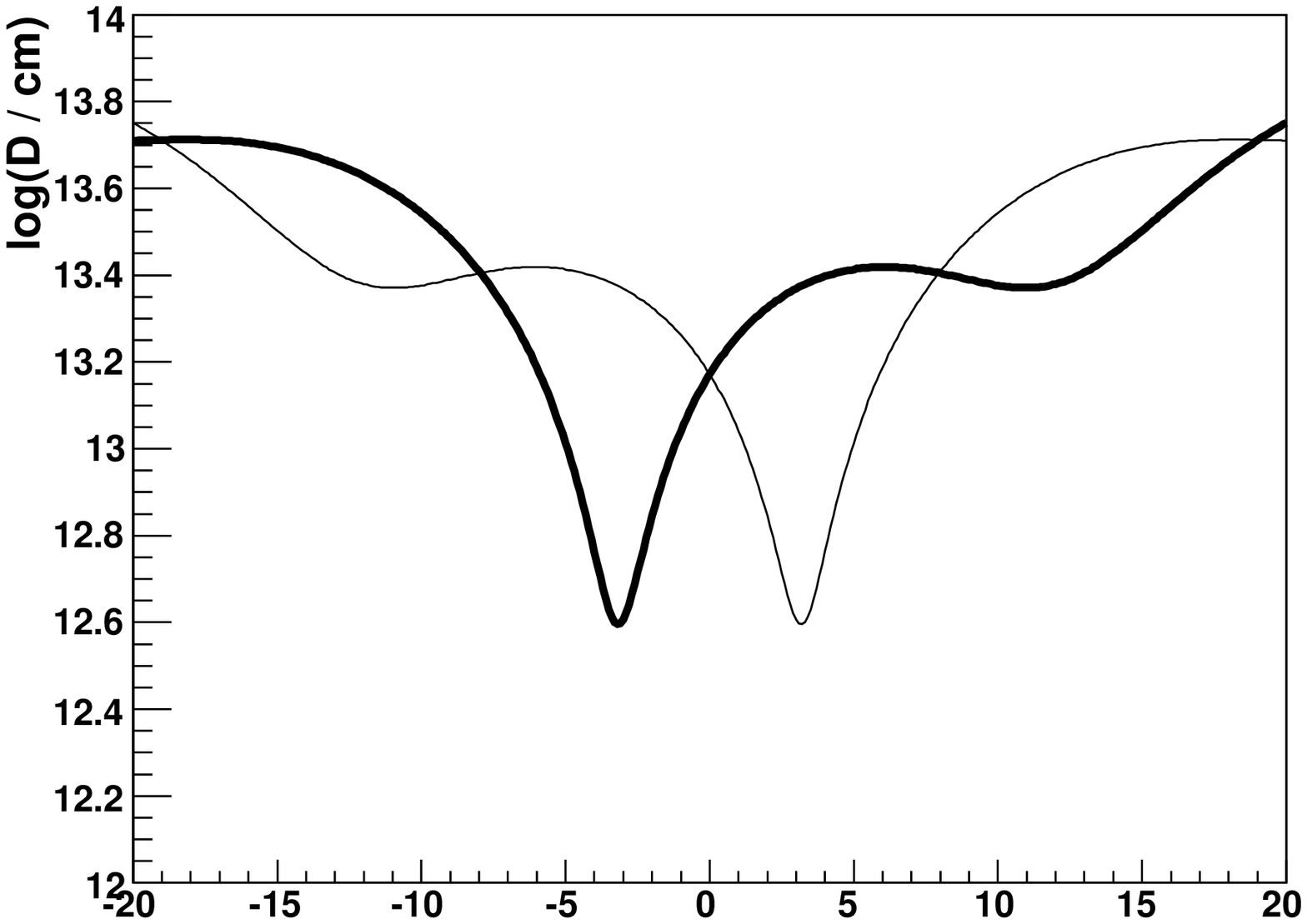}
\includegraphics{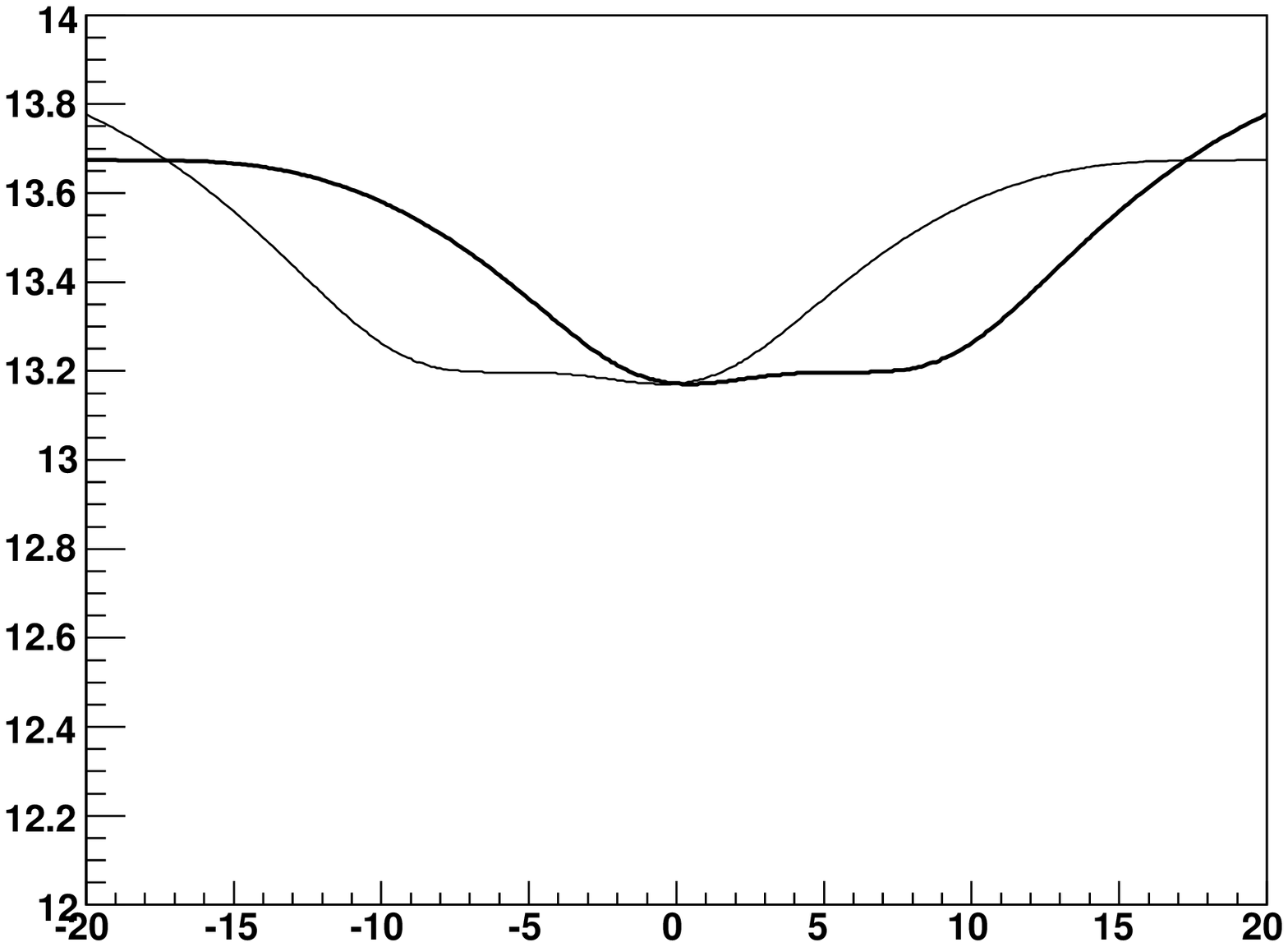}
\includegraphics{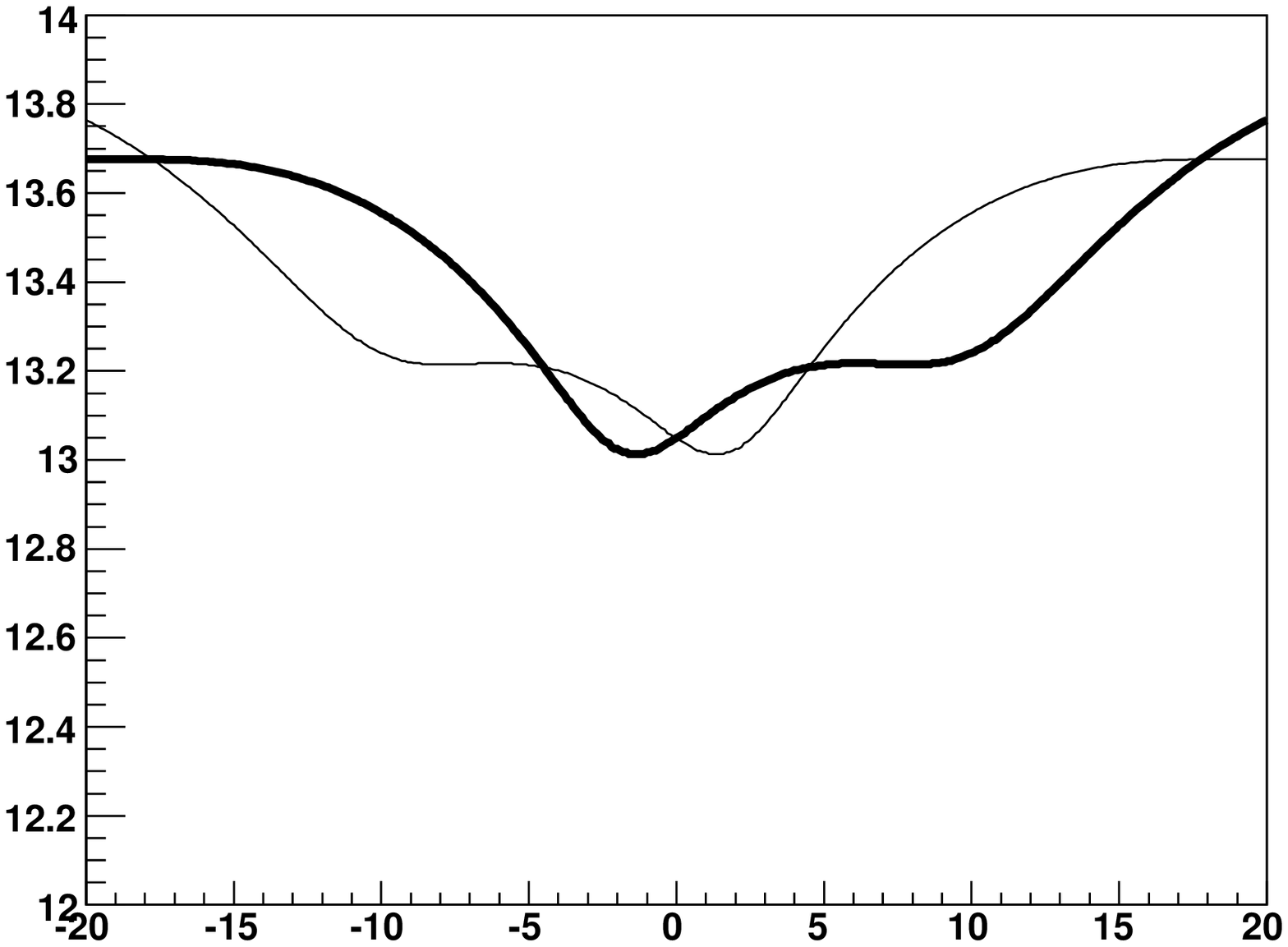}
\includegraphics{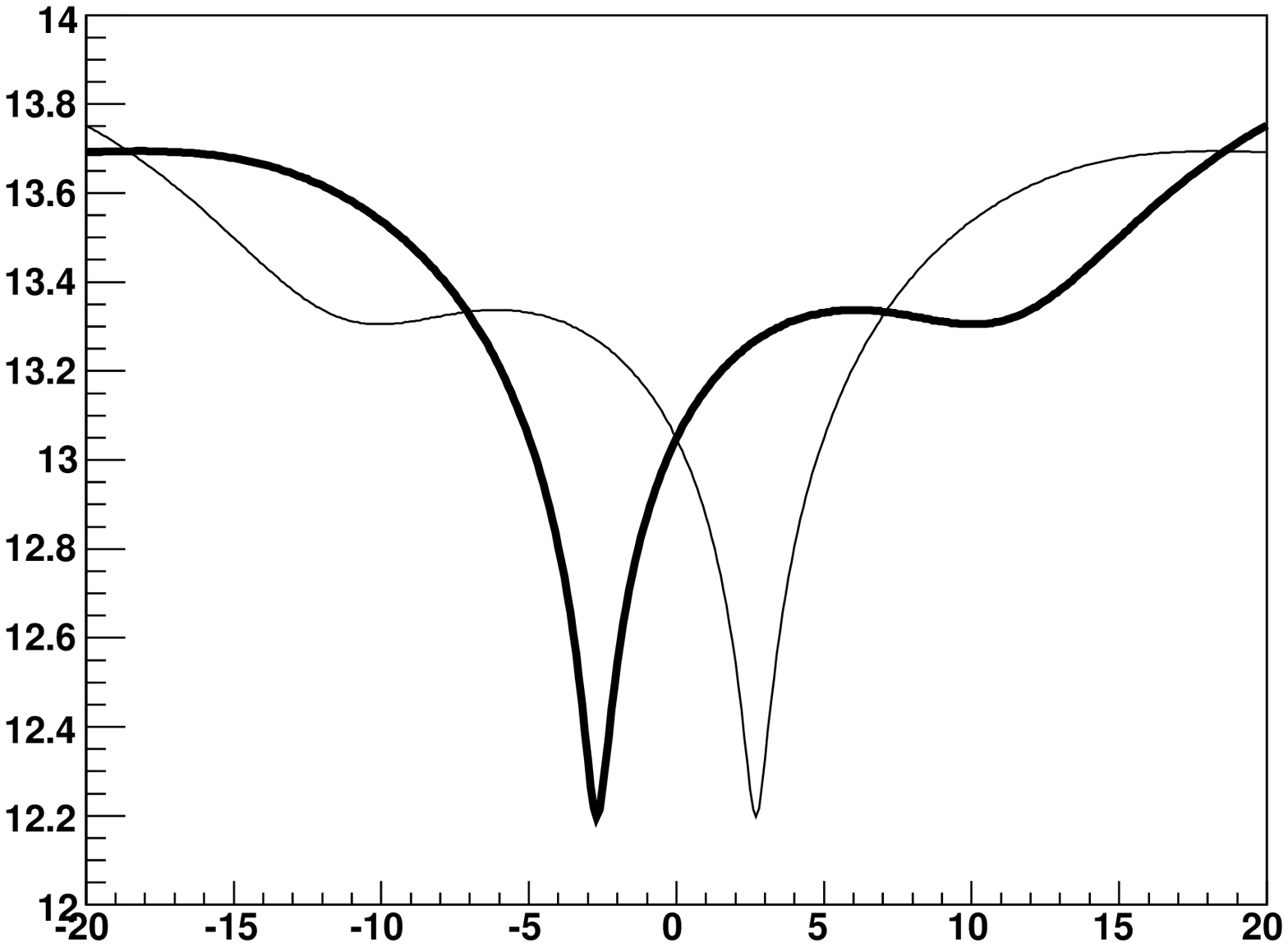}
\includegraphics{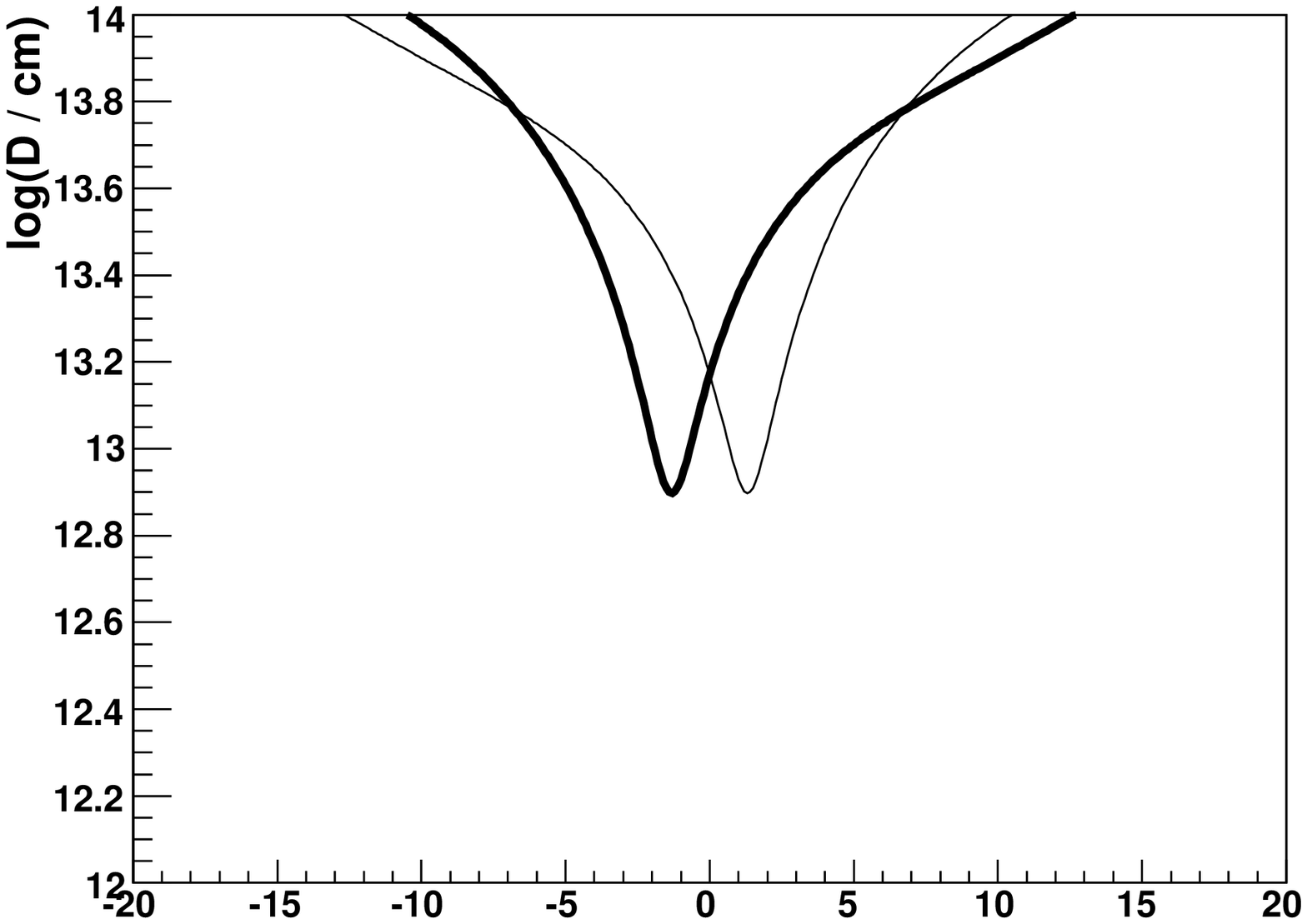}
\includegraphics{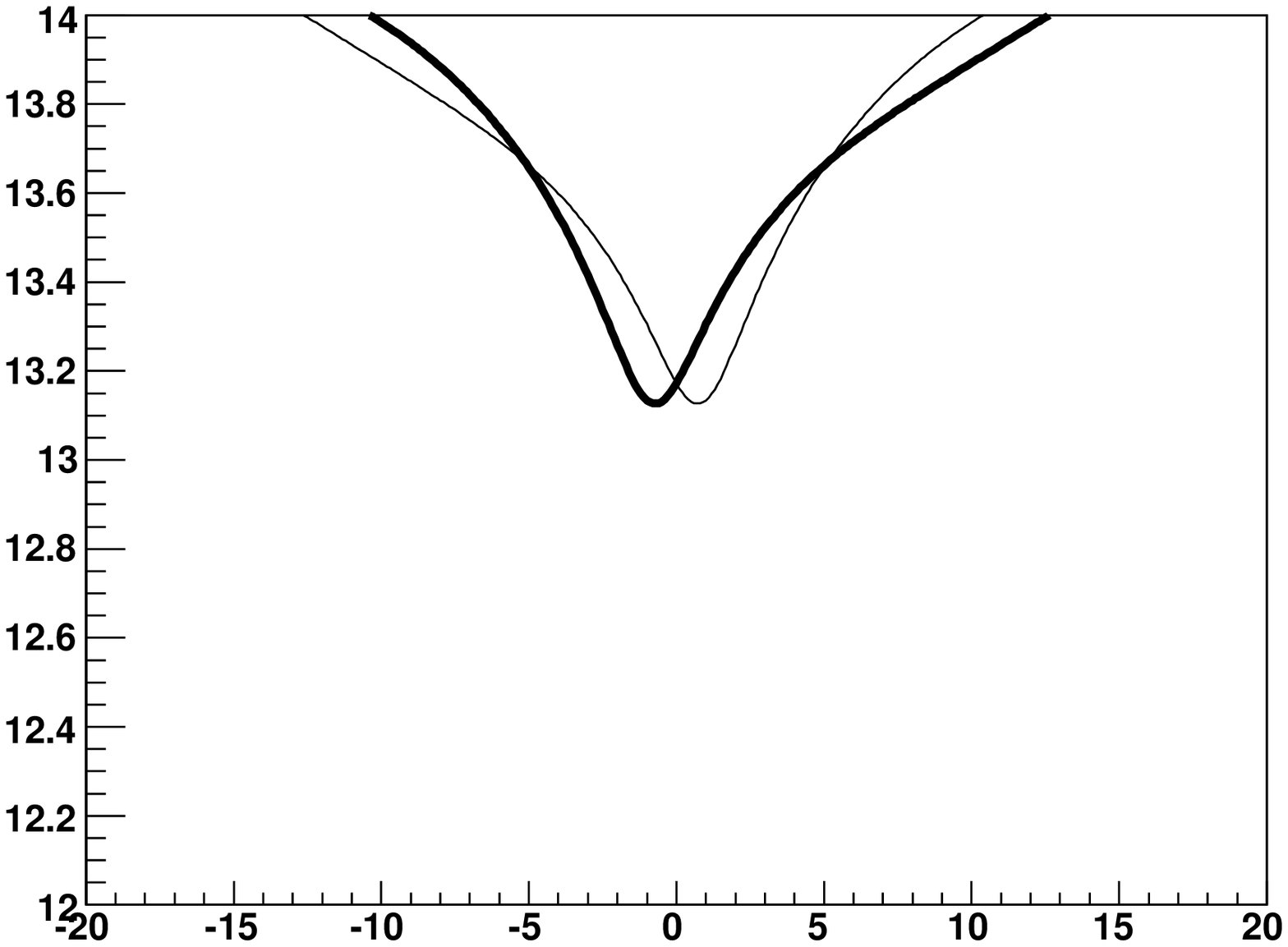}
\includegraphics{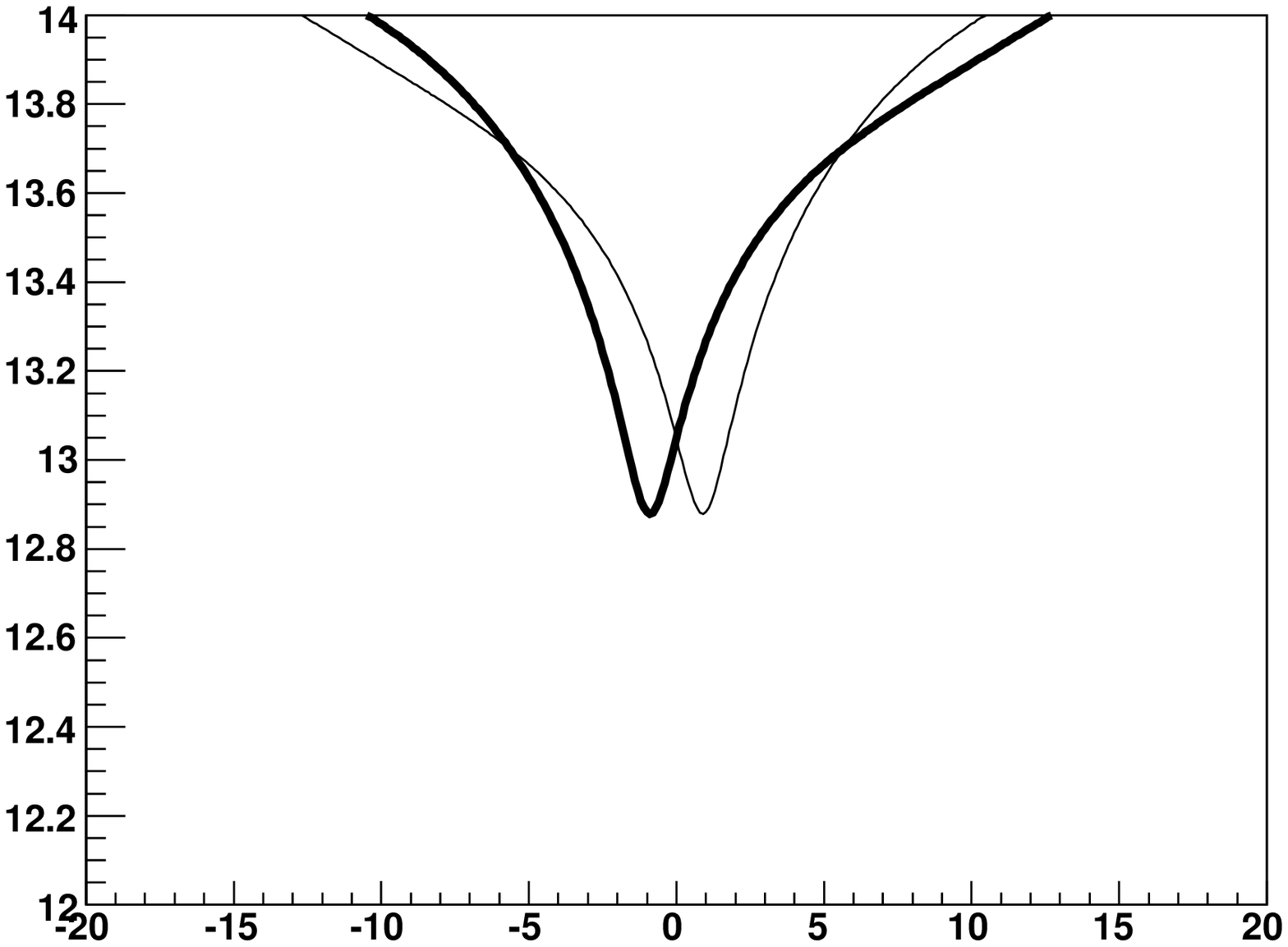}
\includegraphics{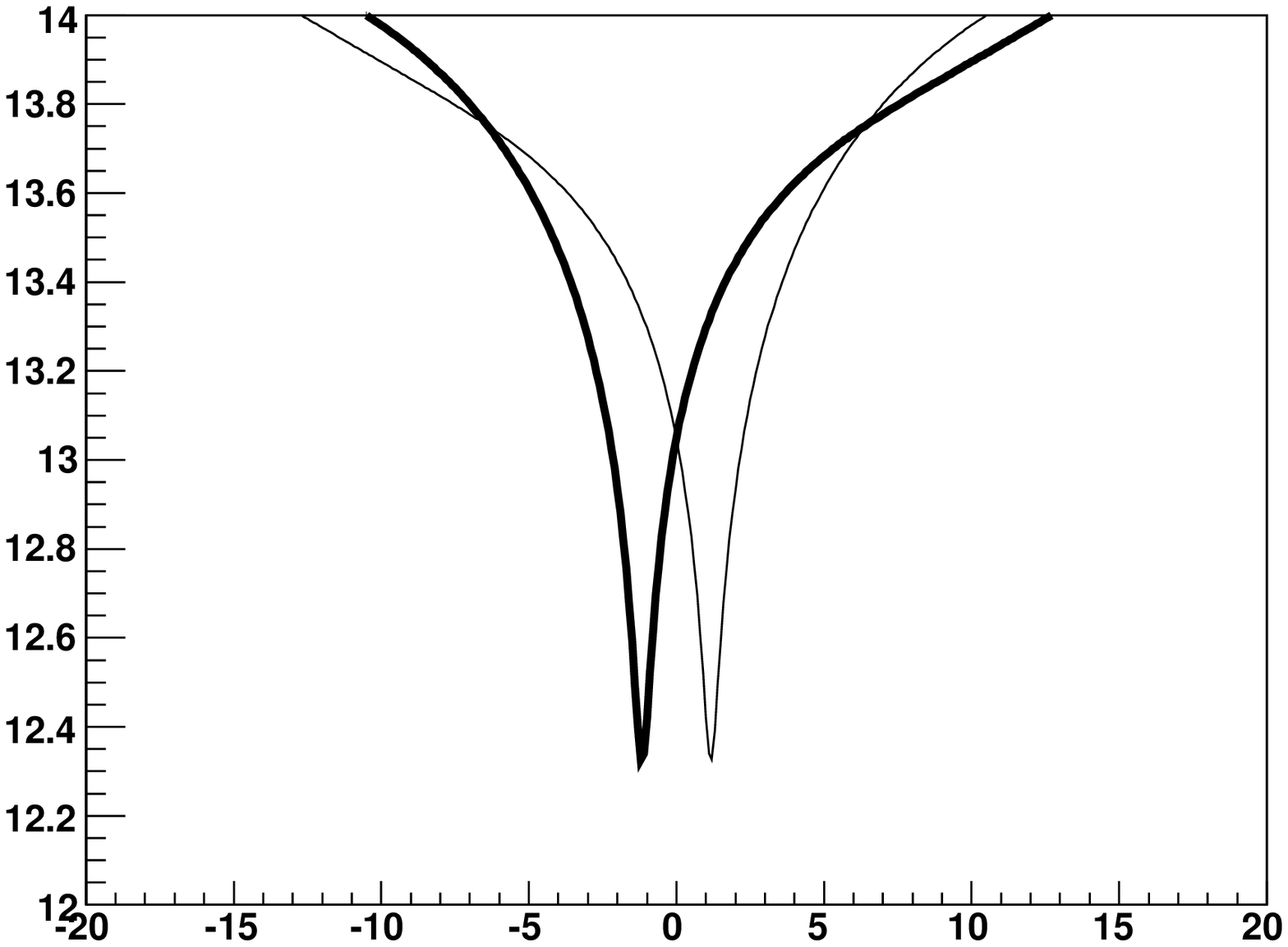}
\includegraphics{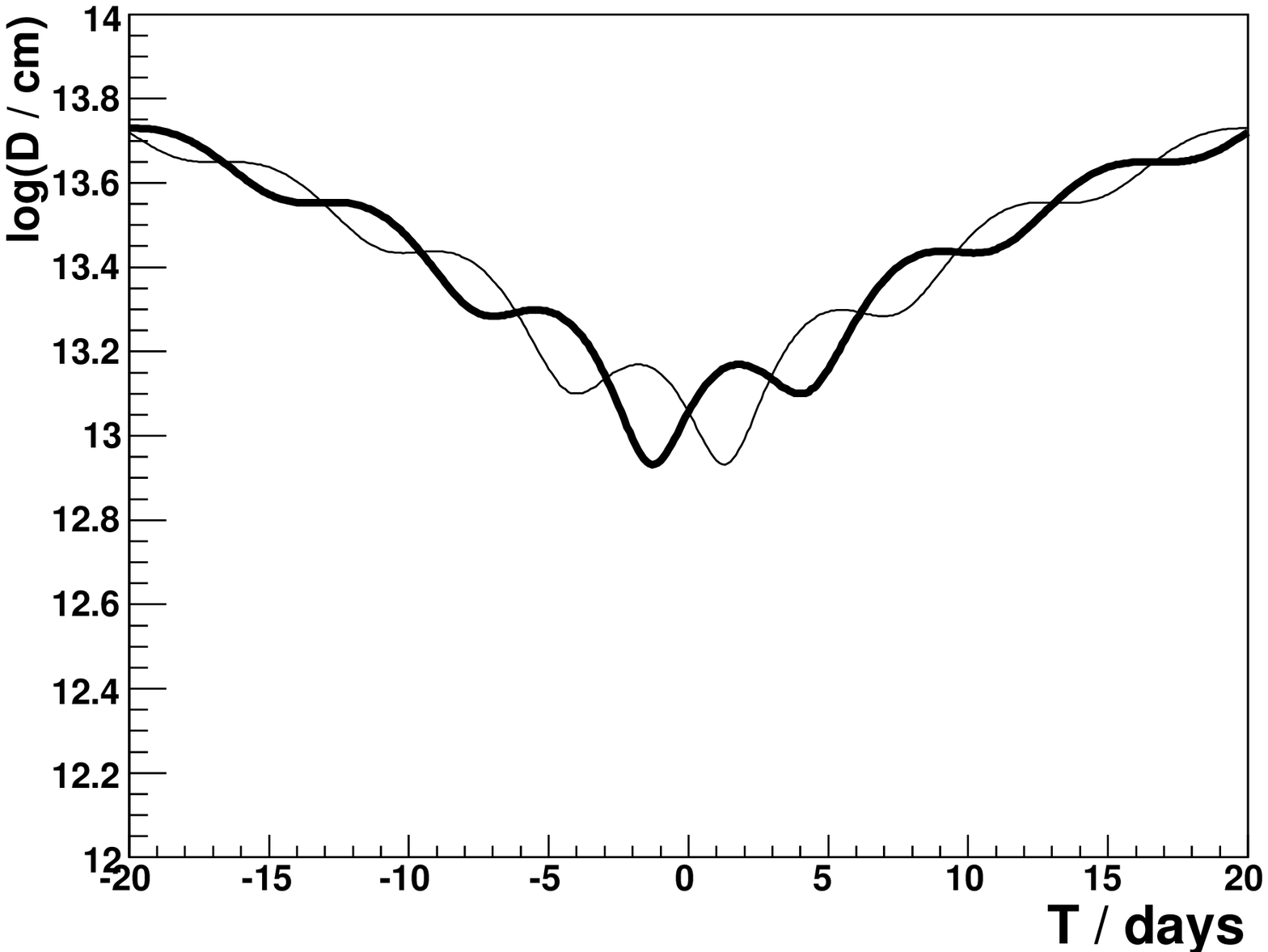}
\includegraphics{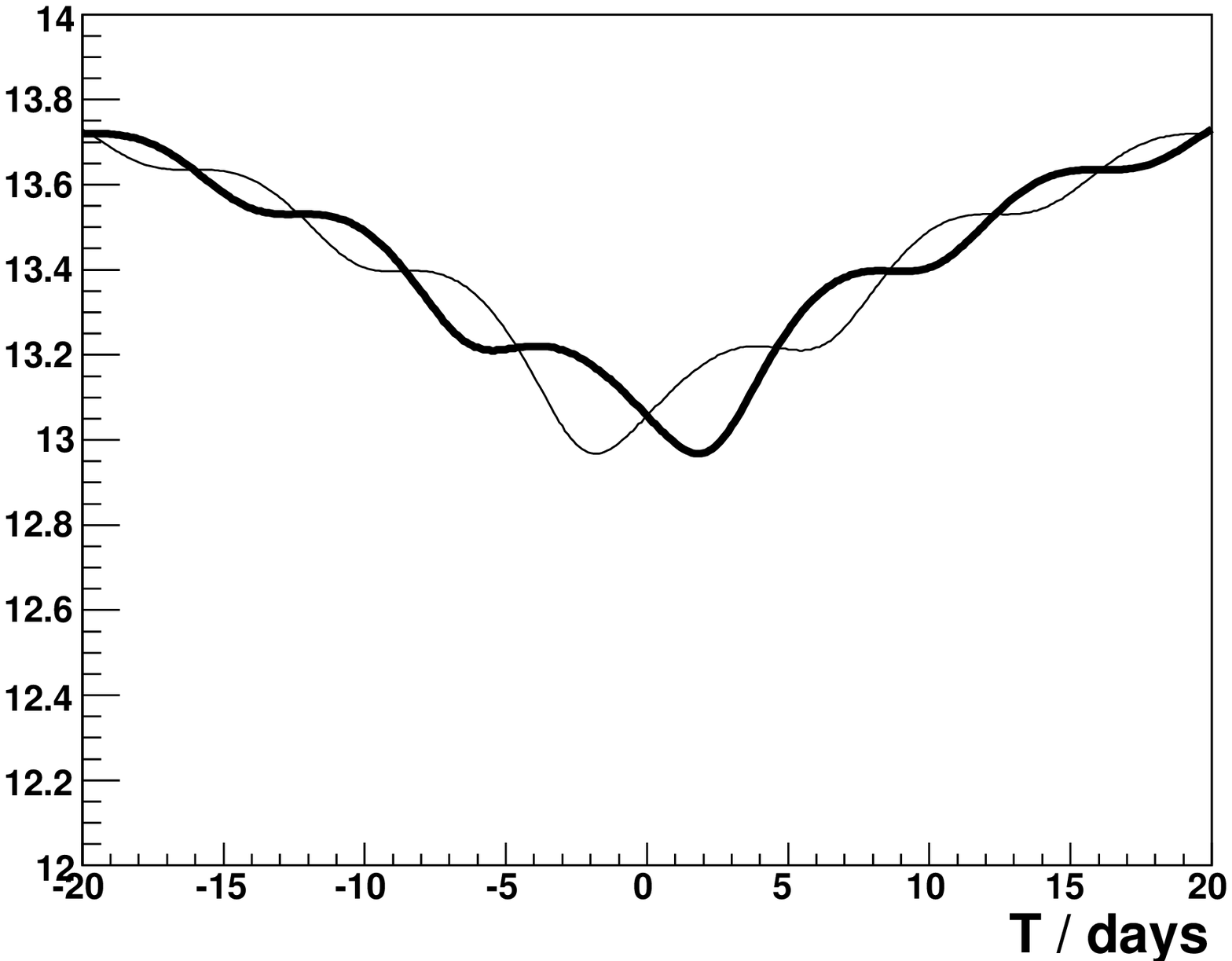}
\includegraphics{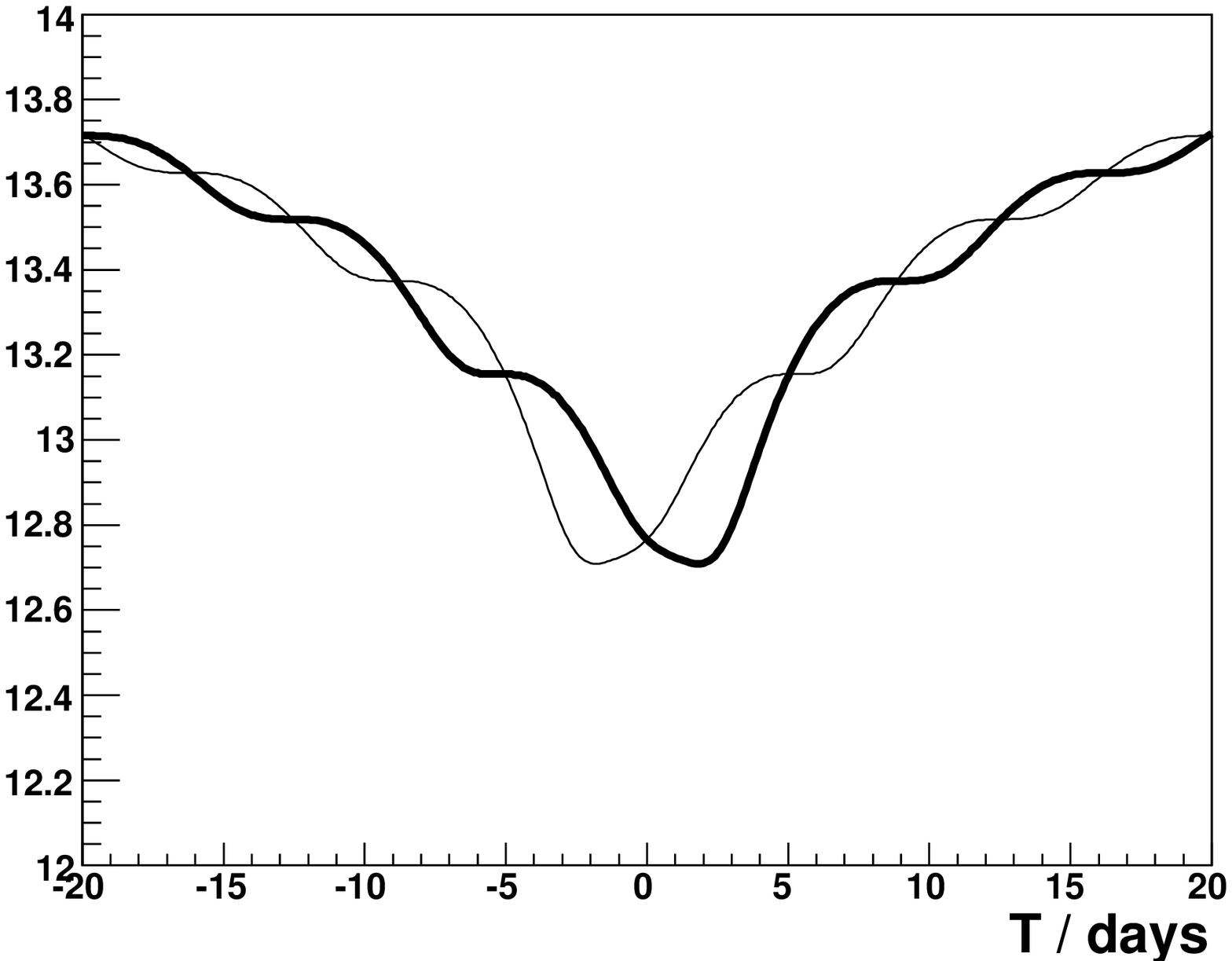}
\includegraphics{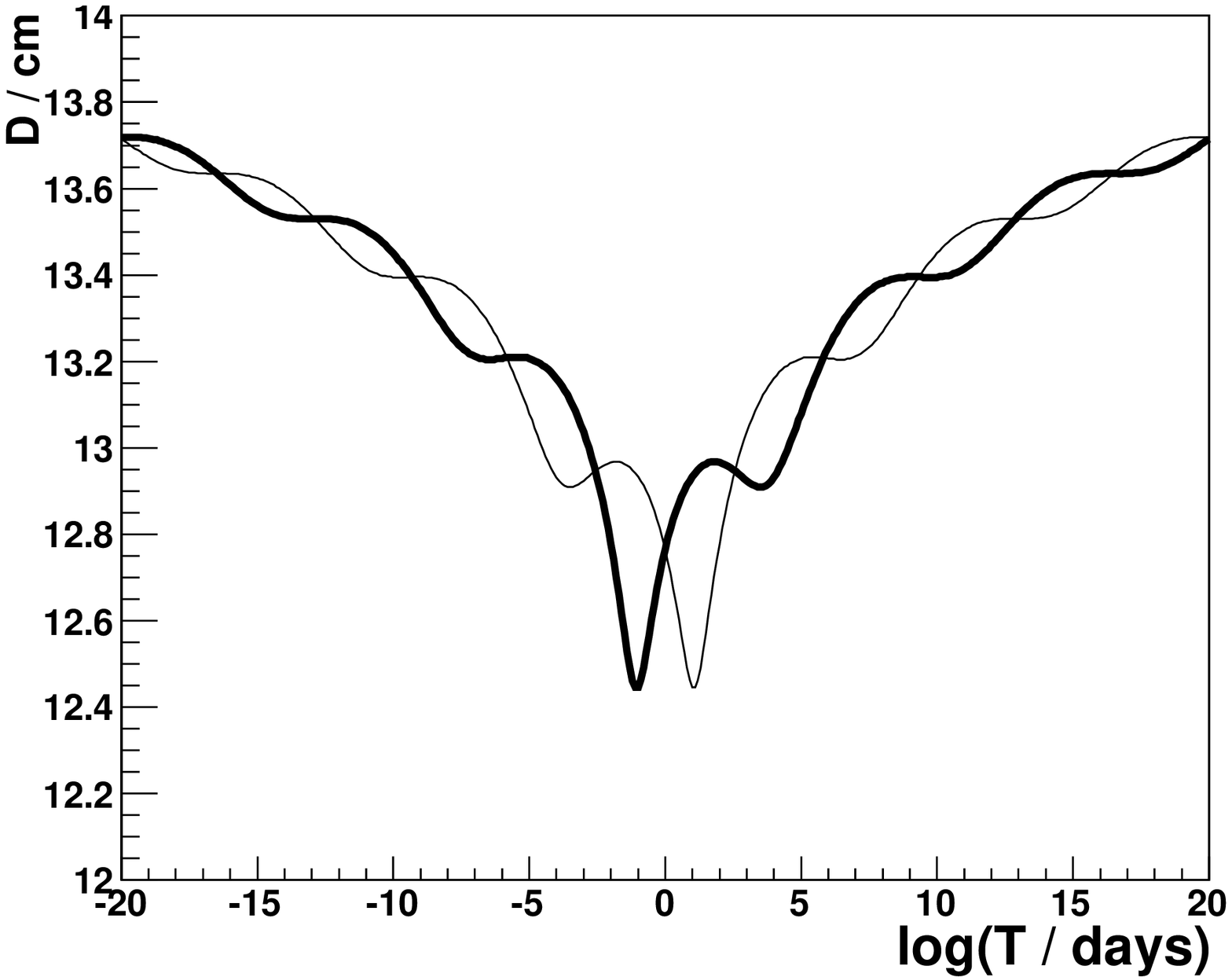}
\caption{Distance, $D$, between the observer's line of sight and the first star (see thick curves) and the
second star (thin curve) forming the binary system. The binary passes with the impact parameter
$D_{\rm bin} = -1.1\times 10^{13}$ cm (left column), $1.1\times 10^{13}$ cm (left-central column, 
$5\times 10^{12}$ cm (right-central), and $-5\times 10^{12}$ cm (right).  
The parameters of the binary system are: radius of stars  $R_\star = 10^{12}$ cm, 
the surface temperature $T_\star = 3\times 10^4$ K, semimajor axis $a = 10^{13}$ cm, and the velocity
of the stars 
$v_\star = 3\times 10^7$ cm s$^{-1}$. The velocity of the binary system is 
$v_{\rm bin} = 3\times 10^7$ cm s$^{-1}$ (upper panel) and $v_{\rm bin} = 10^8$ cm s$^{-1}$ 
(middle panel). The case with the parameters of the upper panel but radius of the binary system 
$a = 3\times 10^{12}$ cm is shown in the bottom panel.
The phase of the stars within the binary system is $\phi = 0^\circ$.} 
\label{fig6}
\end{figure*}
\begin{figure*}
\vskip 7.truecm
\includegraphics{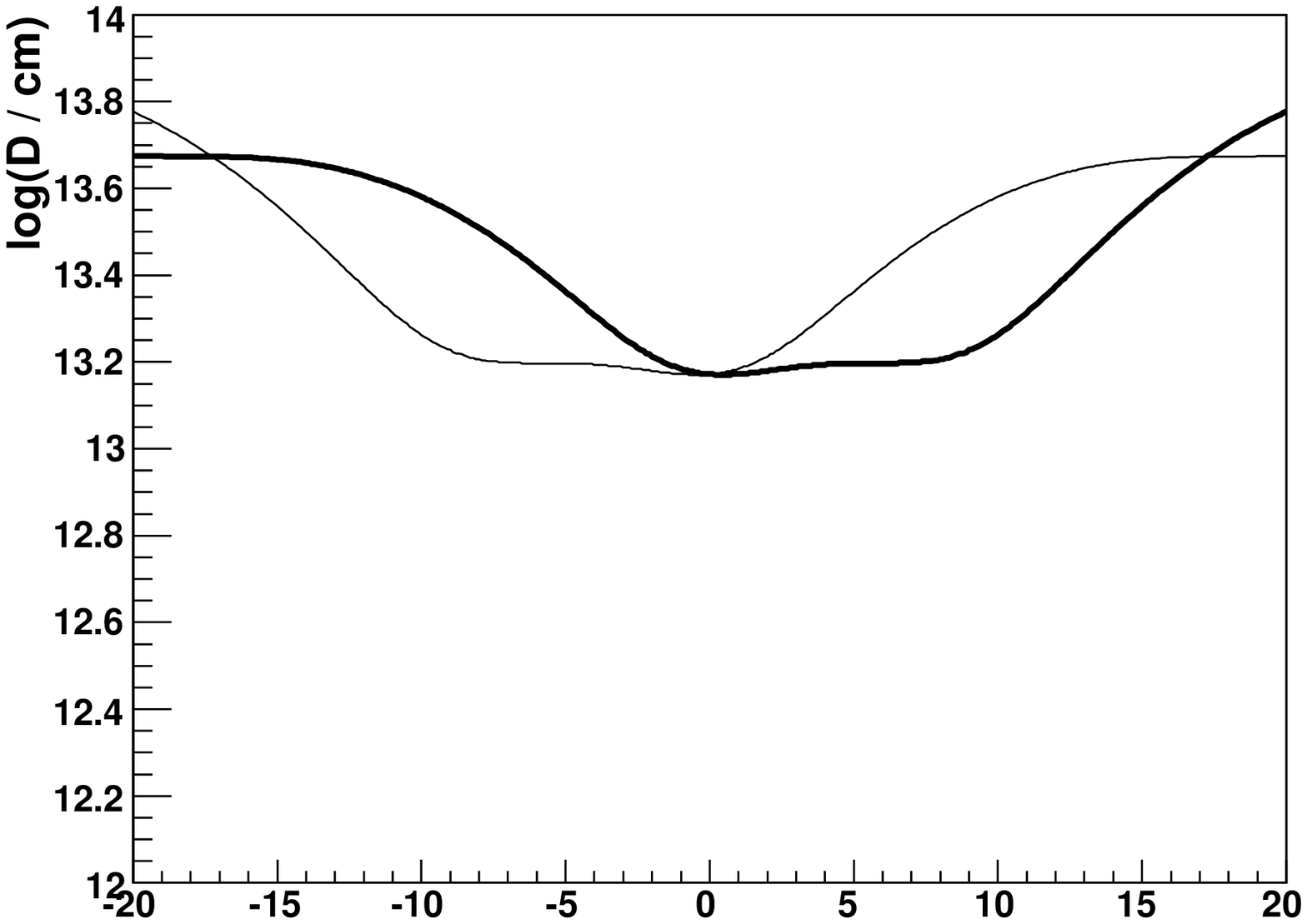}
\includegraphics{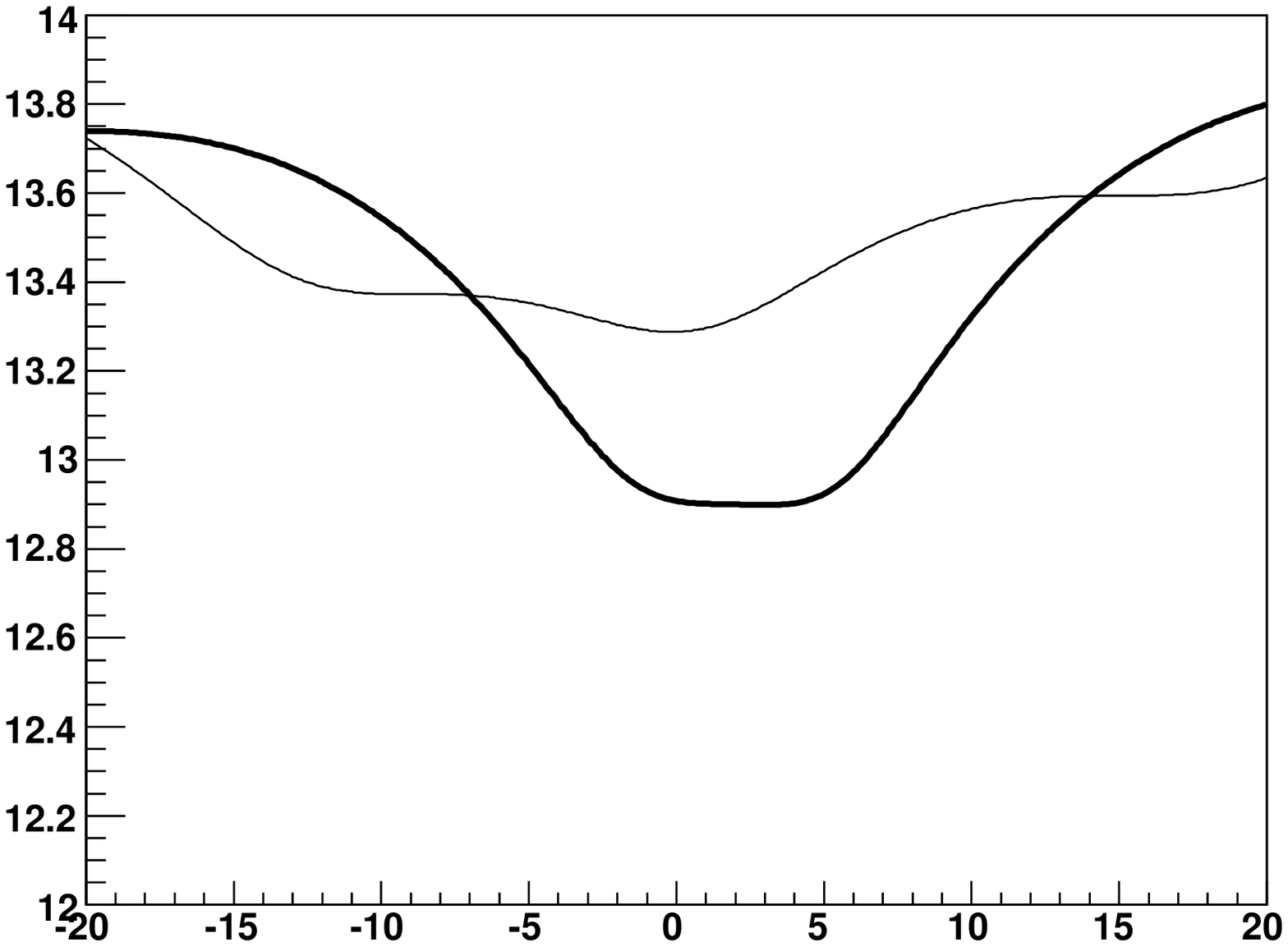}
\includegraphics{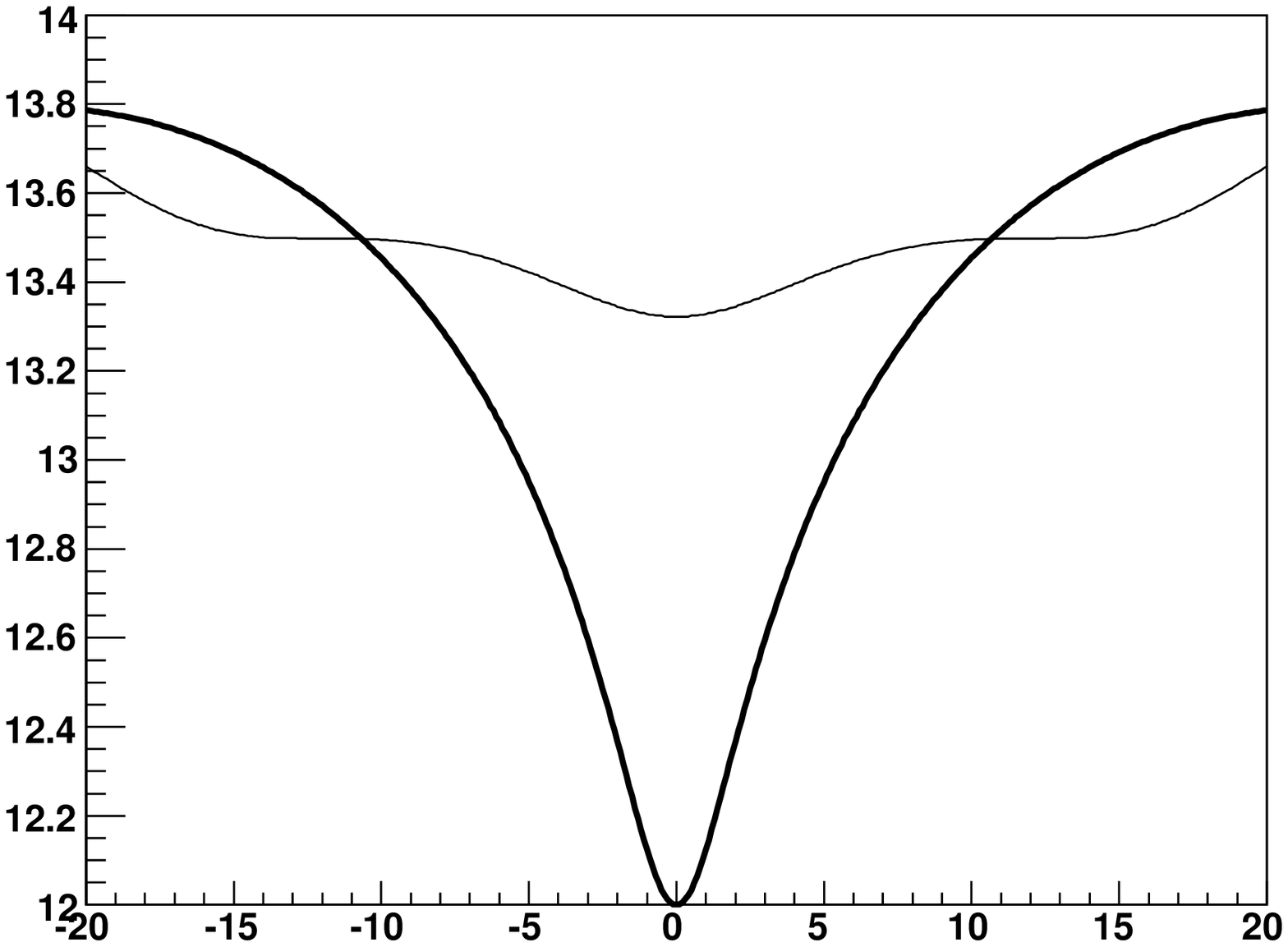}
\includegraphics{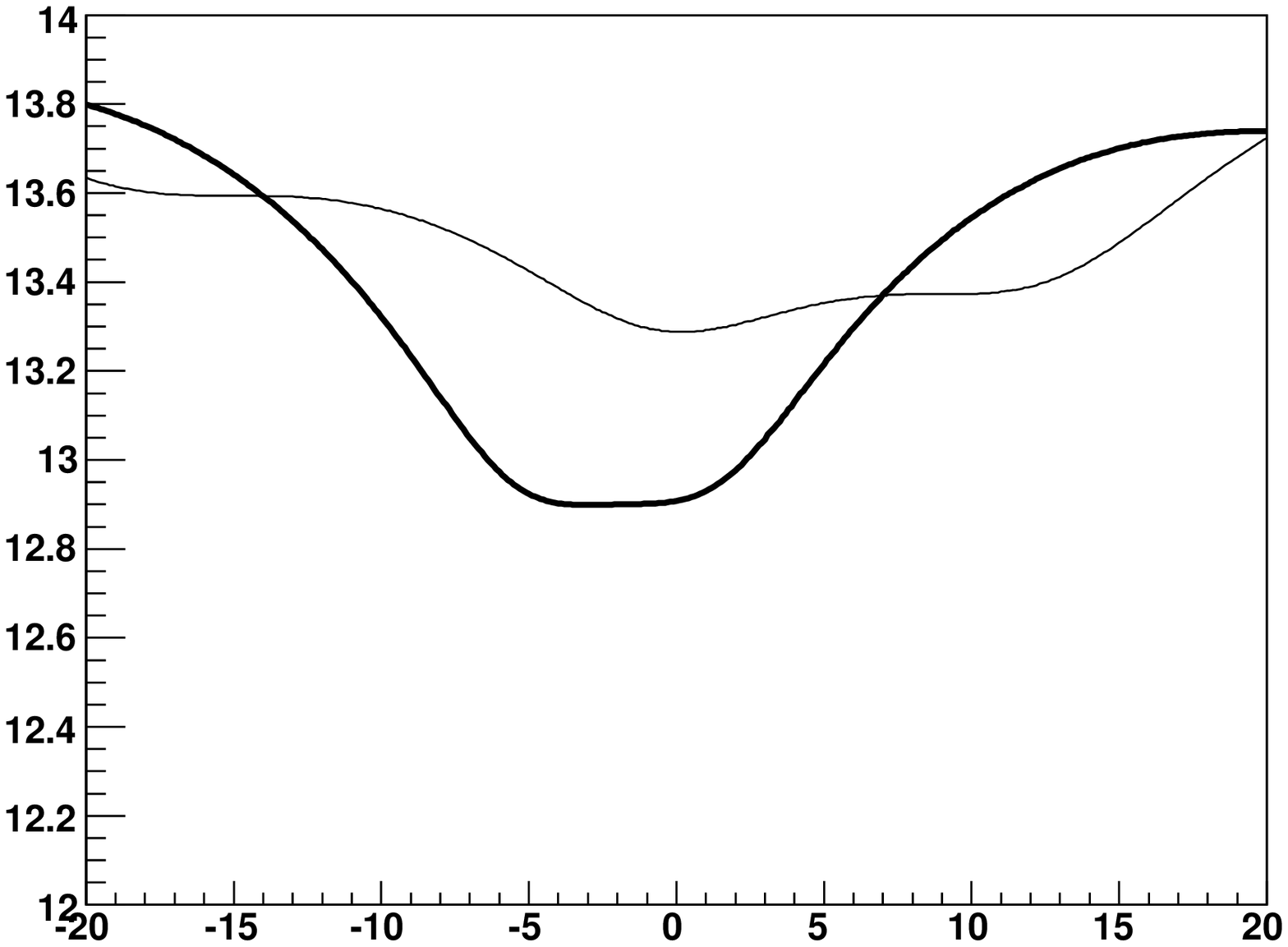}
\includegraphics{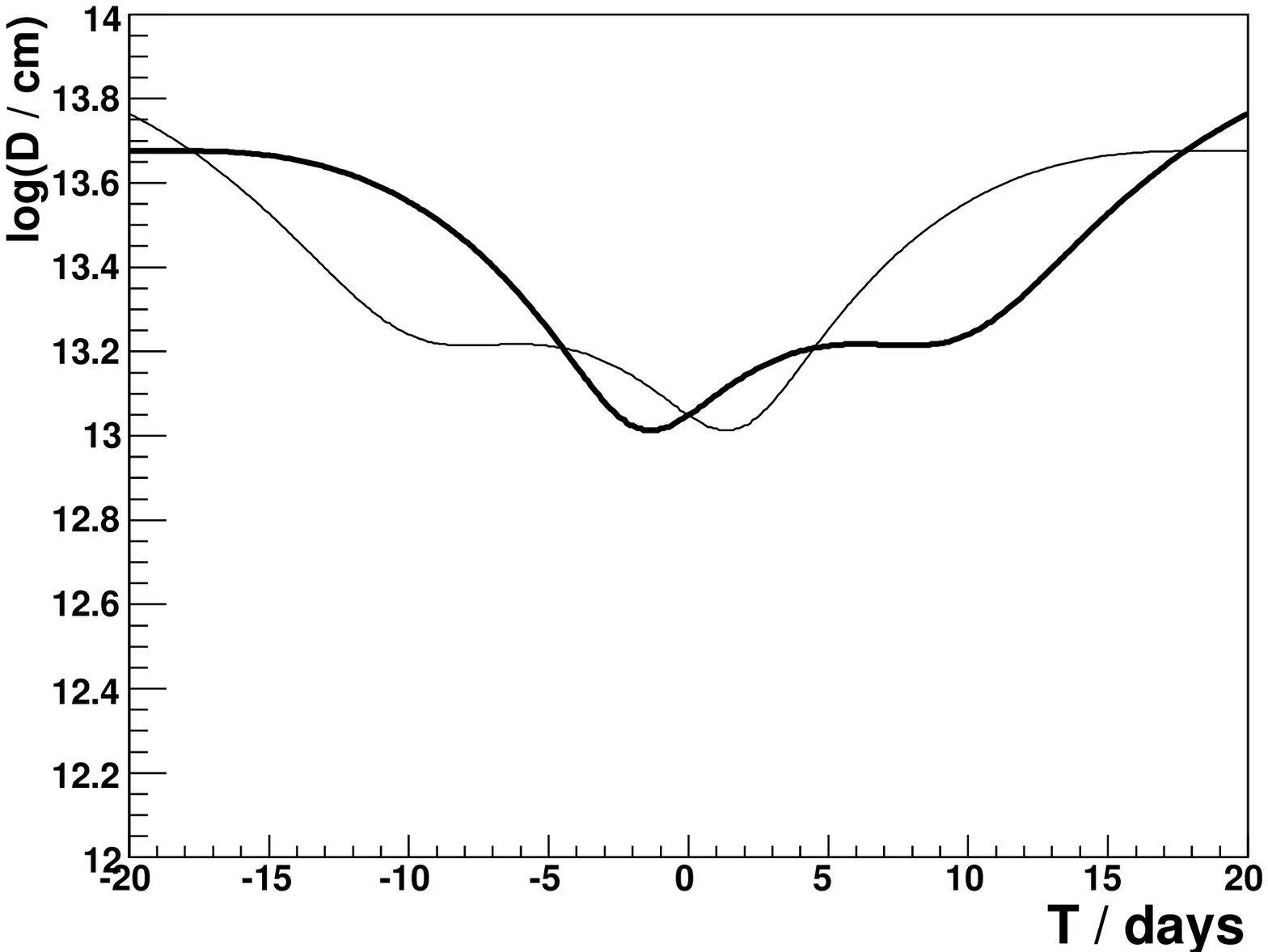}
\includegraphics{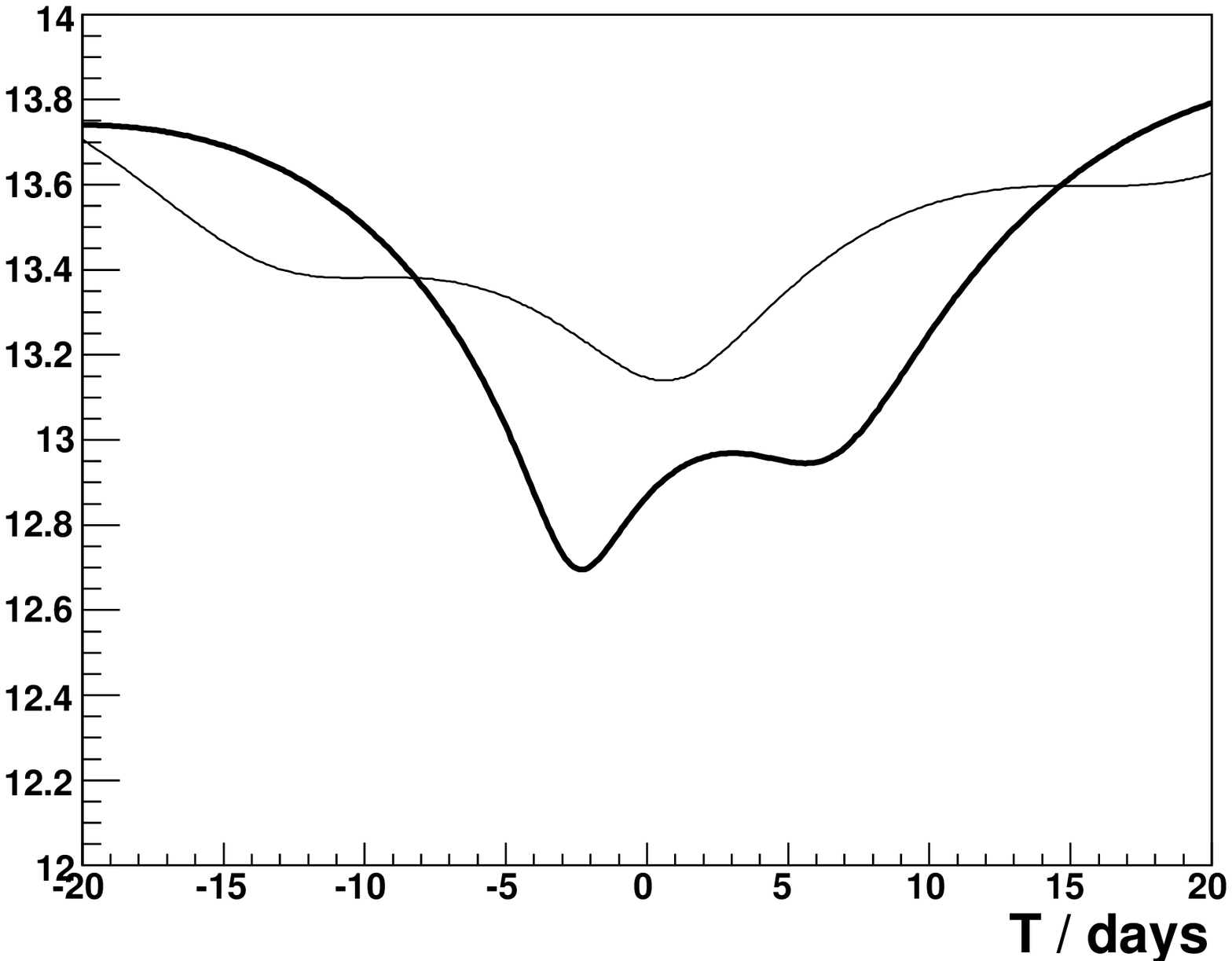}
\includegraphics{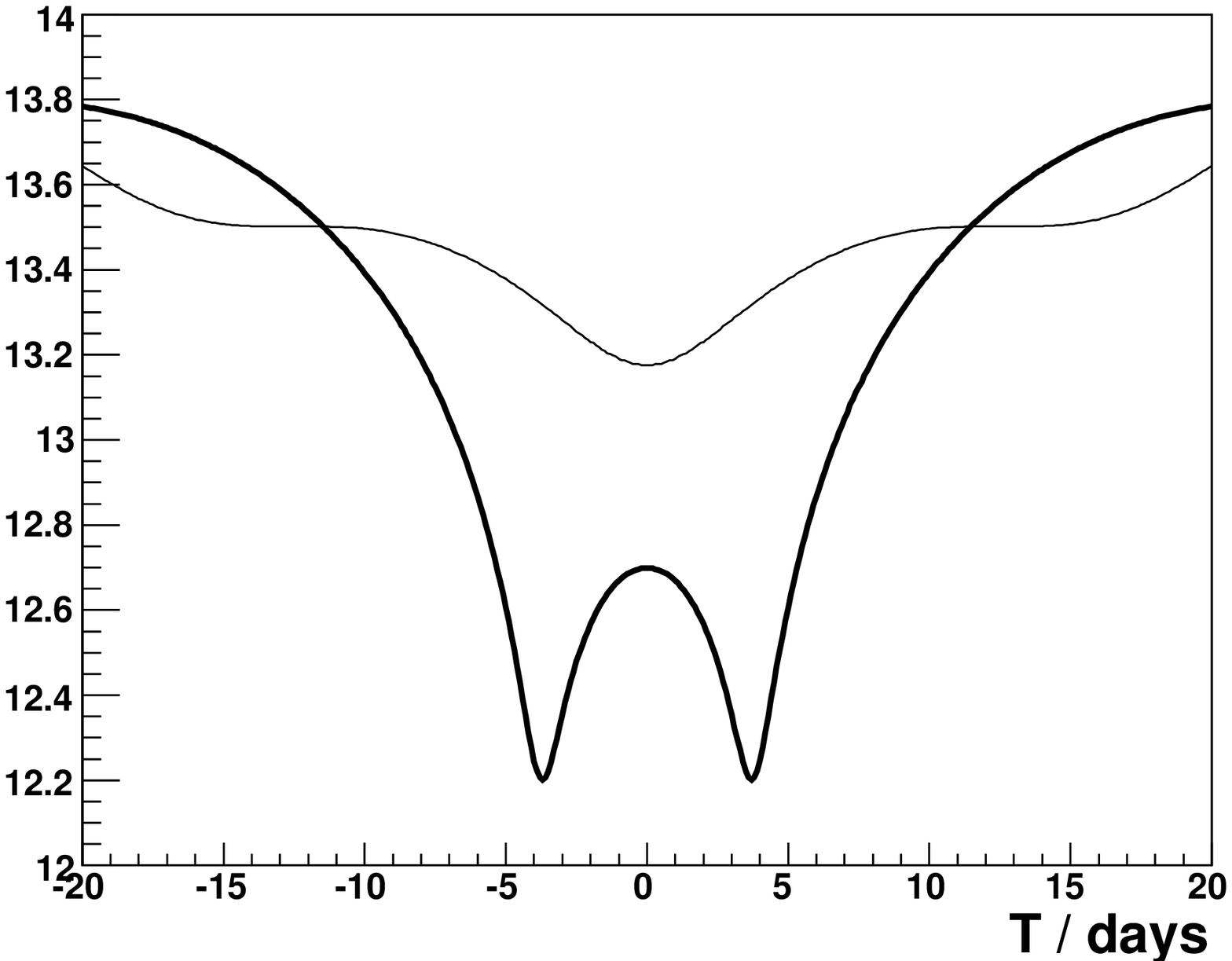}
\includegraphics{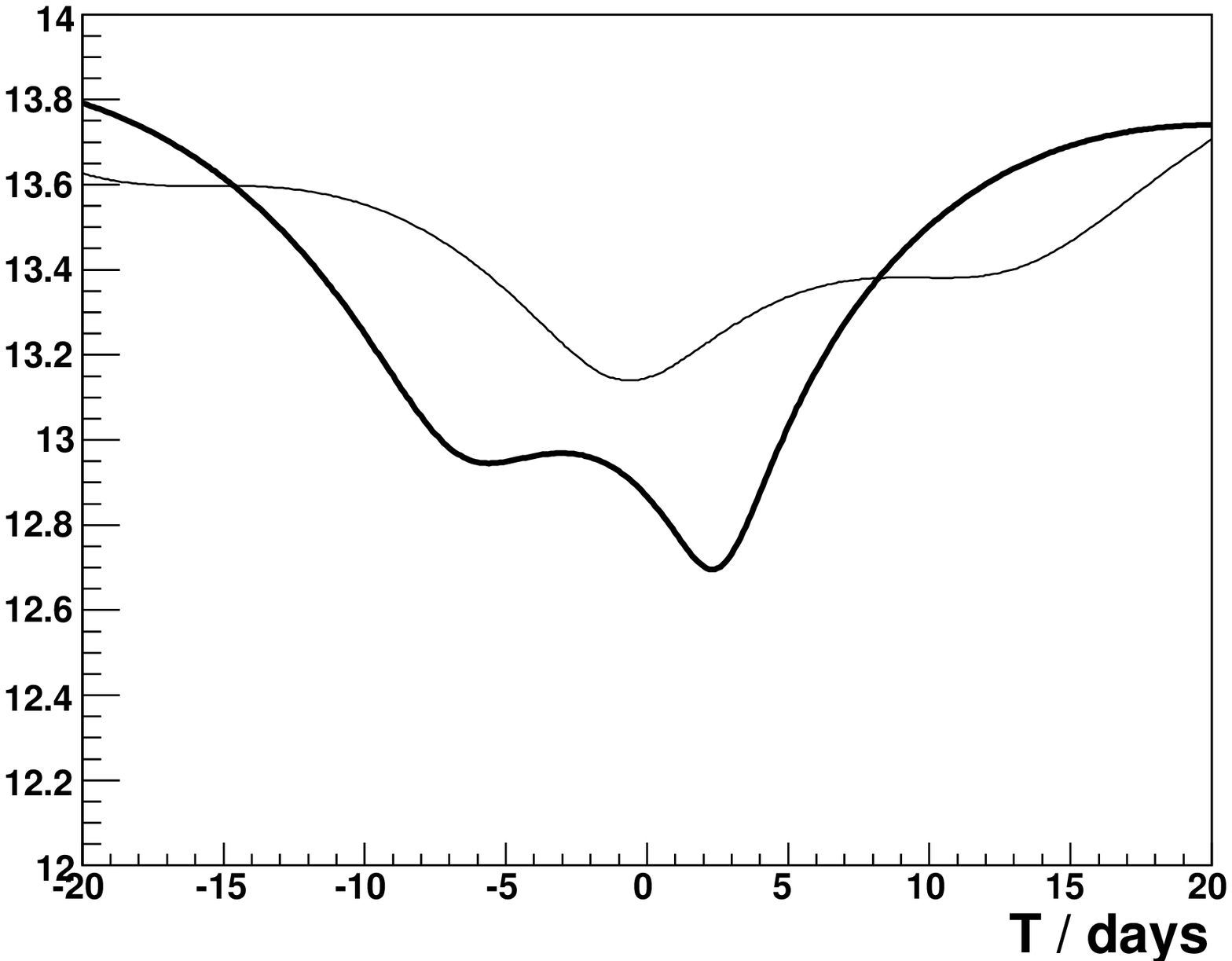}
\caption{Dependence of the distances, $D$, between the stars and the direction to the observer's line of sight
on the phase of the binary system $\phi$ is shown for the parameters as 
in Fig.~6 in the upper panel and the impact parameter $D_{\rm bin} = 1.1\times 10^{13}$ cm (upper panel) 
and  $D_{\rm bin} = 5\times 10^{12}$ cm (bottom panel): $\phi = 0^\circ$ (left column), $45^\circ$ (left-centre),
$90^\circ$ (right-centre), and $135^\circ$ (right).
The parameters of the binary system are $v_{\rm bin} = 3\times 10^7$~cm~s$^{-1}$, 
$v_\star = 3\times 10^7$~cm~s$^{-1}$, $a = 10^{13}$~cm. The binary system is composed from two stars with 
parameters $R_\star = 10^{12}$~cm and $T_\star = 3\times 10^4$~K.} 
\label{fig7}
\end{figure*}

The cumulative effect of absorption of $\gamma$ rays (expressed by the reduction factor $RF$) 
in the radiation of stars in the example transiting binary system
are shown in Fig.~8 for the case of two threshold energies $E_{\rm min} = 30$ GeV (upper figures) and 300 GeV
(lower figures).
The simplest binary system case, namely two identical stars with the orbital plane perpendicular to the direction of the $\gamma$-ray beam, is considered.
We show the absorption effects for the case of four impact parameters,
$D_{\rm bin} = 10^{13}$ cm, $5\times 10^{12}$ cm, $-5\times 10^{12}$ cm, and $-10^{13}$ cm,  
and the initial phase $\phi = 0^\circ$ (see Fig.~5). 
The negative distances of the impact parameters denote opposite rotation direction of the stars than marked in Fig.~5.

The significant reduction of the $\gamma$-ray emission is 
predicted to occur on a time scale of a few tens of days. However, a strong reduction of the $\gamma$-ray flux, 
in the form of characteristic two strong 
absorption dips, is also expected on a time scale of a few days when the observer's line of sight comes close
to the stars within the binary system.

\begin{figure*}
\vskip 7.truecm
\includegraphics{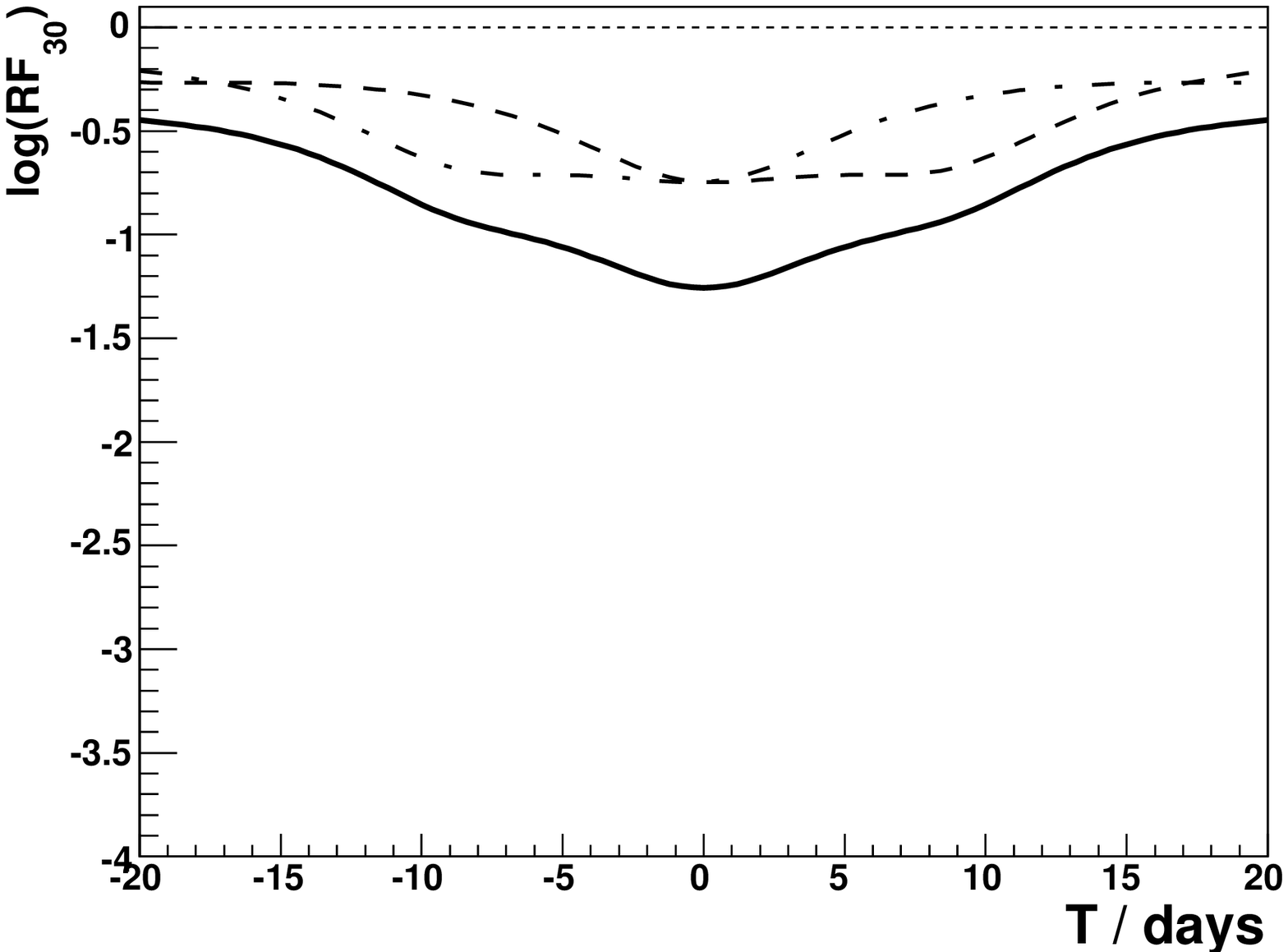}
\includegraphics{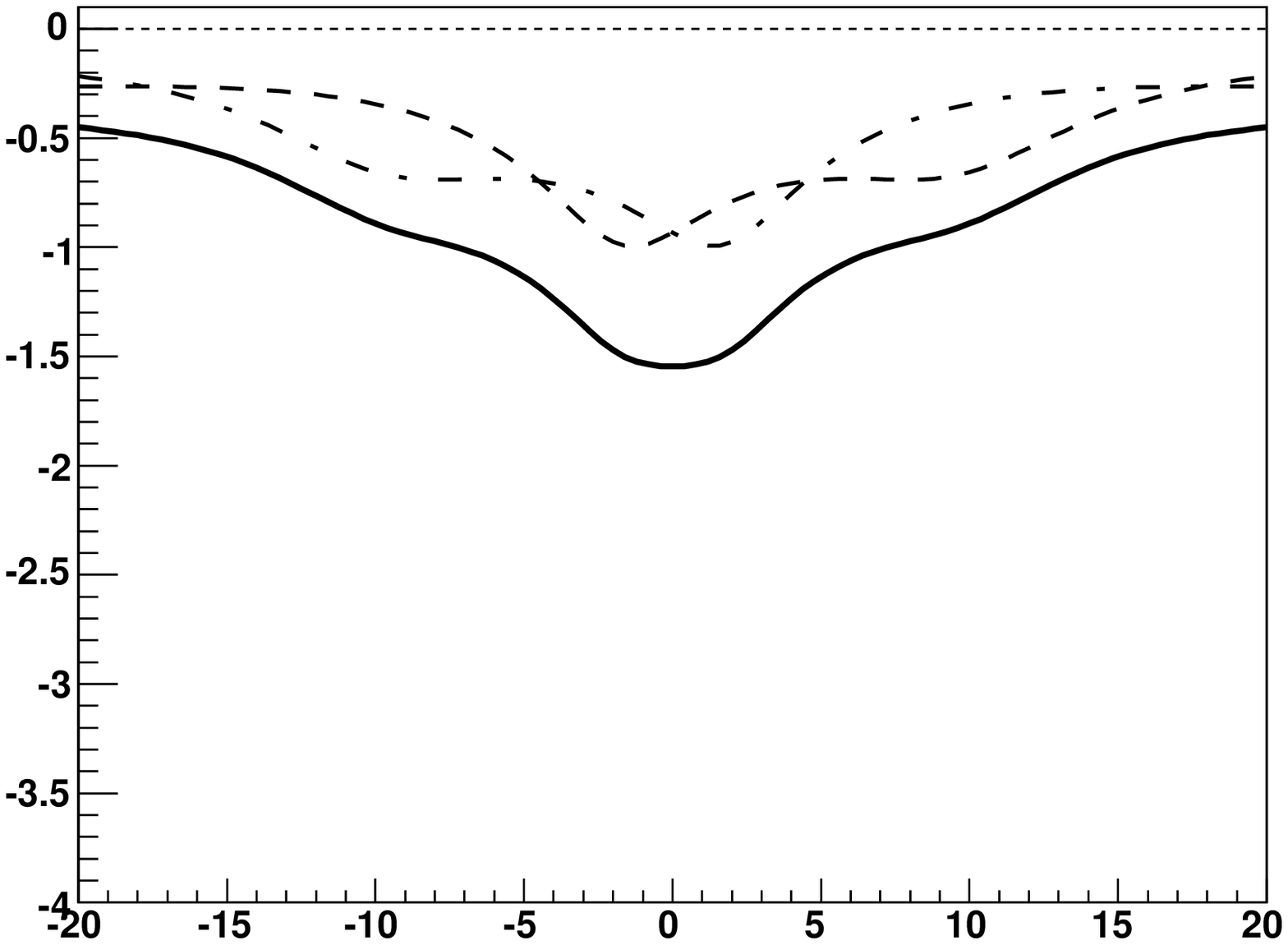}
\includegraphics{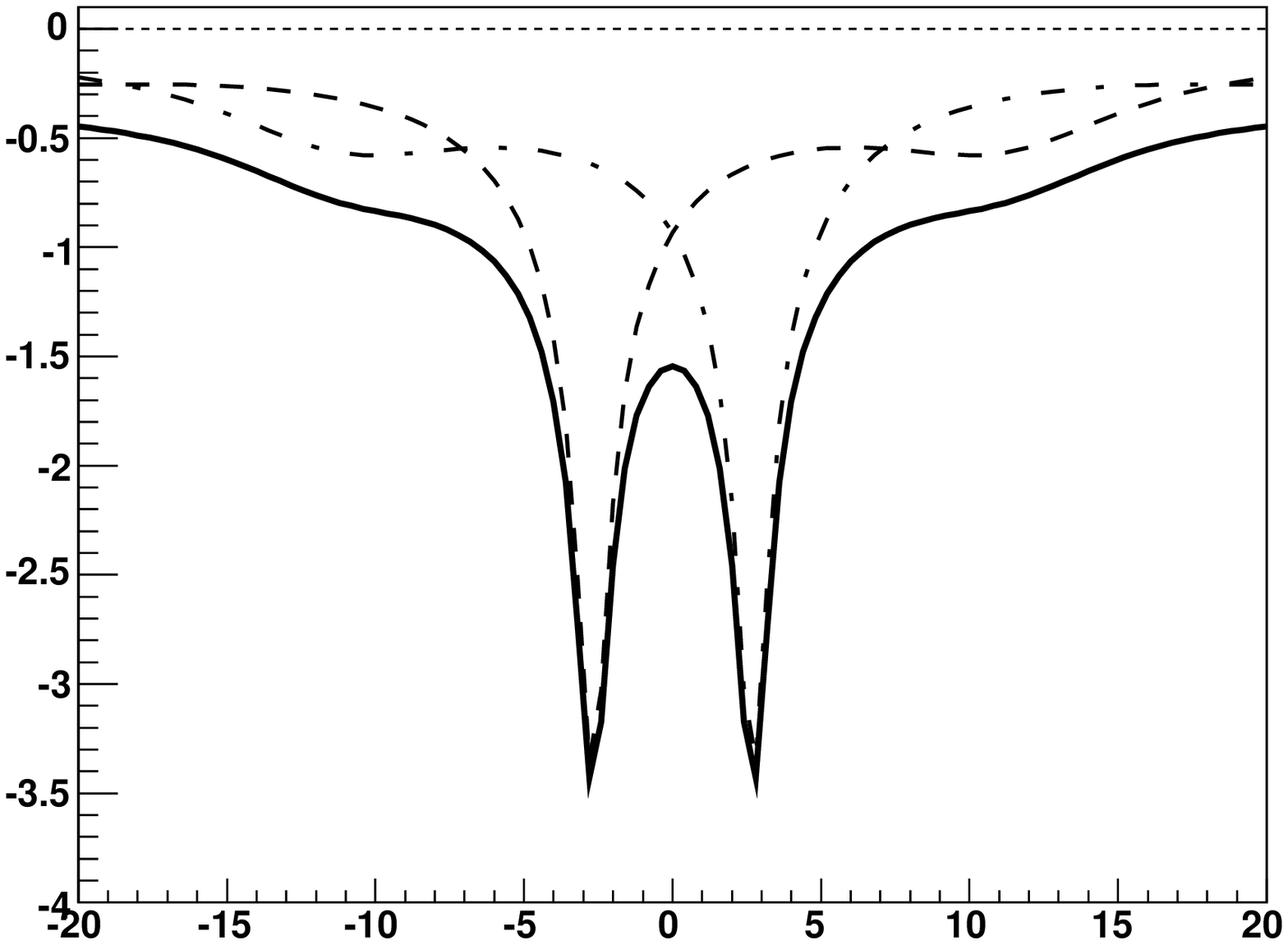}
\includegraphics{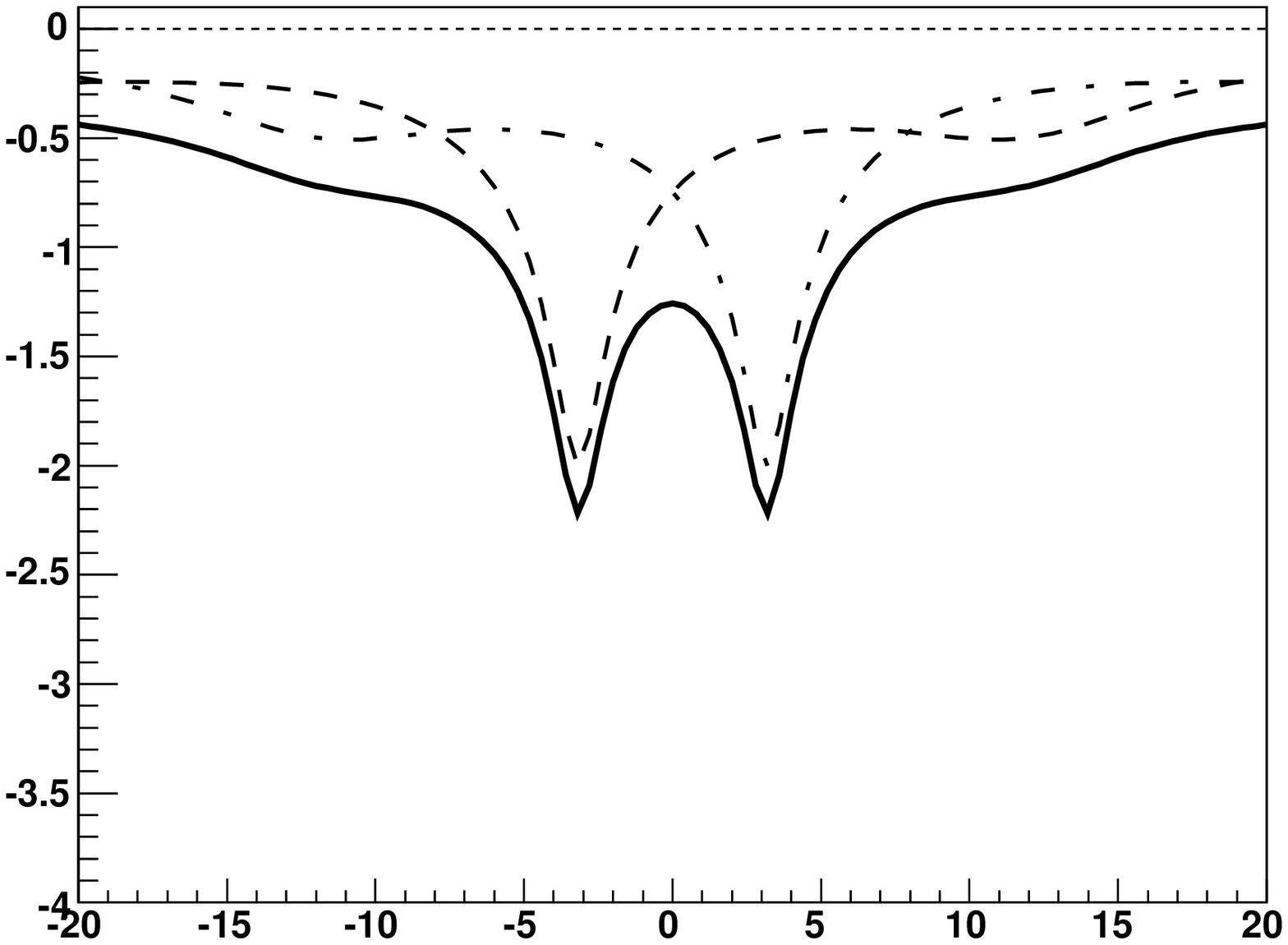}
\includegraphics{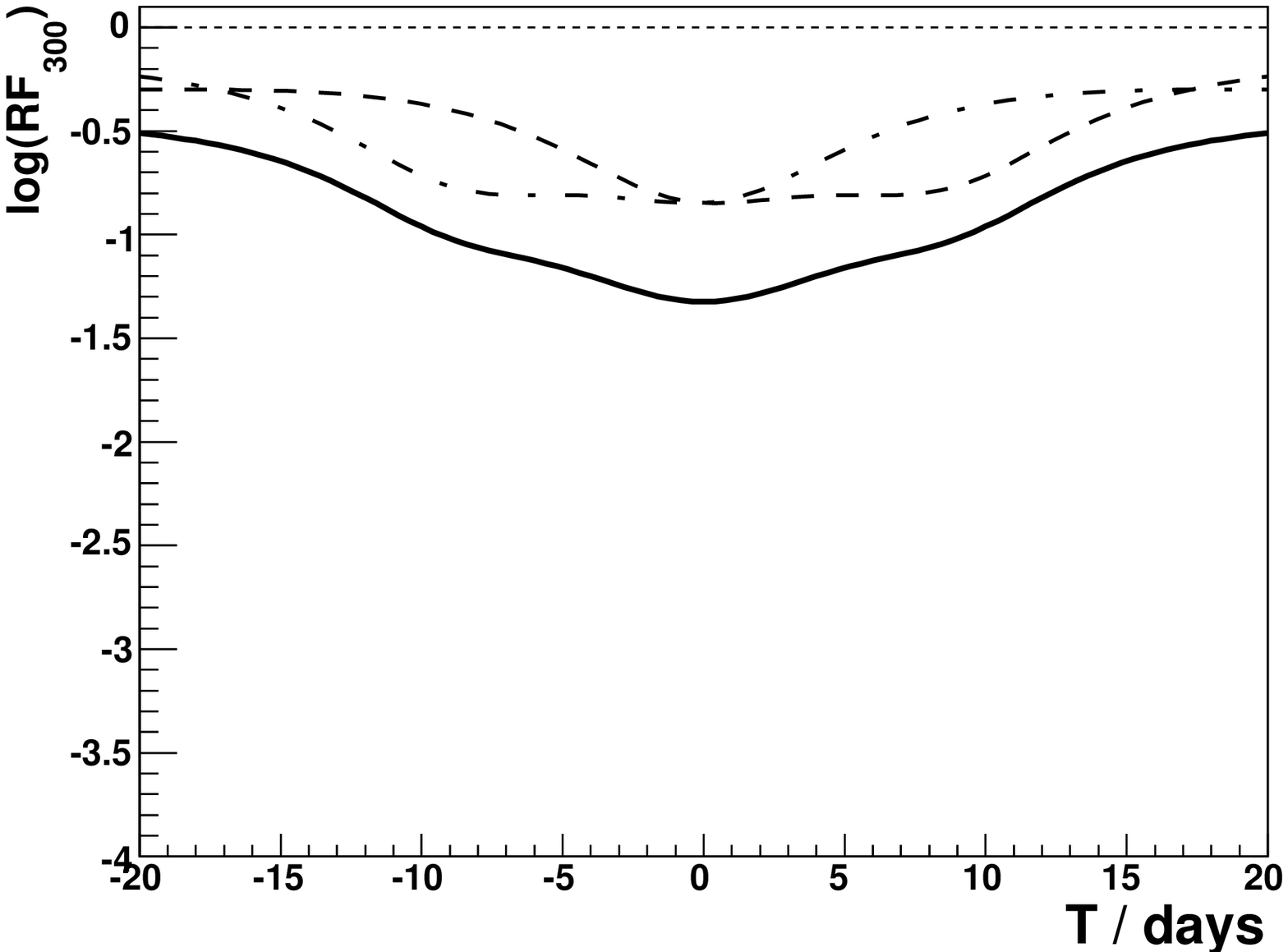}
\includegraphics{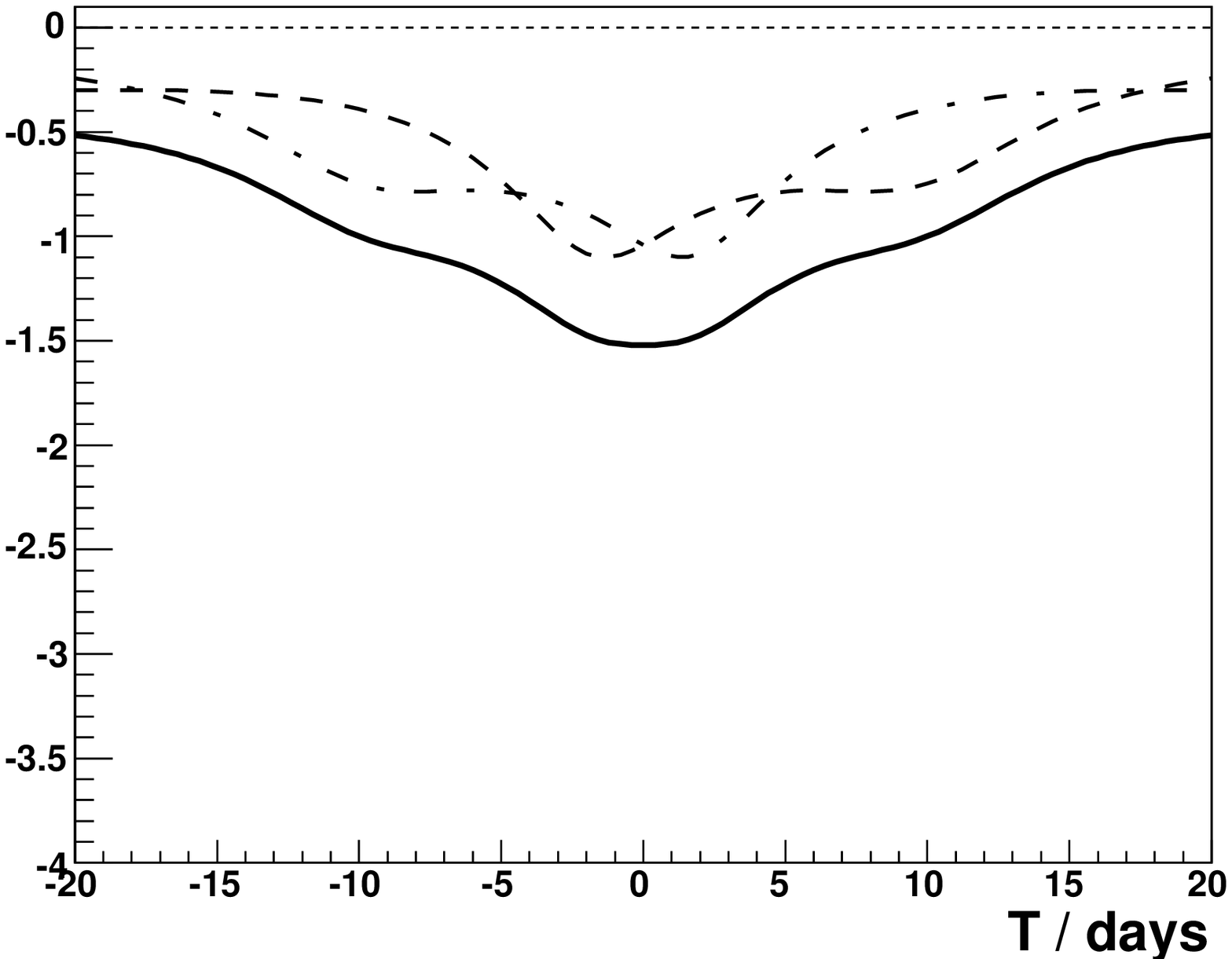}
\includegraphics{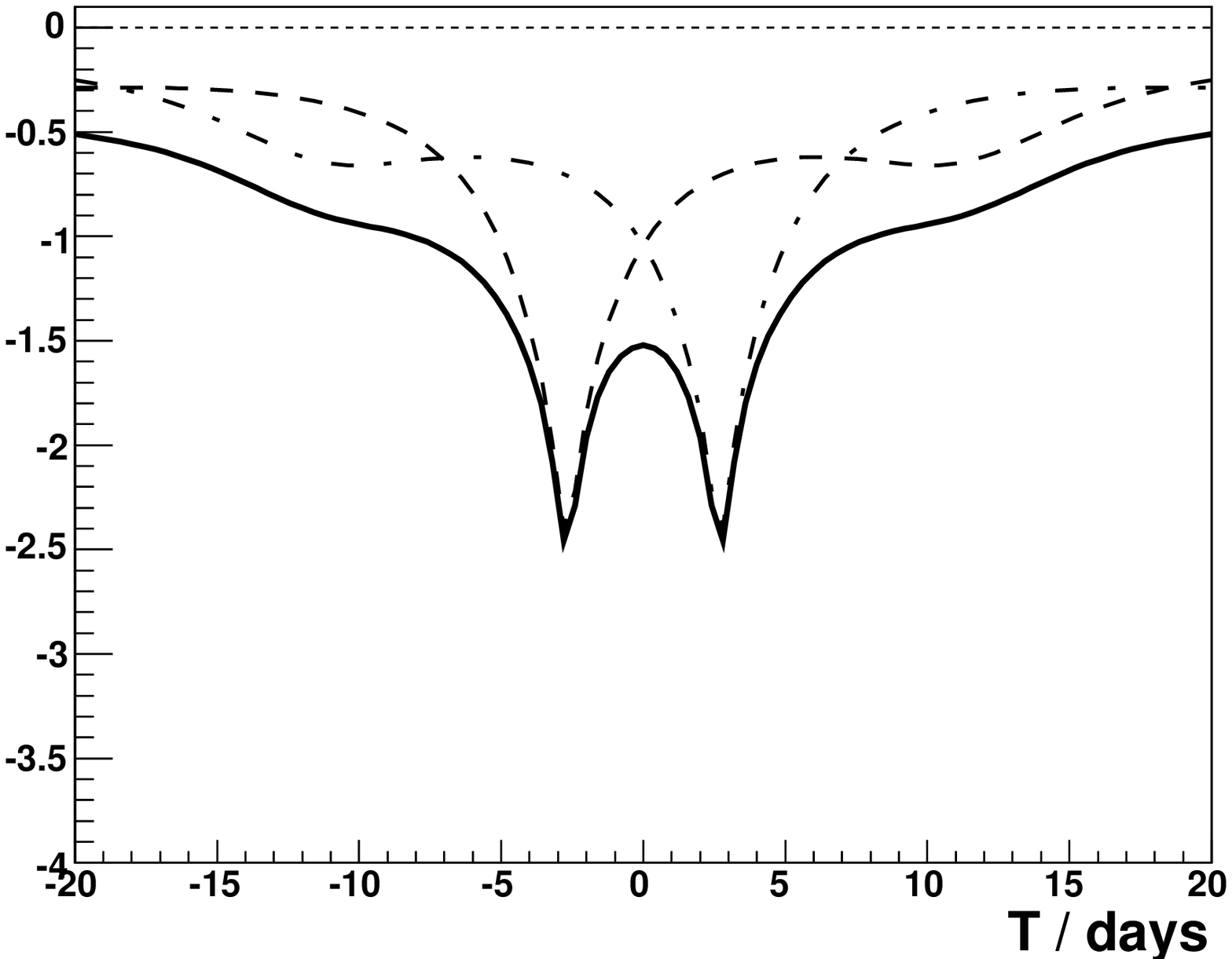}
\includegraphics{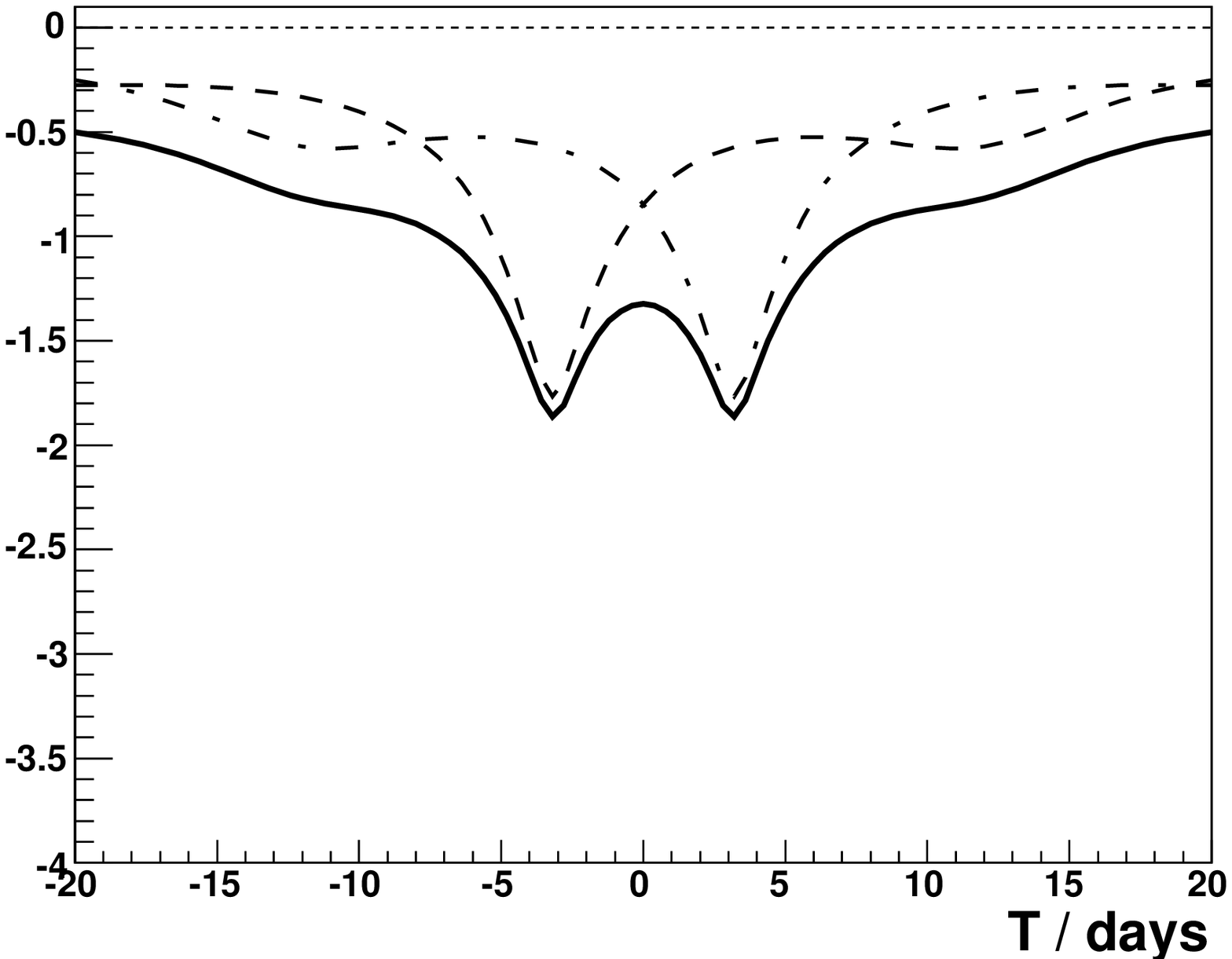}
\caption{Effect of the transition of the stellar binary system on 
the $\gamma$-ray flux above 30 GeV (see log RF$_{30}$ on the upper figures) and above 300 GeV (see log $RF_{300}$ 
on the bottom figures). The parameters of the binary system are the following: stellar radius 
$R_\star = 10^{12}$ cm, surface temperature $T_\star = 3\times 10^4$ K, semimajor axis $a = 10^{13}$ cm, 
stellar velocity on the binary orbit $v_\star = 3\times 10^7$ cm s$^{-1}$). Specific figures show the results 
for different impact parameters $D_{\rm bin} = 1.1\times 10^{13}$ cm (on the left figures), $5\times 10^{12}$ cm 
(left-centre), $-5\times 10^{12}$ cm (right-centre), and $-1.1\times 10^{13}$ cm (right). 
The $\gamma$-ray beam has the differential power law spectrum with the spectral index -2.} 
\label{fig8}
\end{figure*}

\begin{figure*}
\vskip 4.truecm
\includegraphics{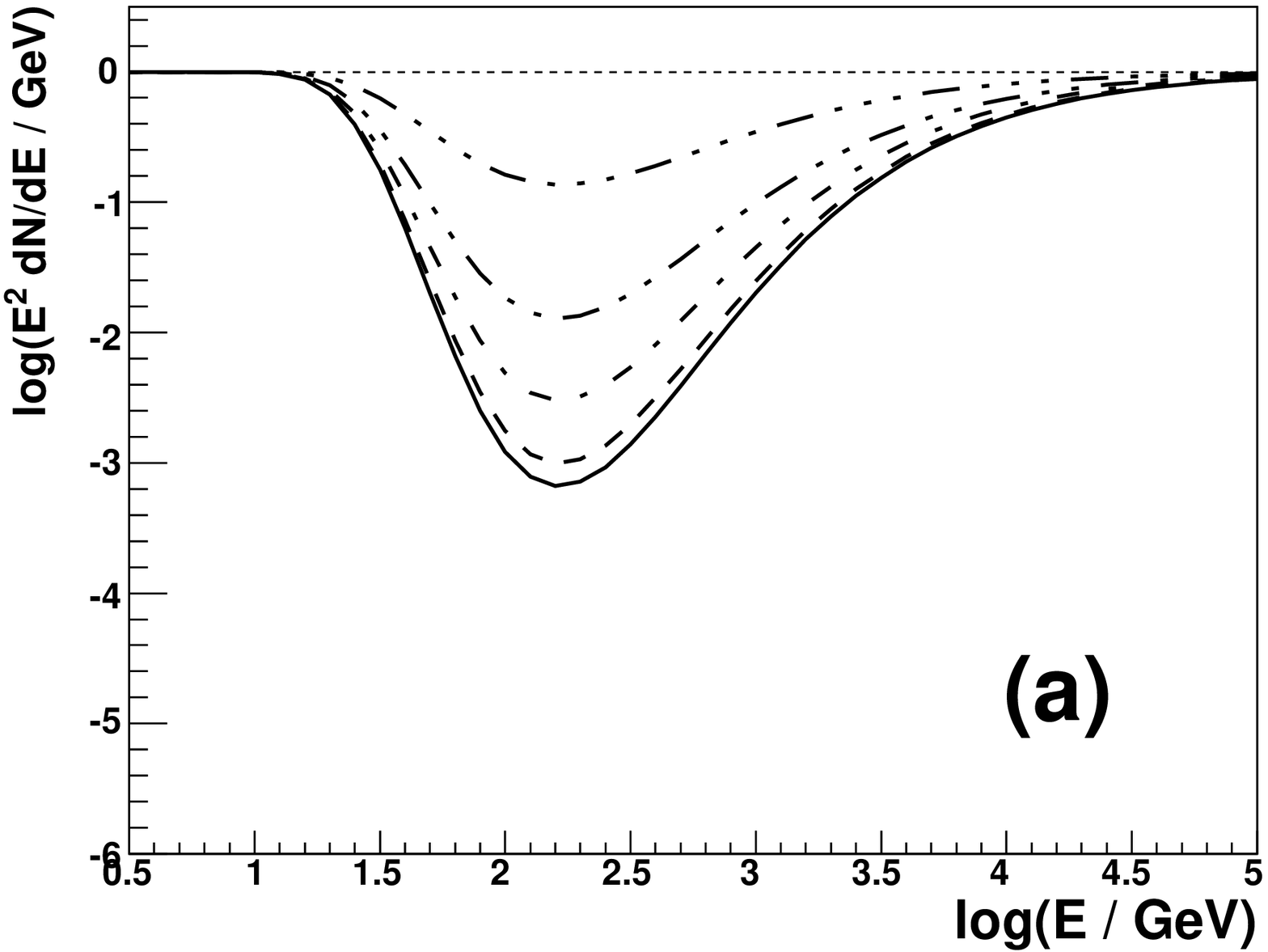}
\includegraphics{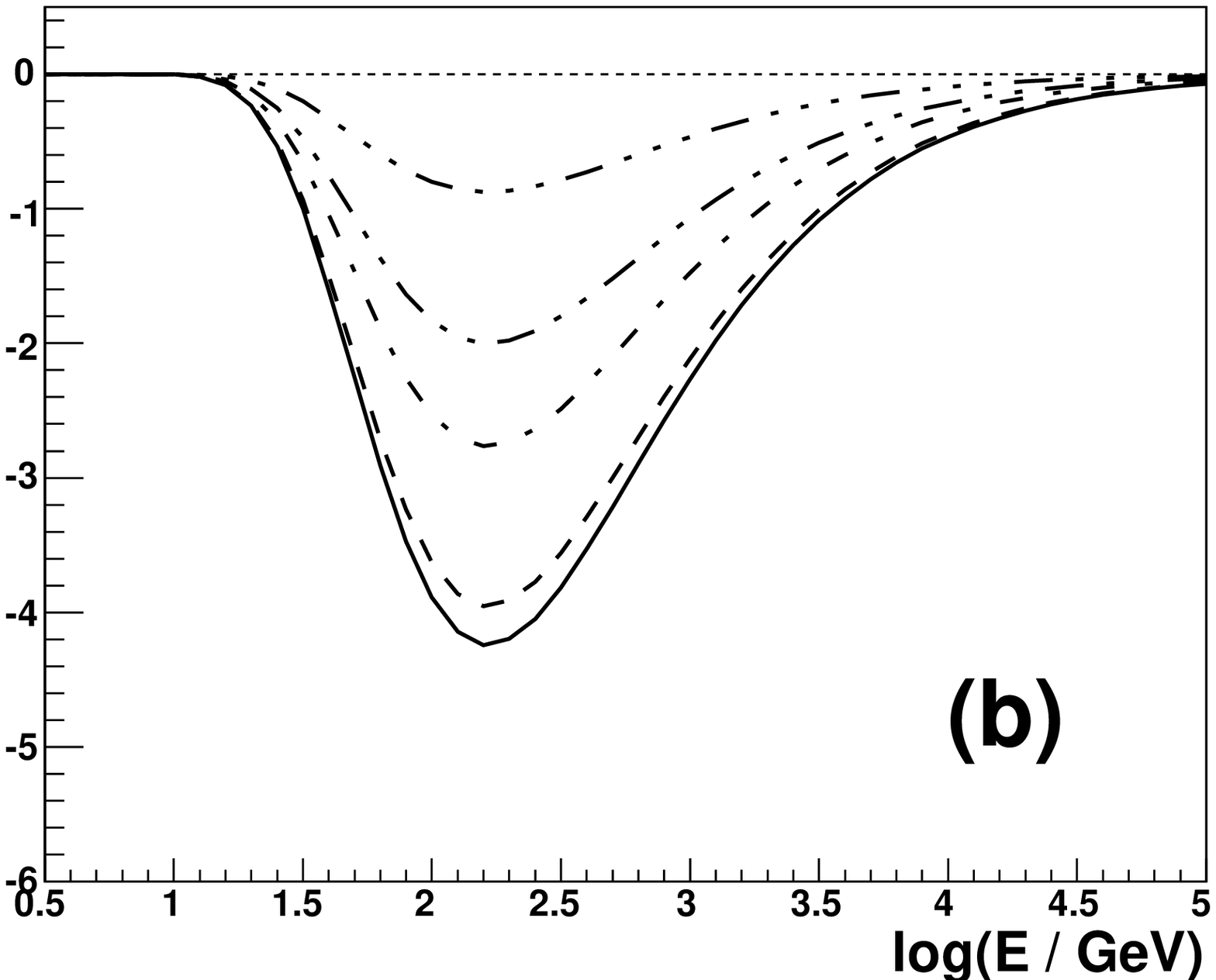}
\includegraphics{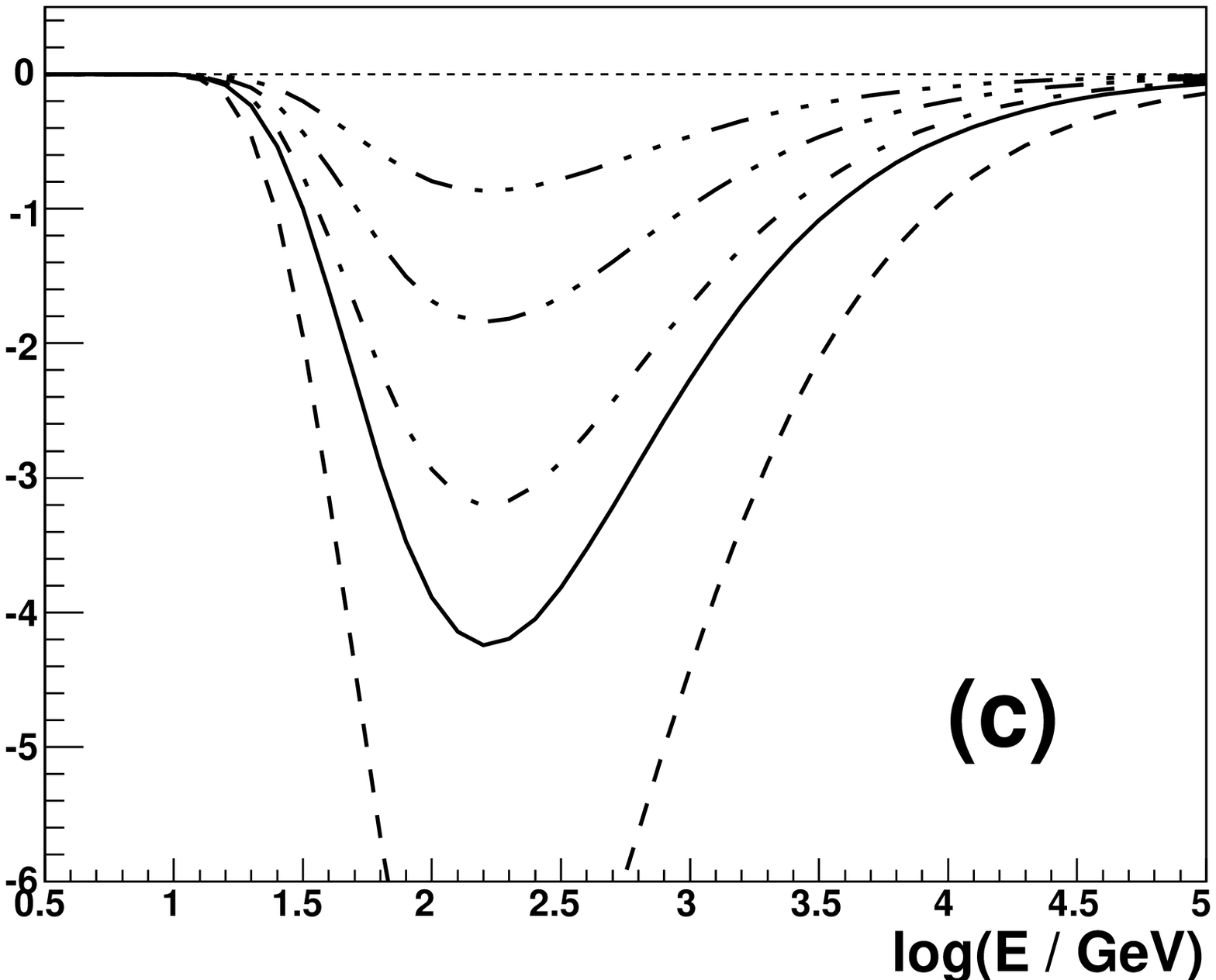}
\includegraphics{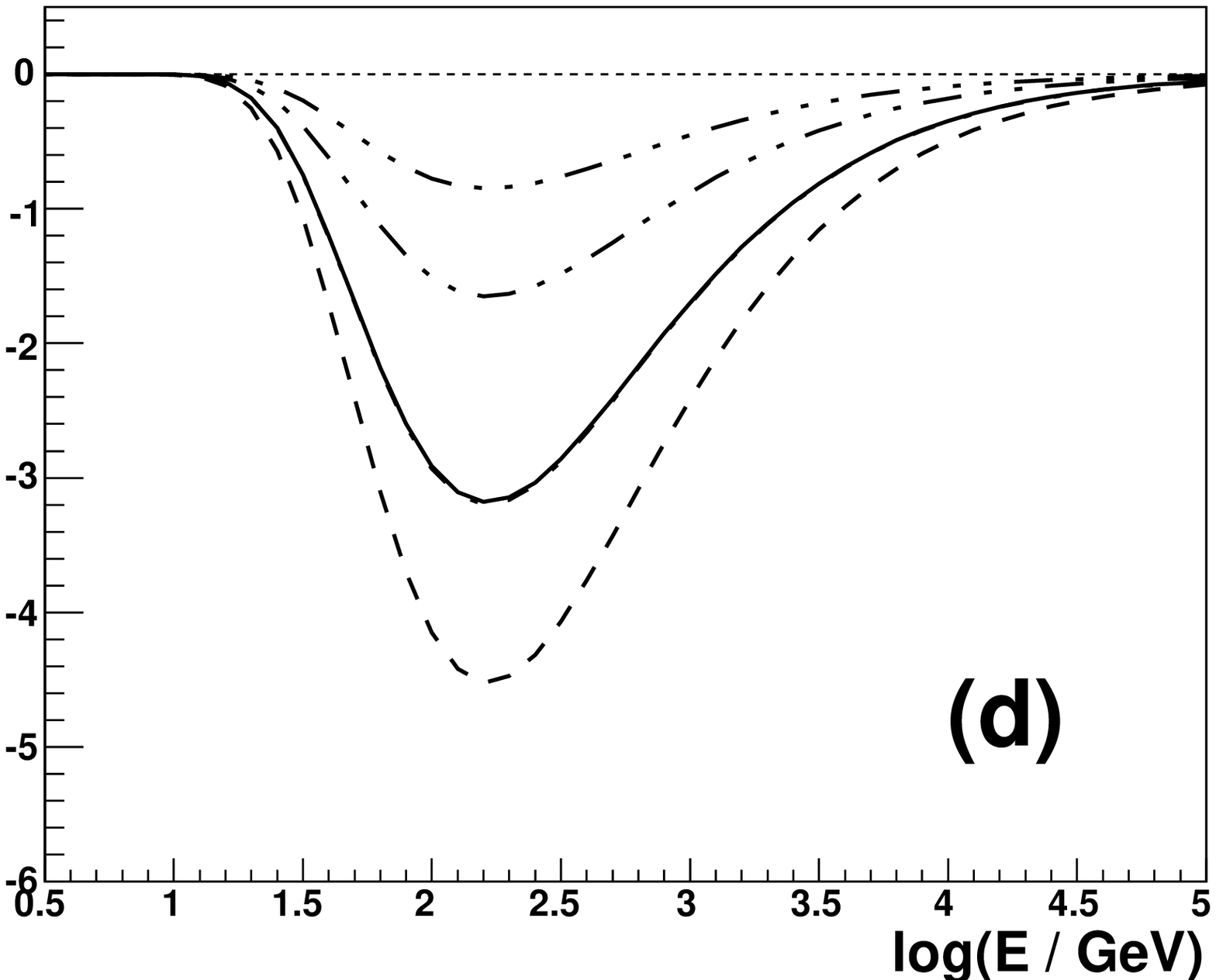}
\caption{Absorption effect on a simple spectrum of the $\gamma$-ray emission
(differential power law spectrum multiplied by energy squared) at different transit times, 
T = -20~days (dot-dot-dot-dashed), -10 days (dot-dot-dashed), -5 days (dot-dashed), -2 days (dashed), 
0 days (solid) for the parameters: (a) the binary system of two stars (with stellar radius 
$R_\star = 10^{12}$ cm, surface temperature $T_\star = 3\times 10^4$ K, semimajor axis 
$a = 10^{13}$ cm, stellar velocity on the binary orbit $v_\star = 3\times 10^7$ cm s$^{-1}$)
passes the $\gamma$-ray beam with the velocity of the binary $v_{\rm bin} = 3\times 10^7$ cm s$^{-1}$,
 and the impact parameter $D_{\rm bin} = 1.1\times 10^{13}$\ cm (see a),  
$D_{\rm bin} = 5\times 10^{12}$ cm (b), $D_{\rm bin} = -5\times 10^{12}$ cm (c), and  $D_{\rm bin} = -1.1\times 10^{13}$\ cm (d). 
See Fig.~5, for the location of the binary in respect to the plane of the sky.} 
\label{fig9}
\end{figure*}

We also calculate the effect of absorption on the spectrum of the $\gamma$-ray beam (see Fig.~\ref{fig9}), 
for the
case of the example binary system considered above. Interesting dependence of the spectrum on the 
transiting time can be observed.  The basic feature, softening of the multi-GeV part  of the spectrum and 
hardening of the sub-TeV part of the spectrum, can appear regularly during the transition event close to
the minimum approach of the observer's line of sight. For the considered transition event, the effects
of absorption are so strong that the $\gamma$-ray flux of the $\gamma$-ray beam can be drastically reduced.
It can easily fall below the sensitivity limits of the $\gamma$-ray telescopes in this energy range.   
Note that considered here effects should strongly depend on the geometry of the binary system in respect 
to the observer's line of sight which greatly enhances the possibility of different absorption effects.
Here we consider only the simplest possible geometry of the binary system, i.e. it is composed of 
two equal-mass stars and its plane lays in the plane 
of the sky. However, in general, the plane of the binary system can be inclined at an arbitrary angle to 
the plane of the sky. Moreover, the binary system can be formed from two luminous stars which 
significantly differ in their basic parameters (stellar mass and temperature, binary radius). Therefore, 
we expect even more complicated structures in the light curve of the transition of binary stars 
through the $\gamma$-ray beam formed close to the central engine of active galaxy.

\section{Effects on specific sources}

As an example, we consider the absorption effects on the $\gamma$-ray spectra due to the transiting 
binary system through the $\gamma$-ray beam in the case of two well known active galaxies from
which GeV-TeV $\gamma$-ray emission have been reported by the \textit{Fermi}-LAT and Cherenkov telescopes. 
The first active galaxy is a BL Lac object: 1ES\ 1959+650.
Recently it was detected in the high state by the \textit{Fermi}-LAT in the multi-GeV energies and by MAGIC 
in sub-TeV energies (Acciari et al. 2020b).
We include the effects of absorption of $\gamma$-ray spectra observed from 
1ES\ 1959+650 in the case of transits of two binary systems with likely binary system and transit parameters 
(see left columns of Fig.~10). The modifications of the $\gamma$-ray spectrum by the transiting binary system 
at a specific
time before minimum distance between the centre of the binary system and the observer's line of sight are
shown for $T=-20$ -- $0$\,days. 
We also consider the effect of a transit of the binary system on the spectrum of the well known radio 
galaxy M87 detected in a low state (Acciari et al. 2020a).
As in the case of 1ES\ 1959+650, the transits of two binary systems with such parameters are considered. 
Expected modifications of the GeV to TeV $\gamma$-ray spectrum are reported in the middle column of Fig.~10.

\begin{figure*}
\includegraphics[width=0.49\textwidth]{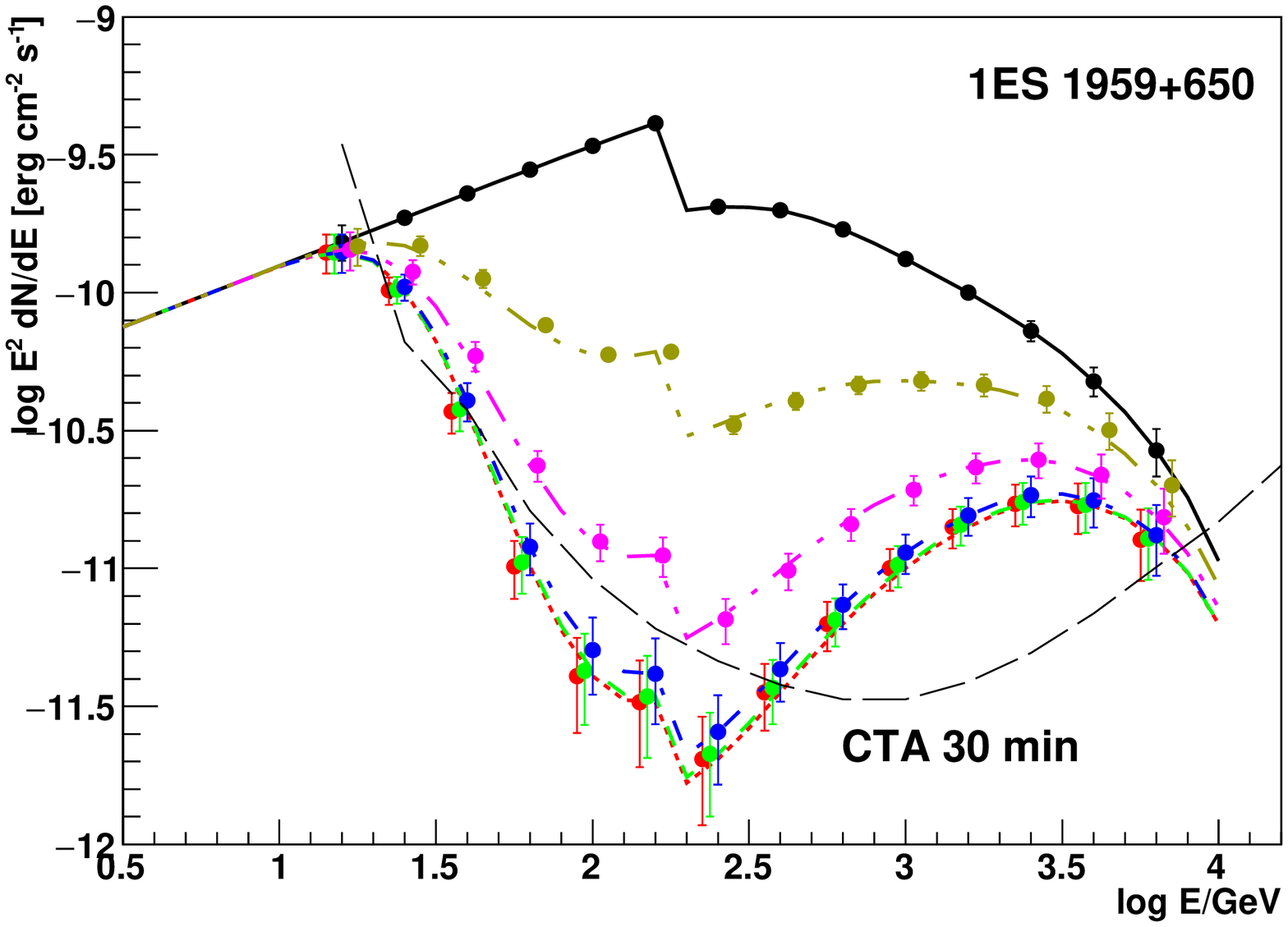}
\includegraphics[width=0.49\textwidth]{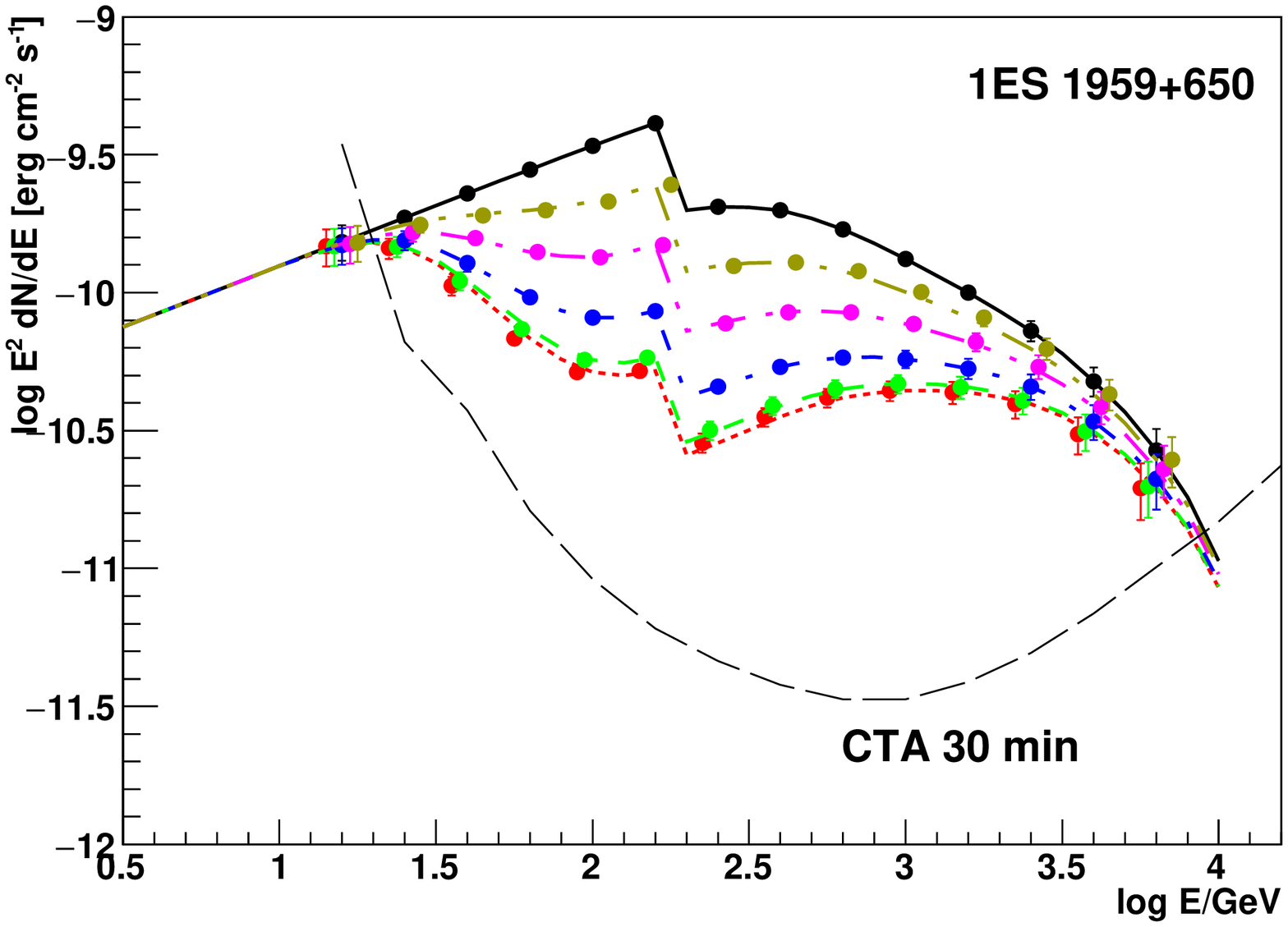}\\
\includegraphics[width=0.49\textwidth]{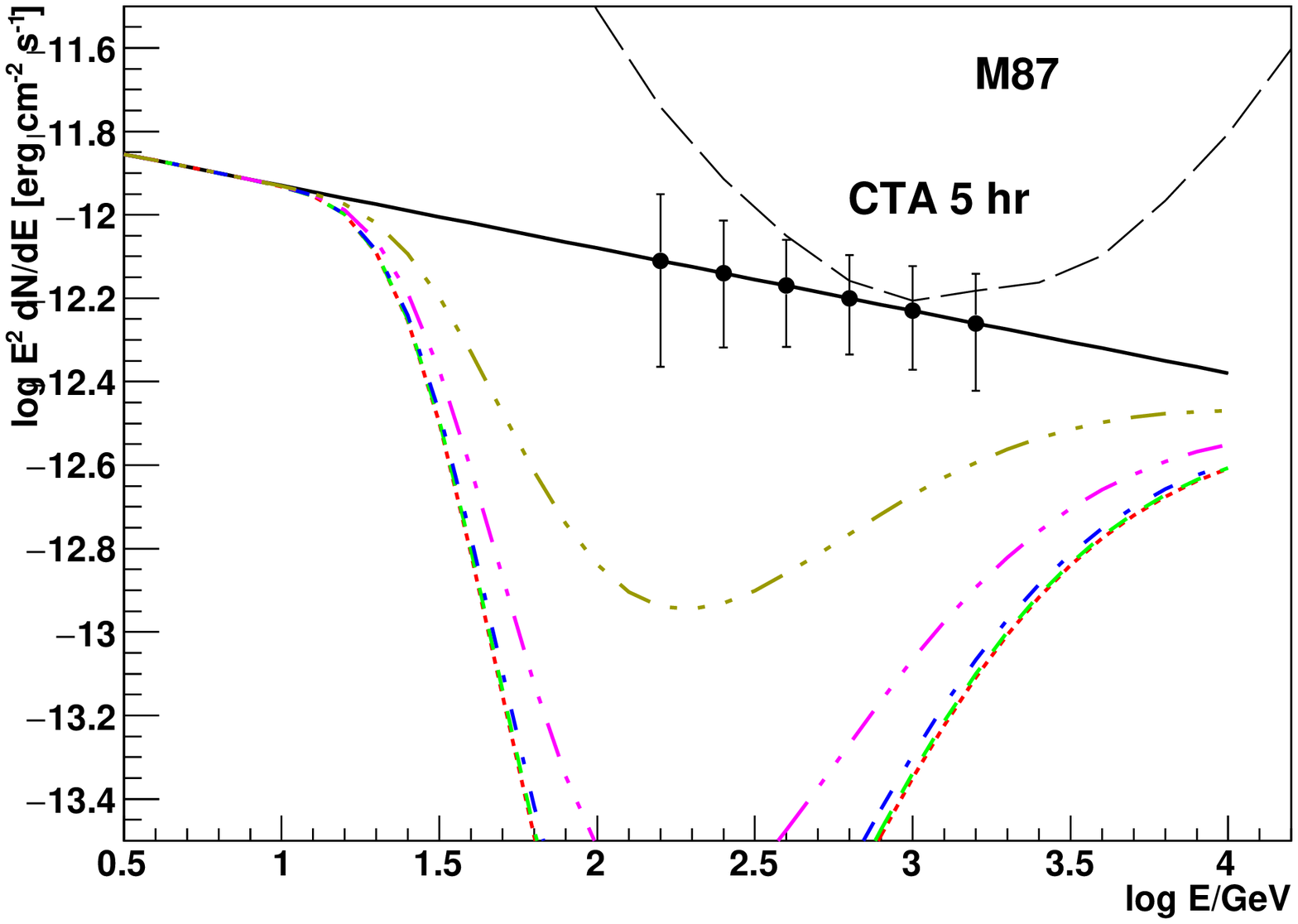}
\includegraphics[width=0.49\textwidth]{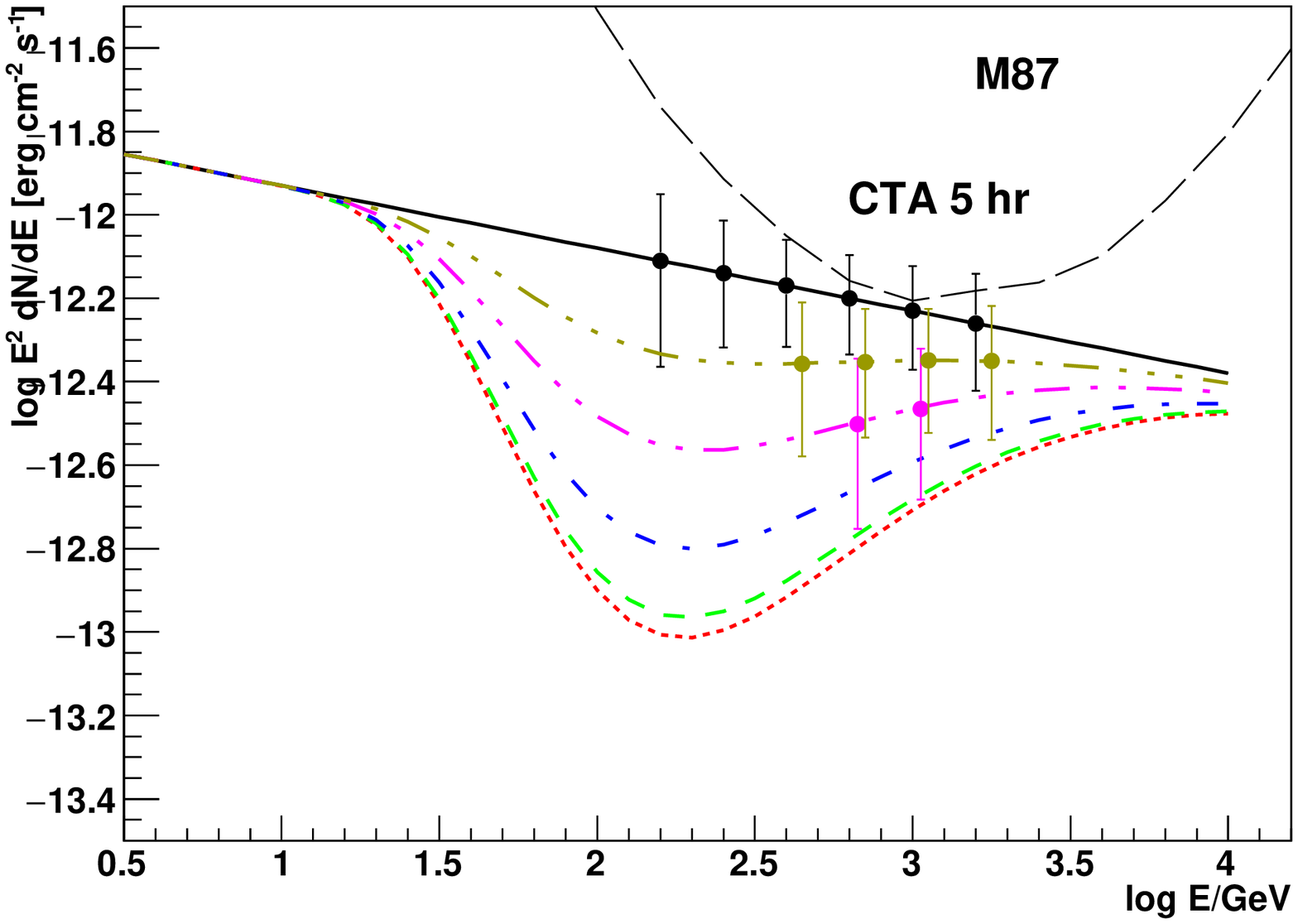}\\
\includegraphics[width=0.49\textwidth]{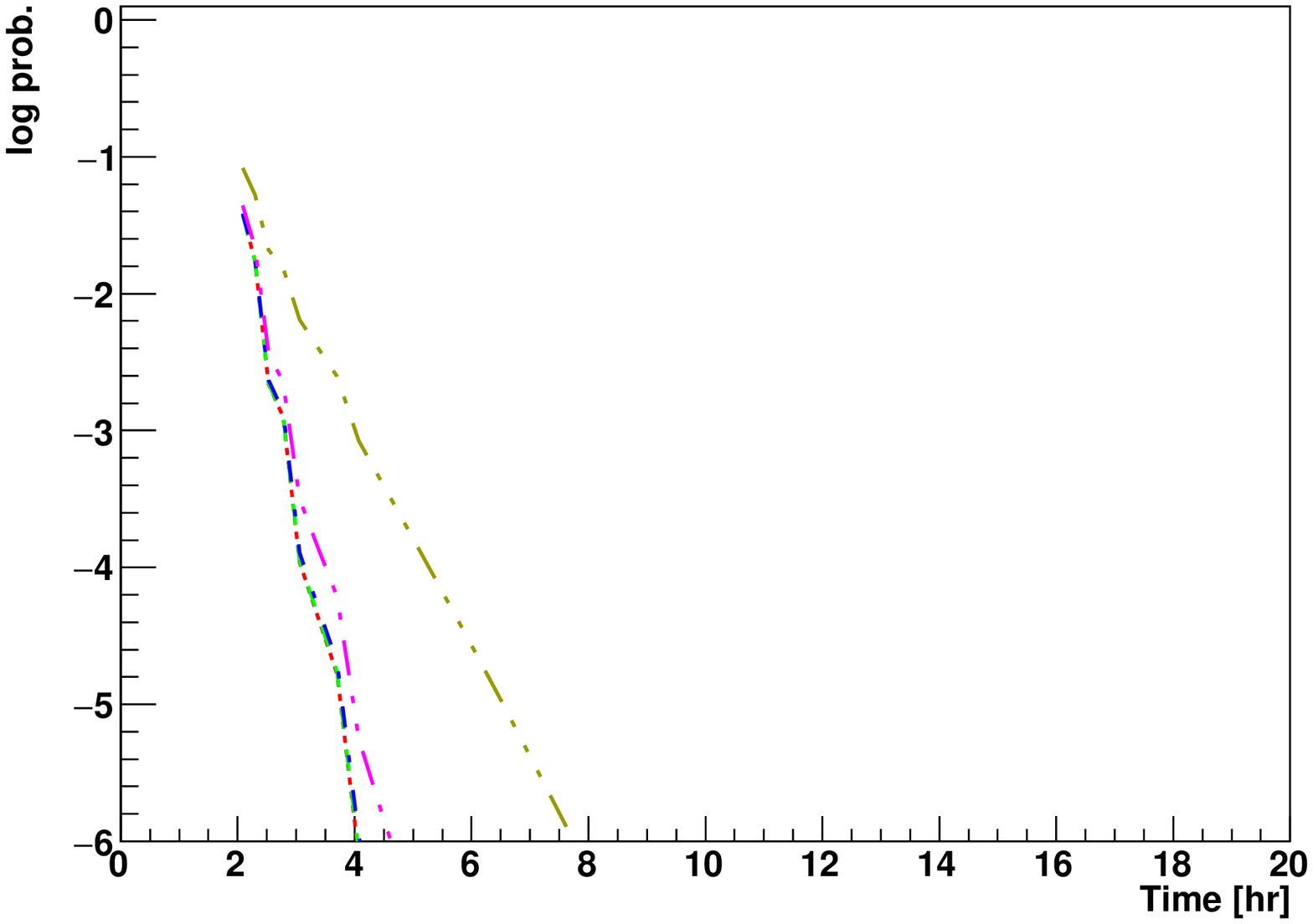}
\includegraphics[width=0.49\textwidth]{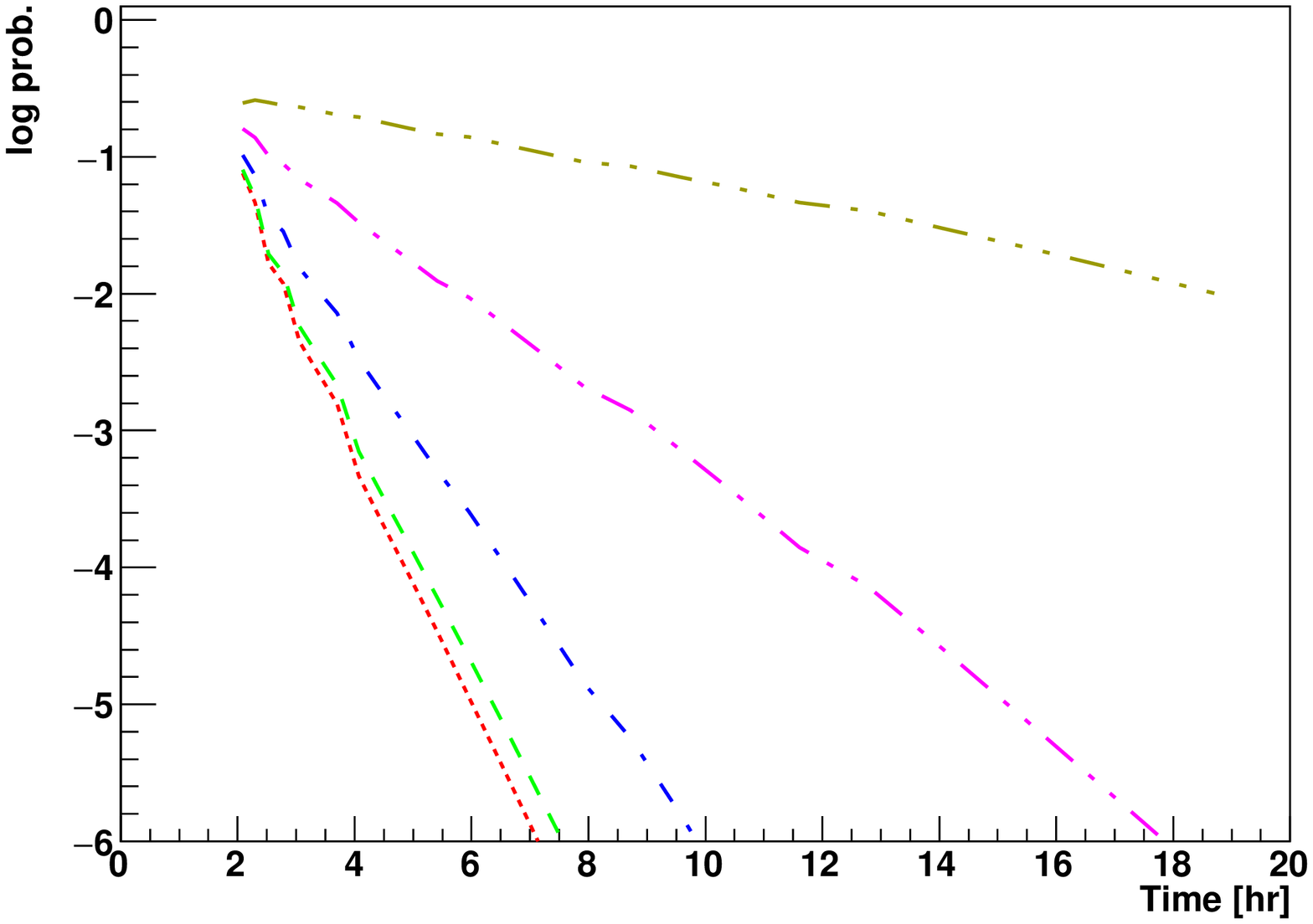}

\caption{Effect of absorption on the $\gamma$-ray spectrum observed from two active galaxies: 
BL Lac 1ES\ 1959+650 (in the high emission state, see Acciari et al. 2020b, including the absorption 
in extragalactic background light using Dom{\'{\i}}nguez et al.\ 2011 model) and the radio galaxy M87 
(in the low state, see Acciari et al. 2020a). The absorption effect 
due to the passing binary system of 
stars with the parameters
$R_\star = 10^{12}$ cm, $T_\star = 3\times 10^4$ K, $a = 10^{13}$ cm, $D_{\rm bin} = 2\times 10^{13}$ cm, $v_\star = 3\times 10^7$ cm s$^{-1}$, and 
$v_{\rm bin} = 3\times 10^7$ cm s$^{-1}$ through the $\gamma$-ray beam is shown for 1ES\ 1959+650 (top left)
and M87 (center left)  for the parameters,
$R_\star = 10^{12}$ cm, $T_\star = 3\times 10^4$ K, $a = 10^{13}$ cm, $D_{\rm bin} = 5\times 10^{13}$ cm, $v_\star = 3\times 10^7$ cm s$^{-1}$. 
The transition events with  
$v_{\rm bin} = 10^8$ cm s$^{-1}$ are shown for 1ES\ 1959+650 (top right) and M87 (center right).
The spectra with absorption features correspond to different time measured in respect to the closest 
distance between the centre of the binary system and the direction to the observer: T = 0 days 
(dotted red curve), 
-2 days (dashed green), -5 days (dot-dashed blue), -10 days (dot-dot-dashed magenta), and -20 days 
(dot-dot-dot-dashed olive).  
The spectra are confronted with the 0.5\ hr (for 1ES\ 1959+650) or 5\ hr  (for M87) sensitivity of CTA 
(see text).
Points and uncertainties show the expected range and accuracy of the reconstructed spectrum by CTA for 
the given observation time with CTA (the points are slightly shifted between each curve for clarity).
The bottom row shows the chance probability corresponding to detection of absorption feature in M87 
for $v_{\rm bin} = 3 \times 10^7$ cm s$^{-1}$ (bottom left) and $v_{\rm bin} = 10^8$ cm s$^{-1}$ (bottom right) 
as a function of observation time. 
} 
\label{fig10}
\end{figure*}

We confront these modified $\gamma$-ray spectra with the sensitivities of the $\gamma$-ray ground-based and 
space instruments.
We used the publicly available instrument response functions (IRF) 
\texttt{prod3b-2}\footnote{\url{https://www.cta-observatory.org/science/cta-performance/}} of the CTA 
Bernl{\"o}hr et al. (2013).
For the stronger source, 1ES1959+650 we use short term (30\,min) sensitivity IRF, while for a weaker M87 
we use the corresponding mid-term (5\,hr) IRF.
Moreover, for 1ES1959+650 we also take into account the absorption in the extragalactic background light 
following Dom{\'{\i}}nguez et al. (2011) model, which affects the observed emission from the source in the multi-TeV range. 
Using a given flux model, the collection area of the instrument, and its migration matrix, we determine 
the expected excess rates in each estimated energy bin.
With such computed expected number of observed $\gamma$ rays, and the rate of background events obtained 
from CTA IRF we estimate the expected uncertainty of the reconstructed flux.
We consider that the flux can be probed at a given energy if the expected uncertainty is below 50\% of 
the flux (i.e. resulting with $>2\sigma$ point) and the expected number of $\gamma$ rays in this energy 
bin is above 10. 
In the case of 1ES\ 1959+650 it is clear that even very short ($\sim 30$\,min) exposure can be used for 
very accurate probing of the source spectrum and reproducing the absorption feature with high details.
In the case of the much weaker M87 emission, while the intrinsic spectrum can be reconstructed without 
the stellar absorption (the emission is at the border of differential sensitivity for 5\,hr exposures), 
the absorption in stellar radiation would render the emission undetectable close to the transit time.
For M87 we perform additional calculations to evaluate at which statistical level the occurrence of 
the absorption can be probed for a given observation time.
For each i-th energy bin that is measurable in the non-absorbed spectrum ($F_{0,i}\pm\Delta F_{0,i}$) 
we compute their flux and uncertainty for a given absorption conditions ($F_{\tau,i}\pm\Delta F_{\tau,i}$).
We then compute:
\begin{equation}
\chi^2 = \sum_{i=1}^N\left(1+\frac{(F_{\tau,i}-F_{0,i})^2}{\Delta F_{\tau,i}^2+\Delta F_{0,i}^2} \right)
\end{equation}
Note that '1' is added in summation to simulate the effect of the fluctuations of the measured fluxes. 
We then compute the corresponding probability as the $\chi^2$ distribution with $N$ degrees of freedom 
integrated above $\chi^2$ (the values for M87 are given in the right columns of Fig.~\ref{fig10}).
Such probability can be interpreted as the chance probability related to detection of an absorption 
feature by future CTA observations.
Curiously, it is easier to detect strong absorption as a sudden lack of detectable flux by CTA, 
however in this case the ability to extract the physical parameters of the absorbing star/system from 
the sub-TeV data would be clearly reduced. 

We investigate if the softening of the $\gamma$-ray spectrum at energies of tens of GeV can be observed 
by \textit{Fermi}-LAT Atwood et al. (2009).
We used the publicly available \emph{Pass8} collection 
area\footnote{\url{https://www.slac.stanford.edu/exp/glast/groups/canda/lat_Performance.htm}} to judge 
the achieved performance for a 10-day exposure of the \textit{Fermi}-LAT instrument.
We find that even for a high flux level comparable to the 1ES\ 1959+650 high state, even without 
the absorption of the $\gamma$ rays in the stellar radiation field, the sensitivity of the instrument 
does not allow to significantly probe the emission of the source above $\sim 10$\,GeV, where most of 
the effect is expected.
Nevertheless, simultaneous observations in GeV range by satellite experiment to 
the observations in the sub-TeV range by Cherenkov telescopes would be still strongly desirable, 
since they can be used to constrain intrinsic source variability. 

It is concluded that
the CTA should easily detect the broad absorption feature between $\sim$30 GeV and $\sim$1 TeV 
(i.e. steepening of the part of the spectrum at lower energies and its hardening at larger energies) 
even during short observations. For bright sources, such effects might be even detected already with 
the current generation of Cherenkov telescopes. 
While, the detection of such effects in the low state emission of radio galaxies is also in reach of CTA, 
it would require to monitor source with deep (a few hours per night) exposures.

\section{Discussion and Conclusion}

We considered the effects of transition of luminous stars (single or within the binary system) 
through the $\gamma$-ray beam produced in the direct surrounding of the SMBH, i.e. either in the SMBH magnetosphere
or in the inner part of the jet. It is assumed that from time to time stars pass close to the line of sight of 
a distant observer.
If the transition is closer than hundreds of stellar radii of a luminous star, the $\gamma$ rays 
from primary $\gamma$-ray beam are partially absorbed. As a result, a broad dip at multi-GeV to sub-TeV 
range should appear in the continuous $\gamma$-ray spectral energy distribution produced around SMBH. 
This dip should be observed by the Cherenkov telescopes as a significant steepening of the spectrum above 
$\sim$10 GeV and as a hardening of the spectrum below $\sim$TeV energies.
Such a combined feature is characteristic, and hence can be easily disentangled from a possible intrinsic 
variability of the $\gamma$-ray beam. 
The time scale of such a transition event should depend on the distance 
of the transiting star from the SMBH. It is predicted to take typically a few to a few tens of days.
Therefore, it is expected that such events might be detected with the help of the current and future Cherenkov 
telescopes. 

As an example, we apply our calculations to two well known active galaxies. 
In the case of the  BL Lac type active galaxy 1ES\ 1959+650, we show that in its high state the effect of 
$\gamma$-ray absorption can be easily observed by the CTA telescopes (under construction) in the sub-TeV 
energy range during a few days 
(see Fig.~10) even with very short nightly exposures. 
We also investigate the case of radio galaxy M87 in the low emission state. 
With enough exposure, CTA would be able to observe a clear hardening of the $\gamma$-ray spectrum at sub-TeV energies.
In fact, for bright sources detection of such an absorption feature might be even possible with the present generation 
of Cherenkov telescopes (H.E.S.S., MAGIC and VERITAS) which sensitivity is nearly an order of magnitude lower than CTA.
Unfortunately, in neither of the two simulated cases \textit{Fermi}-LAT is sensitive enough to detect such 
absorption feature.

Detection of such transiting events (or their lack) should provide interesting constraints on the parameters of 
the central stellar clusters in active galaxies and on the production site of the 
$\gamma$-ray emission in active galaxies. Note that the duration of transition depends on the velocity of 
transiting stars which depends on the distance from SMBH, provided that its mass is known. On the other hand, 
the frequency of detected transiting events should allow us to constrain the distribution function of luminous 
stars in the central stellar cluster.
The temperature of the star in turn affects strongly the depth of the dip in the spectral energy distribution and its position.
In the case of transiting binary systems of two luminous stars, the time structure of transition should
have curious features containing double peak structure with very fast change of the absorption efficiency 
(see Fig.~8).
Observation of such double peaked absorption structures will allow us to constrain the surviving 
frequency of binary stellar systems in the direct vicinity of the SMBH.
Therefore, the observations of transition events via sub-TeV absorption can be used to study the basic parameters 
of the stars in extreme conditions of neighbourhood of SMBH. 

In principle, more than a single star (or a single binary system) can accidentally appear close to the observer's line of sight 
within the jet. Then, the effects considered in the previous section will become much more complicated. Even two absorption features, 
depending on the surface temperatures of the transiting stars, might appear in the gamma-rays spectrum observed from active galaxy. 
Also, predicted $\gamma$-ray light curves will become very complicated. However, we expect that such multiple encounters happen 
very rarely since the stars have to not only stay within the jet but also relatively close to the observer's line of sight.   

As a result of absorption of $\gamma$ rays in the stellar radiation, secondary leptons with sub-TeV energies will
appear in the volume around luminous stars. We expect that those leptons will be effectively removed from the 
direction towards the observer. Their typical Larmor radii are $R_{\rm L} = 3\times 10^8\gamma_{55}/B_{\rm G}$~cm,
where $\gamma_\pm = 3\times 10^5\gamma_{55}$ is the Lorentz factor of leptons, and $B = 1B_{\rm G}$~G  is the magnetic 
field 
strength at the location of the star. In our case the typical Lorentz factors of leptons are expected to be of 
the order of $3\times 10^5$ (note the location of the absorption dip).
The mean free path of leptons on  the synchrotron process in the magnetic field is 
$\lambda_{\rm syn}\approx 7.5\times 10^{12}/(B_{\rm G}^2\gamma_{55})$~cm. 
Their mean free path for energy losses on the IC process in the Thomson regime is 
$\lambda_{\rm IC}\approx 3\times 10^9r_1^2/(T_{4.5}^4\gamma_{55})$~cm, 
where $r_1 = r/10$ with $r$ defined below Eq.~1.
The magnetic field strengths in jets of AGNs at sub-parsec 
distances from the SMBHs are expected to be of the order of a Gauss.  Similar order of the magnetic fields
are also expected in the vicinity of massive stars. In fact, WR and O type stars have the surface magnetic 
fields as strong as $\sim (1-3)\times 10^3$ G on the surface. When we apply such parameters in our calculations, 
then
the Larmor radius of secondary leptons becomes shorter than their typical mean free paths for energy losses on 
the synchrotron and the IC processes. Therefore, the radiation from secondary leptons should be isotropised,
i.e. mainly produced outside the direction towards the observer. Small excess of X-ray synchrotron emission 
should be expected in other directions with typical energies around,
\begin{eqnarray}
\varepsilon\approx m_{\rm e}c^2 (B/B_{\rm cr})\gamma^2\approx 1.2B_{\rm G}\gamma_{55}^2~~~{\rm keV},
\label{eq7}
\end{eqnarray}
\noindent
where $B_{\rm cr} = 4.4\times 10^{13}$ G is the critical magnetic field strength.
On the other hand, energies of $\gamma$ rays from the IC process of the secondary leptons should be comparable 
to those of $\gamma$ rays absorbed in the primary $\gamma$-ray beam, i.e. in the GeV-TeV energy range. 
The small excess of secondary radiation, 
emitted in much broad solid angle, will be rather difficult to 
distinguish from the primary beam of non-thermal emission from the central region of active galaxy.

In principle, considered above absorption features can also appear at about an order of magnitude larger energies 
(TeV $\gamma$-ray energy range) in the case of transiting red hyper- and super-giants. Their typical radii and surface 
temperatures are $R_{\rm RG}\sim 10^3$~R$_\odot$ and $T_{\rm RG}\sim 3000$~K. Then, the optical depth becomes
of the order of unity already at the distance from the star $D\sim 10~R_{\rm RG}$ (see Eq.~1). 
In fact, large number of red giants
is expected around SMHBs. For example, in the case of Cen A the number of red giants around SMBH is estimated on
$\sim 10^6$ (see Wykes et al.~2014 and the estimate in Banasi\'nski et al.~2016).
The $\gamma$-ray emission of blazars is detected at the angle a factor of a few larger than the intrinsic opening 
angle of the jet (e.g. Pushkarev et al.~2009). 
Therefore, red giants can transit outskirts of such a $\gamma$-ray beam, avoiding disruption by the jet pressure 
(as considered by e.g. Barkov et al.~2010), but affecting the broad $\gamma$-ray beam at the TeV energy range 
with their thermal radiation field.

\section*{Acknowledgments}
We would like to thank the referee for useful comments.
This work is supported by the grant through the Polish National Research Centre No. 2019/33/B/ST9/01904.
This research has made use of the CTA instrument response functions provided by the CTA Consortium and 
Observatory, see \url{https://www.cta-observatory.org/science/cta-performance/} (version prod3b-v2) for more details.

\section*{Data Availability}
The simulated data underlying this article will be shared on
reasonable request to the corresponding author.


\label{lastpage}

\end{document}